\documentclass[a4paper,11pt,
superscriptaddress,hidelinks,amsmath,amssymb,aps,nofootinbib,floatfix,prd,raggedbottom,raggedfooter]{revtex4-2}
\usepackage[body={17.5cm, 22cm},right=2cm]{geometry}
\usepackage{color}
\usepackage{xcolor}
\usepackage{amssymb,graphicx}
\usepackage{amsmath}
\usepackage{epsfig}
\usepackage[bookmarks=true,pdfborder={0 0 0}]{hyperref}

\usepackage{mathtools,empheq,slashed,braket,array}
\usepackage{setspace}
\usepackage{subfigure}
\usepackage{hyperref}
\usepackage{tabularx}
\usepackage{multirow}
\usepackage[makeroom]{cancel}

\usepackage{float}

\newcommand\cmt[1]{}
\usepackage{accents}

\usepackage{physics}  
\usepackage{environ}      
\usepackage[section]{placeins}
\newcommand{\leri}[1]{\left(#1 \right)}

\usepackage{ulem}

\newlength{\imagewidth} 
\newlength{\imagelength} 
\newcommand{\includetrimmedgraphics}[6]{
\settowidth{\imagewidth}{\includegraphics{#6}}
\settoheight{\imagelength}{\includegraphics{#6}}
\includegraphics[trim = {#1\imagewidth{}} {#2\imagelength{}} {#3\imagewidth{}} {#4\imagelength{}}, clip=true, scale = #5]{#6}} 

\usepackage[nolist]{acronym}
\begin{acronym}
	\acro{SM}{Standard Model}
	\acro{BSM}{beyond the Standard Model}
	\acro{COM}{center of mass}
	\acro{QED}{Quantum Electrodynamics}
	\acro{EFT}{Effective Field Theory}
	\acro{RHS}{right hand side}
	\acro{LHS}{left hand side}
	\acro{EM}{Electromagnetism}
	\acro{SNR}{signal to noise ratio}
	\acro{CL}{confidence level}
	\acro{VEV}{vacuum expectation value}
	\acro{EWSB}{Electroweak Symmetry Breaking}
	\acro{ULDM}{ultralight dark matter}
	\acro{LFV}{Lepton Flavor Violation}
	\acro{FU}{Flavor Universality}
	\acro{ML}{Machine Learning}
	\acro{NN}{Neural Network}
	\acro{PDF}{probability density function}
	\acro{MLE}{maximum likelihood estimator}
	\acro{NDF}{number density function}
	\acro{MC}{Monte Carlo}
	\acro{NP}{New Physics}
	\acro{LHC}{Large Hadron Collider}
	\acro{NP}{New Physics}
	\acro{ML}{Machine Learning}
	\acro{AI}{Artificial Intelligence}
	\acro{FU}{Flavor Universality}
        \acro{LU}{Lepton Universality}
	\acro{LFUV}{Lepton Flavor Universality Violation}
	\acro{GMFV}{General Minimal Flavor Violation}
	\acro{MFV}{Minimal Flavor Violation}
	\acro{FN}{Froggatt-Nielsen}
	\acro{FCNC}{flavor changing neutral current}
	\acro{ULDM}{ultra-light dark matter}
	\acro{NN}{neural network}
	\acro{PDF}{probability density function}
	\acro{MLE}{maximum likelihood estimator}
	\acro{NDF}{number density function}
	\acro{MC}{Monte Carlo}
        \acro{NPLM}{New Physics Learning Machine}
        \acro{DDP}{Data-Directed Paradigm}
        \acro{BR}{branching ratio}
        \acro{DOF}{degrees of freedom}
\end{acronym}

\begin{document}

\title{Learning New Physics from Data -- a Symmetrized Approach}
\author{Shikma Bressler}
\email{shikma.bressler@weizmann.ac.il}
\affiliation{Department of Particle Physics and Astrophysics, Weizmann Institute of Science, Rehovot 7610001, Israel}

\author{Inbar Savoray}
\email{inbar.savoray@berkeley.edu}
\affiliation{Berkeley Center for Theoretical Physics, University of California, Berkeley, CA 94720, USA}
\affiliation{Physics Division, Lawrence Berkeley National Laboratory, Berkeley, CA 94720, USA}
\author{Yuval Zurgil}
\email{yuval.zurgil@weizmann.ac.il}
\affiliation{Department of Particle Physics and Astrophysics, Weizmann Institute of Science, Rehovot 7610001, Israel}

\begin{abstract}
\begin{acronym}
	\acro{SM}{Standard Model}
	\acro{BSM}{beyond the Standard Model}
	\acro{COM}{center of mass}
	\acro{QED}{Quantum Electrodynamics}
	\acro{EFT}{Effective Field Theory}
	\acro{RHS}{right hand side}
	\acro{LHS}{left hand side}
	\acro{EM}{Electromagnetism}
	\acro{SNR}{signal to noise ratio}
	\acro{CL}{confidence level}
	\acro{VEV}{vacuum expectation value}
	\acro{EWSB}{Electroweak Symmetry Breaking}
	\acro{ULDM}{ultralight dark matter}
	\acro{LFV}{lepton flavor violation}
	\acro{FU}{flavor universality}
	\acro{ML}{machine learning}
	\acro{NN}{neural network}
	\acro{PDF}{probability density function}
	\acro{MLE}{maximum likelihood estimator}
	\acro{NDF}{number density function}
	\acro{MC}{Monte Carlo}
	\acro{NP}{New Physics}
	\acro{LHC}{Large Hadron Collider}
	\acro{AI}{artificial intelligence}
        \acro{LU}{lepton universality}
	\acro{LFUV}{lepton flavor universality violation}
	\acro{PDF}{probability density function}
	\acro{MLE}{maximum likelihood estimator}
	\acro{NDF}{number density function}
        \acro{NPLM}{New Physics Learning Machine}
        \acro{DDP}{Data-Directed Paradigm}
        \acro{BR}{branching ratio}
        \acro{DOF}{degrees of freedom}
\end{acronym}
Thousands of person-years have been invested in searches for \ac{NP}, the majority of them motivated by theoretical considerations. Yet, no evidence of \ac{BSM} physics has been found. This suggests that model-agnostic searches might be an important key to explore \ac{NP}, and help discover unexpected phenomena which can inspire future theoretical developments. A possible strategy for such searches is identifying asymmetries between data samples that are expected to be symmetric within the \ac{SM}. We propose exploiting \acp{NN} to quickly fit and statistically test the differences between two samples. Our method is based on an earlier work, originally designed for inferring the deviations of an observed dataset from that of a much larger reference dataset. We present a symmetric formalism, generalizing the original one; avoiding fine-tuning of the \ac{NN} parameters and any constraints on the relative sizes of the samples. Our formalism could be used to detect small symmetry violations, extending the discovery potential of current and future particle physics experiments.
\end{abstract}

\maketitle

\section{Introduction}

\acresetall
While the \ac{SM} of particle physics has had great empirical success, there are reasons to believe it is incomplete. Among them are the unknown origin of dark matter, neutrino masses and the matter-antimatter asymmetry, as well as examples in which the model parameters seem fine-tuned (e.g. the ``Higgs Hierarchy Problem", the ``Flavor Puzzle" and the ``Strong CP Problem"). This has been a major driving force for the search for \ac{NP} -- new particles and interactions that could naturally solve these problems. While many dedicated searches for theoretically motivated \ac{NP} in particle colliders have yet to identify phenomena \ac{BSM} conclusively, the potential of collider data is far from being thoroughly exhausted. First, there is still much to uncover in the theory space of \ac{NP}, and it is not unlikely that the one ``true" model has not been developed yet. Second, focusing on specific final states, driven by particular models or other considerations, guarantees that many final states will remain unexplored.

The volume, richness and complexity of collider data and its possible theoretical interpretations necessitate a complementary strategy to the dedicated model-specific analyses. These challenges are addressed by, e.g., the \ac{DDP} \cite{Volkovich:2021txe} for model-independent searches. By efficiently scanning the data for deviations from \ac{SM}-predicted properties, model-agnostic searches could identify possible starting points for detailed studies (e.g.~\cite{D_Agnolo_2019,Volkovich:2021txe,Birman:2022xzu}). In particular, symmetries respected by the \ac{SM} set selection rules for, and relations between, different observable processes. Therefore, as previously discussed in~\cite{PhysRevD.90.015025,Lester_2017}, \ac{SM} symmetries could be exploited to identify \ac{NP} directly from data.

The elementary matter particles can be divided into four sectors based on their color charges and electromagnetic charges, with three ``flavors" of particles belonging to each sector, only differing in their masses - up-type quarks ($u,c,t$), down-type quarks ($d,s,b$), charged leptons ($e,\mu,\tau$) and neutral leptons (the three neutrinos). 
The \ac{SM} is approximately ``flavor-blind", resulting in similar predictions for particles of the same sector. Within the \ac{SM}, this symmetry -- referred to as \ac{FU}, is only violated by the Higgs Yukawa interactions. This yields a particular pattern for \ac{SM}-\ac{FU} violation, which is suppressed in scenarios when the mass differences of the different flavors are negligible. The motivation to search for violations of \ac{FU} in collider data is threefold. First -- within the \ac{SM}, \ac{FU} is expected to be maintained to very high accuracies at high energies, thus providing a relatively clean test for \ac{NP}. Second -- it is naturally broken in many extensions of the \ac{SM}. Third -- the observation of neutrino oscillations (see review in~\cite{Workman:2022ynf}) already provides an empirical hint for the violation of \ac{FU} \ac{BSM}\footnote{That cannot be explained by Yukawa interactions or phase space effects.} in the lepton sector.

A few proposals for model-agnostic searches for \ac{LFUV} have been introduced in the literature, with some already implemented by ATLAS~\cite{ATLAS:2016joj,ATLAS:2021tar}. A recent test of \ac{LFUV} has been presented in~\cite{Birman:2022xzu}, where the data is divided into small sub-matrices, and a similarity $N_\sigma$ score of the equivalent sub-matrices in the two samples is calculated. The $N_\sigma$ test is fast and simple to calculate, and has a well-known distribution under the symmetric hypothesis for a large enough number of events, thus not requiring detailed simulations. It is therefore a computationally-cheap search, and easily applicable to large datasets. However, as this method relies on analyzing small sub-sections of the data, it suffers from a large trial factor and, by definition, omits possibly valuable information in other pieces of the data, which are artificially disconnected. It is also unclear how the sub-selections should be chosen in practice, and what implications these would have on performance. We are then interested in exploring other methods for a model-independent analysis, possibly improving sensitivity by simultaneously accessing all data and reducing the number of free parameters. 

\Ac{ML} and \ac{AI} tools are uniquely flexible. They have already been shown to perform well in noisy environments and with partial prior information (e.g.~\cite{dAgnolo:2021aun,Karagiorgi:2022qnh} and references within), and are therefore well-suited for model-agnostic searches. Ref.~\cite{Birman:2022xzu} discusses a complementary method for identifying \ac{LFUV}, using a classifier \ac{NN}, which is trained to minimize the cross-entropy of the mixed sample. On the other hand, likelihood ratio-based statistical tests have the advantage of being optimal tests according to the Neyman-Pearson Lemma while also exhibiting a simple asymptotic behavior of the background-only distribution, as was shown by Wilks~\cite{Wilks:1938dza} and Wald~\cite{10.2307/1990256}. 
Therefore, a rather promising straightforward extension of the methods studied in~\cite{Birman:2022xzu} would be fitting the asymmetries in the data with a \ac{ML}-generated function found by maximizing some likelihood ratio. 

A similar idea for using \ac{ML} for \ac{NP} detection, dubbed ``\ac{NPLM}", has been proposed initially in~\cite{D_Agnolo_2019} and further explored in~\cite{DAgnolo:2019vbw,dAgnolo:2021aun,Letizia:2022xbe}. In these works, an observed dataset is directly compared to a reference dataset that follows the \ac{SM} predictions. \ac{ML} is used for estimating the deviations of the underlying distribution of the observed data from that of the reference data, and the result's significance is determined from the likelihood ratio of the two distributions. A fundamental assumption of the \ac{NPLM} method is that the reference dataset is much larger than the observed dataset, and thus, its statistical fluctuations could be neglected. A large ratio between the sizes of the samples being compared is easily achievable if one of them is generated from a \ac{MC} simulation. However, relying on simulations is not always possible and is generically undesirable when designing a robust tool aimed at scanning large portions of the data. We would then like to examine whether this method could be applied to an entirely data-driven search. As our focus here is on searches for small symmetry violations, we are mainly interested in comparing observed samples that are inherently close in size and associated with comparable uncertainties.

In Section~\ref{subsec:Learning New Physics from a Machine}, we outline the \ac{NPLM} procedure developed in~\cite{D_Agnolo_2019}, and explain its main challenges concerning the task at hand. In Section~\ref{subsec:The Symmetrized Formalism} we present our proposed ``symmetrized formalism", modifying this procedure to make it appropriate for analyzing approximately balanced datasets, as well as improving performance even when the datasets are of different sizes. We demonstrate the performance of the symmetrized formalism on concrete case studies as detailed in Section~\ref{sec:methods}. Our results are presented in Section~\ref{sec:results}, followed by our conclusions in Section~\ref{sec:Conclusions}. Alternative approaches and open questions are addressed in Section~\ref{sec:AAOQ}.

\section{Framework and formalism}

Our main goal is to search for \ac{NP} contributions to some observed processes in a model-agnostic fashion, solely by identifying possible violations of symmetries predicted by the \ac{SM} in data. In particular, we are interested in detecting asymmetries between two (or more) samples, which within the \ac{SM} are expected to have been generated from the same underlying distribution, and thus should only differ by statistical fluctuations. 

Let us consider two datasets, $\mathbf{A}$ and $\mathbf{B}$, of sizes $\tilde{N}_{\mathbf{A}}$ and $\tilde{N}_{\mathbf{B}}$, respectively, consisting of measurements of a $d$-dimensional observable $x$. We assume that each measurement within a dataset is drawn from some common \ac{PDF} $p(x)$, while the total number of measurements is Poisson-distributed\footnote{Although a Poisson-distributed number of events is not necessary for the application of this method, the exact form of the likelihood could vary for different assumptions. See Sec.~\ref{subsubsec:cross_entropy} and Ref.~\cite{Nachman:2021yvi} for a brief discussion.} with an expectation value $N$. The expected \ac{NDF}, $n(x)$, is the \ac{PDF} scaled by the total number of expected events, namely $n(x)\equiv N p(x)$, such that
\begin{align}
N_\mathbf{A}=\int n_\mathbf{A}\leri{x} dx\equiv\int p_\mathbf{A}\leri{x}N_{\mathbf{A}} dx\,,~\label{eq:N_A}\\
N_\mathbf{B}=\int n_\mathbf{B}\leri{x} dx\equiv\int p_\mathbf{B}\leri{x}N_{\mathbf{B}} dx\,.~\label{eq:N_B}
\end{align}
We would like to determine whether $\mathbf{A}$ and $\mathbf{B}$ are drawn from the same distribution, the \textit{symmetric case}, or from different distributions, the \textit{asymmetric case}. Therefore, we are interested in testing the symmetric null hypothesis $p_\mathbf{A}\leri{x} = p_\mathbf{B}\leri{x}$, against the asymmetric alternative hypothesis $p_\mathbf{A}\leri{x} \neq p_\mathbf{B}\leri{x}$. These could also be expressed in terms of the \acp{NDF}, $n_\mathbf{A}\leri{x}$ and $n_\mathbf{B}\leri{x}$, allowing for the incorporation of some additional information on the expected sizes, or ratio of sizes, of the two samples. 

We wish to construct a reasonably quick method for learning these asymmetries directly from data, relying as little as possible on performing dedicated simulations. Ideally, we would like our method to be robust against miss-modelling of the \ac{SM} background, but also expressive enough to identify small signals of various shapes. In Section~\ref{subsubsec:NPLM_procedure} we explore the \ac{NPLM} proposal~\cite{D_Agnolo_2019}, which has been shown to satisfy these requirements when searching for deviations of an observed sample compared to a much larger reference sample. The challenges of this method, stemming from the treatment of the larger sample as an almost exact representation of the background distribution common to the two samples, are discussed in Section~\ref{subsec:NPLM_challenges}. In Sec.~\ref{subsec:The Symmetrized Formalism}, we present our symmetrized formalism tackling these challenges. 

\subsection{Learning New Physics from a Machine}\label{subsec:Learning New Physics from a Machine}

\subsubsection{Procedure}\label{subsubsec:NPLM_procedure}
Within the \ac{NPLM} framework~\cite{D_Agnolo_2019}, sample $\mathbf{A}$ is considered as some observation of an experiment, which one is interested in determining whether it contains evidence for \ac{BSM} physics. The sample $\mathbf{B}$ is assumed to be much larger, $N_\mathbf{A}\ll N_\mathbf{B}$, and is a result of a \ac{MC} simulation of the same process within the \ac{SM}, or deduced using other background estimation techniques. Assuming that the difference between the \acp{PDF} generating $\mathbf{A}$ and $\mathbf{B}$ is small, Ref.~\cite{D_Agnolo_2019} proposes to test for the deviations of their ratio from unity.
For this purpose, one may construct a maximum-likelihood test for explaining the observed data $\mathbf{A}$, testing the null hypothesis $\mathcal{H}_0$ under which the \ac{NDF} $n_\mathbf{A}$ is
\begin{align}
\mathcal{H}_0 : \quad \quad \quad n_\mathbf{A}\leri{x|\mathcal{H}_0} &= \frac{N_\mathbf{A}}{N_\mathbf{B}}n_\mathbf{B}\leri{x,\nu}\,,~\label{eq:H_0_A}
\end{align}
against the alternative hypothesis $\mathcal{H}_1$ under which $n_\mathbf{A}$ is
\begin{align}
\mathcal{H}_1 : \quad \quad \quad n_\mathbf{A}\leri{x|\mathcal{H}_1} &= \frac{N_\mathbf{A}}{N_\mathbf{B}}e^{f\leri{x,\mu}} n_\mathbf{B}(x,\nu)\,,~\label{eq:H_1_A}
\end{align}
where $\nu$ represents the nuisance parameters, and $\mu$ represents the parameters of interest. Note that $\mathcal{H}_0$ and $\mathcal{H}_1$ are nested, since $n_\mathbf{A}\leri{x|\mathcal{H}_0}=n_\mathbf{A}\leri{x|\mathcal{H}_1,f=0}$. The corresponding test statistic is then
\begin{align}
t=2\log\leri{\frac{\text{max}_{\nu,\mu}\leri{\mathcal{L}\leri{\mathcal{H}_1|\mathbf{A}}}}{\text{max}_{\nu}\leri{\mathcal{L}\leri{\mathcal{H}_0|\mathbf{A}}}}}\,,
\end{align}
where the likelihoods are given by the extended likelihood functions
\begin{align}
\mathcal{L}\leri{\mathcal{H}|\mathbf{A}}=\frac{e^{-N_\mathbf{A}\leri{\mathcal{H}}}}{\tilde{N}_\mathbf{A}!}\prod_{x\in\mathbf{A}} n_\mathbf{A}(x|\mathcal{H})\,,~\label{eq:ext_likelihood}
\end{align}
with
\begin{align}
N_\mathbf{A}\leri{\mathcal{H}}=\int n_\mathbf{A}(x|\mathcal{H}) dx\,.
\end{align}
Denoting the \acp{MLE} parameters by $\hat{ \boldsymbol{\cdot}}$, then
\begin{align}
t=2\leri{\hat{N}_\mathbf{A}\leri{\mathcal{H}_0}-\hat{N}_\mathbf{A}\leri{\mathcal{H}_1}+\log\leri{\prod_{x\in\mathbf{A}} \frac{\hat{n}_\mathbf{A}(x|\mathcal{H}_1)}{\hat{n}_\mathbf{A}(x|\mathcal{H}_0)}}}\,.
\end{align}
One may parameterize 
\begin{align}
\hat{n}_\mathbf{A}(x|\mathcal{H}_1) = e^{\hat{f}\leri{x}} \hat{n}_\mathbf{A}(x|\mathcal{H}_0)\,,
\end{align}
and obtain using Eq.~\eqref{eq:N_A}
\begin{align}
t=2\leri{-\int \leri {e^{\hat{f}\leri{x}}-1} \hat{n}_\mathbf{A}(x|\mathcal{H}_0) dx +\sum_{x\in\mathbf{A}} \hat{f}\leri{x}}\,.~\label{eq:t_int}
\end{align}
Recall that the null hypothesis suggests that the observed sample $\mathbf{A}$ and the reference sample $\mathbf{B}$ were drawn from the same distribution. However, sample $\mathbf{B}$ was assumed to be much larger than $\mathbf{A}$, and thus statistically more reliable. Therefore, the authors of~\cite{D_Agnolo_2019} replaced the weighted integral over $\hat{n}(x|\mathcal{H}_0) dx$ by an empirical summation over the reference sample $\mathbf{B}$, yielding
\begin{align}
t=t_\mathbf{B}\leri{\mathbf{A}} \equiv-2\leri{\frac{\hat{N}_{\mathbf{A}}\leri{\mathcal{H}_0}}{\tilde{N}_{\mathbf{B}}}\sum_{x\in \mathbf{B}} \leri {e^{\hat{f}\leri{x}}-1} -\sum_{x\in\mathbf{A}} \hat{f}\leri{x}}\,,\label{eq:t_nref}
\end{align}
where $\hat{N}_{\mathbf{A}}\leri{\mathcal{H}_0}$ is the \ac{MLE} for the number of events in sample $\mathbf{A}$ under the null hypothesis (which can either be a parameter of the fit, or constrained to some expected value).

 The function $f$ could be parameterized in many different ways, for example by a polynomial of some degree, or a bin-wise function. A useful parameterization of $f$, as explained by the authors of~\cite{D_Agnolo_2019}, is via the output of a \ac{NN}. This parameterization has the advantage of being highly expressive, while also being continuous and smooth. In this work, we follow~\cite{D_Agnolo_2019} and consider a fully-connected \ac{NN}, with one hidden layer of $N_\text{neu}$ neurons, which accepts a one-dimensional variable $x$ and outputs a one-dimensional value for the function $f\leri{x}$. The resulting function $f\leri{x}$ is given by
 \begin{align}
 f\leri{x} = b_{\rm out}+\sum_{\alpha = 1}^{N_\text{neu}} w^\alpha_{\rm out} \sigma\leri{w_\alpha x +b_\alpha}\,,\label{eq:fNN}
\end{align}
where $\sigma\leri{z} = 1/\leri{1+e^{-z}}$ is the logistic sigmoid function. For sufficiently large values of the weights $w$\,, the sigmoid approaches a step function, with a gradient roughly set by $w^\alpha_{\rm out}w_\alpha$. A sum of two sigmoids, as shown in~\cite{D_Agnolo_2019}, can produce a ``bump". This is particularly suitable for resonant \ac{NP} detection, or for marking an area in $x$ space that contains some asymmetry. While a more complex or less complex \ac{NN} (or other types of \ac{ML} functions, as in~\cite{Letizia:2022xbe,Grosso:2023ltd}) could be used, here we stick to this simple choice for concreteness, and note it could be optimized for different problems. 

The likelihood ratio test statistic $t$ is obtained from Eq.~\eqref{eq:t_nref}, by taking the term in its parentheses as a loss function, and training the \ac{NN} to find the parameters $w$ and $b$ that minimize the loss. Then, the parameters of the \ac{NN} at the end of the training would set the function $f\leri{x}$, which indicates the deviations of the data from the \ac{SM} expectation. Since the loss at the end of the training is simply $-0.5 t$, it is used to calculate the significance of the result, once the distribution of $t$ under the null hypothesis is known.

Strictly speaking, the null hypothesis distribution of the test statistic should be obtained from a large number of toy datasets sampled from the null hypothesis. However, the likelihood-ratio test score is known to exhibit an asymptotic behavior at the limit of infinite data. In particular, under the null hypothesis the distribution of $t$ is known to asymptote a $\chi^2_{n}$ distribution, regardless of the background-only $p(x)$. The number of \ac{DOF} $n$ is the number of free parameters in the alternative hypothesis minus the number of free parameters in the null hypothesis~\cite{Wilks:1938dza}. We discuss the number of \ac{DOF} and the expected asymptotic distribution in detail in Sec.~\ref{subsec:dof}. This asymptotic property is extremely beneficial when designing a robust and computationally efficient test that could be applied to many different regions of the data, as it would not require generating different simulations to interpret the test result.

\subsubsection{Challenges}\label{subsec:NPLM_challenges}

\begin{figure}
\centering
\includetrimmedgraphics{0.038}{0.02}{0.47}{0.4}{0.274}{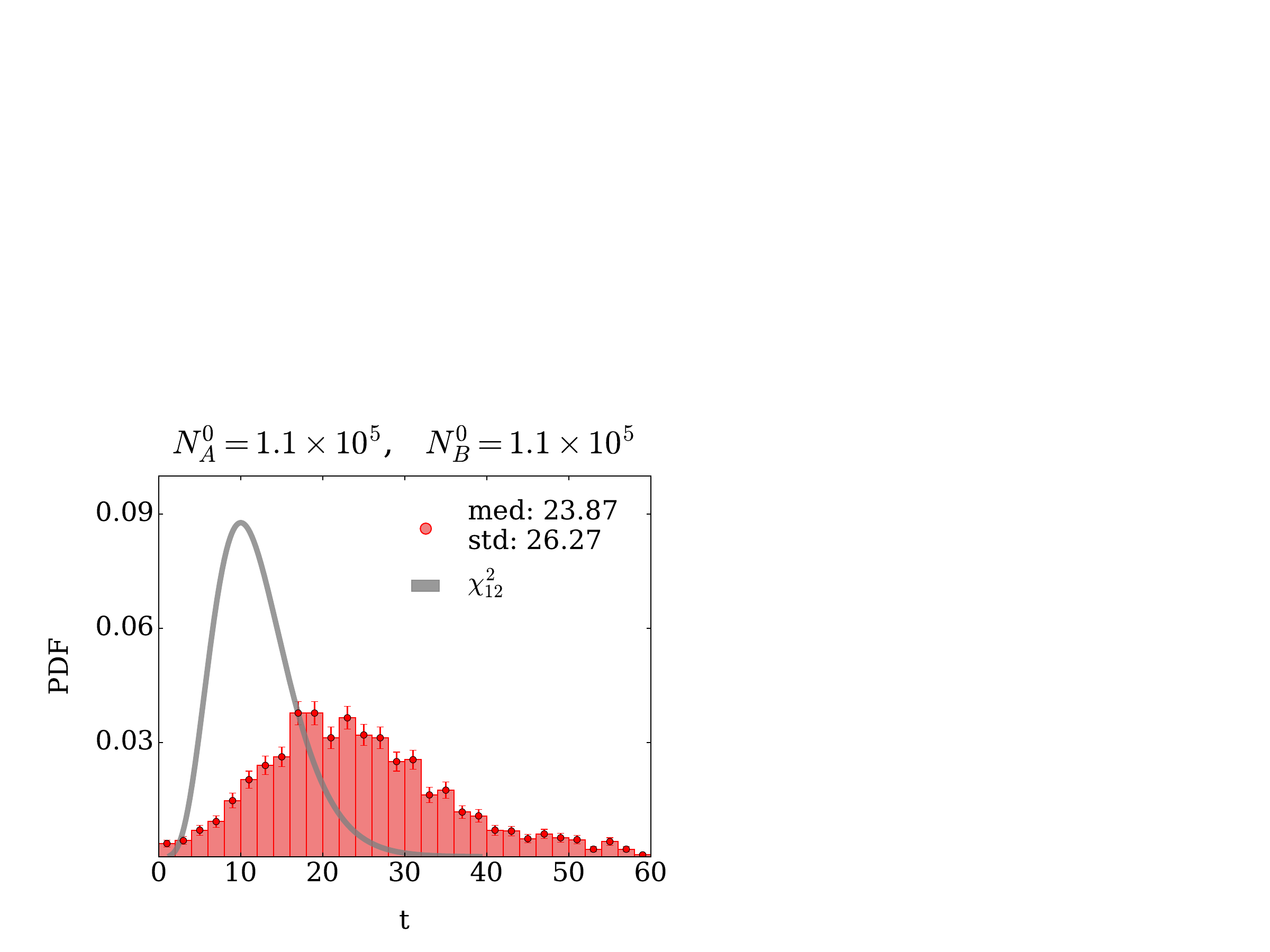}
\includetrimmedgraphics{0.038}{0.02}{0.47}{0.4}{0.274}{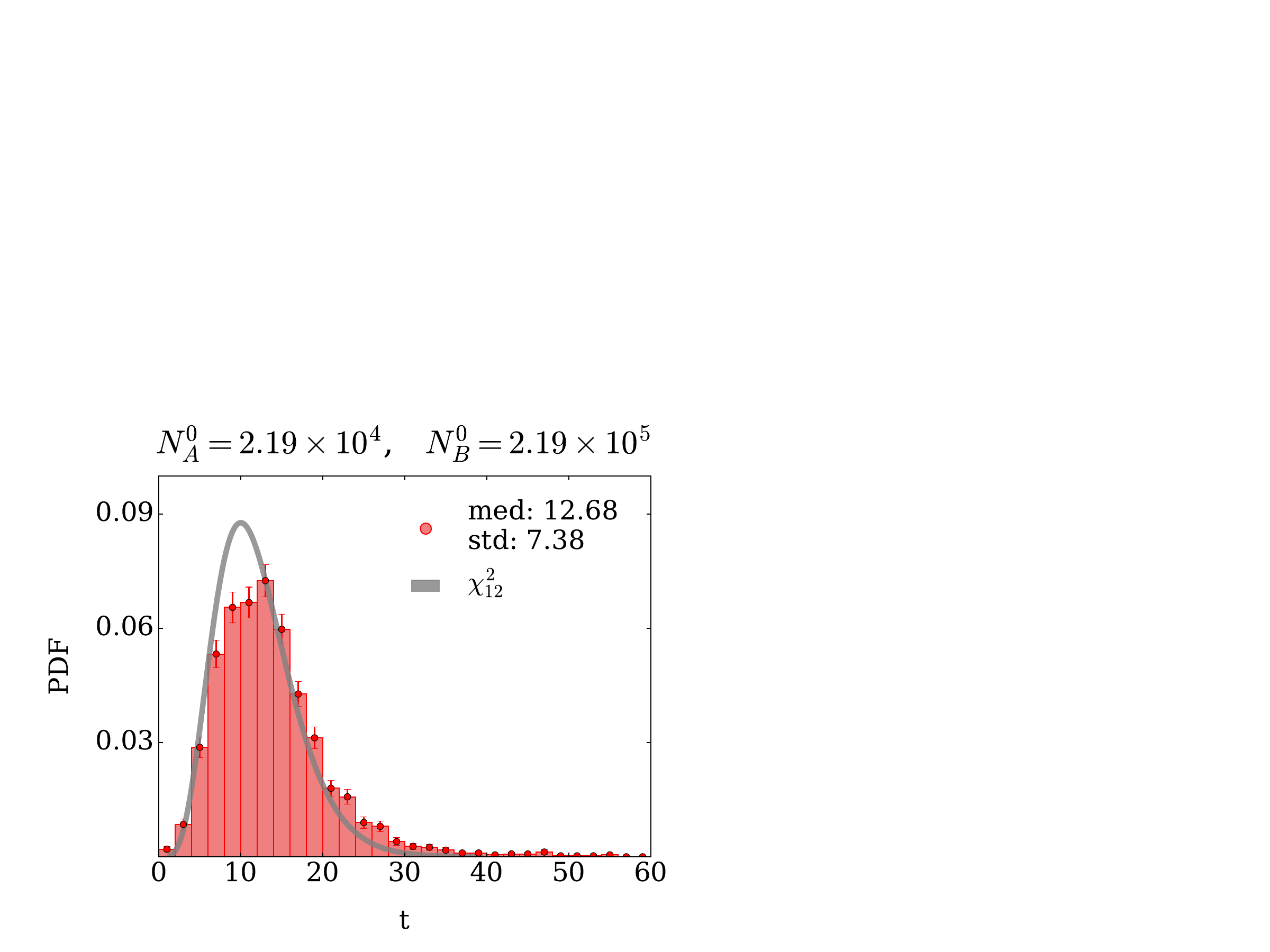}
\includetrimmedgraphics{0.038}{0.02}{0.47}{0.4}{0.274}{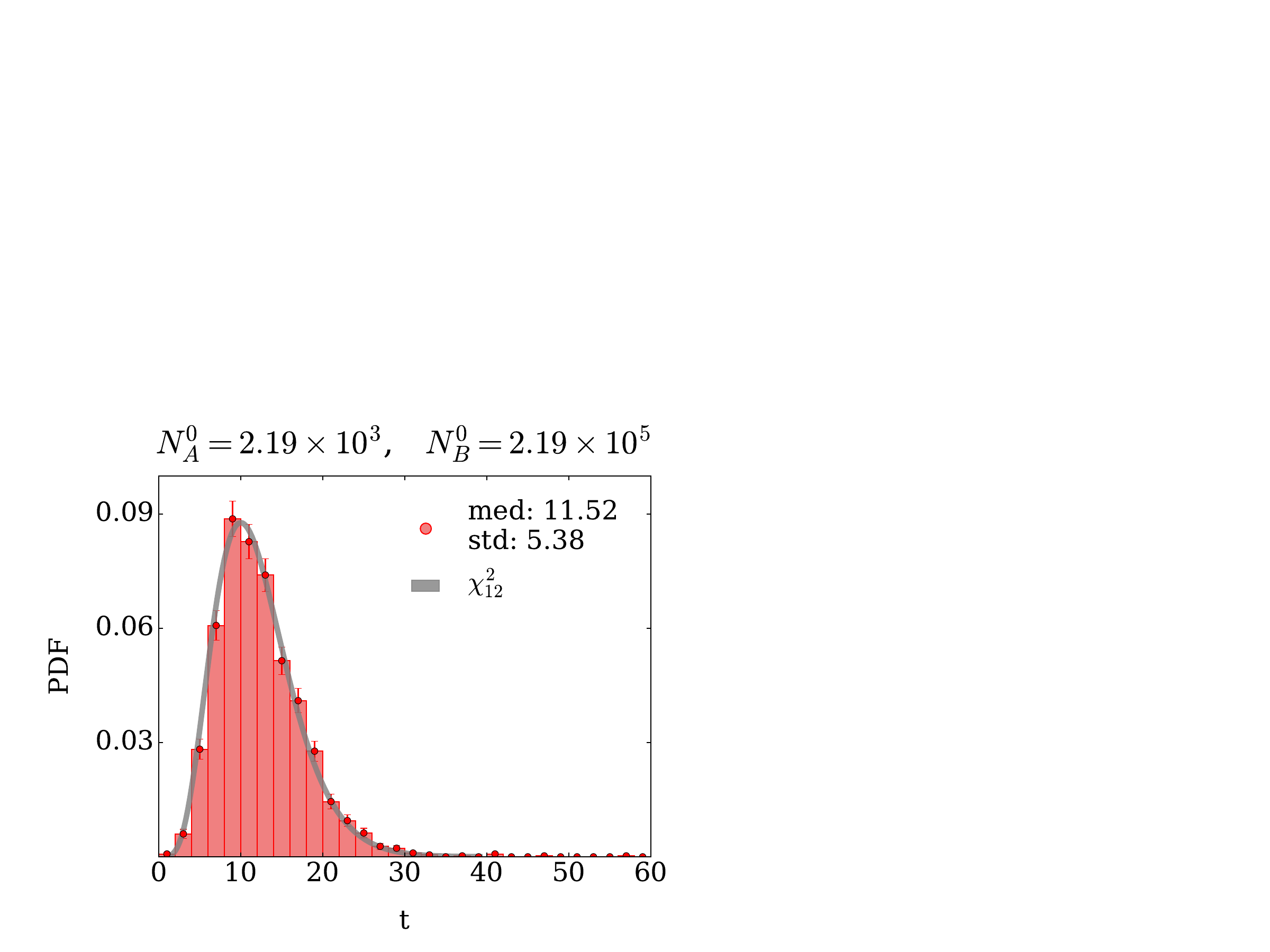}
\\
\includetrimmedgraphics{0.038}{0.02}{0.47}{0.4}{0.274}{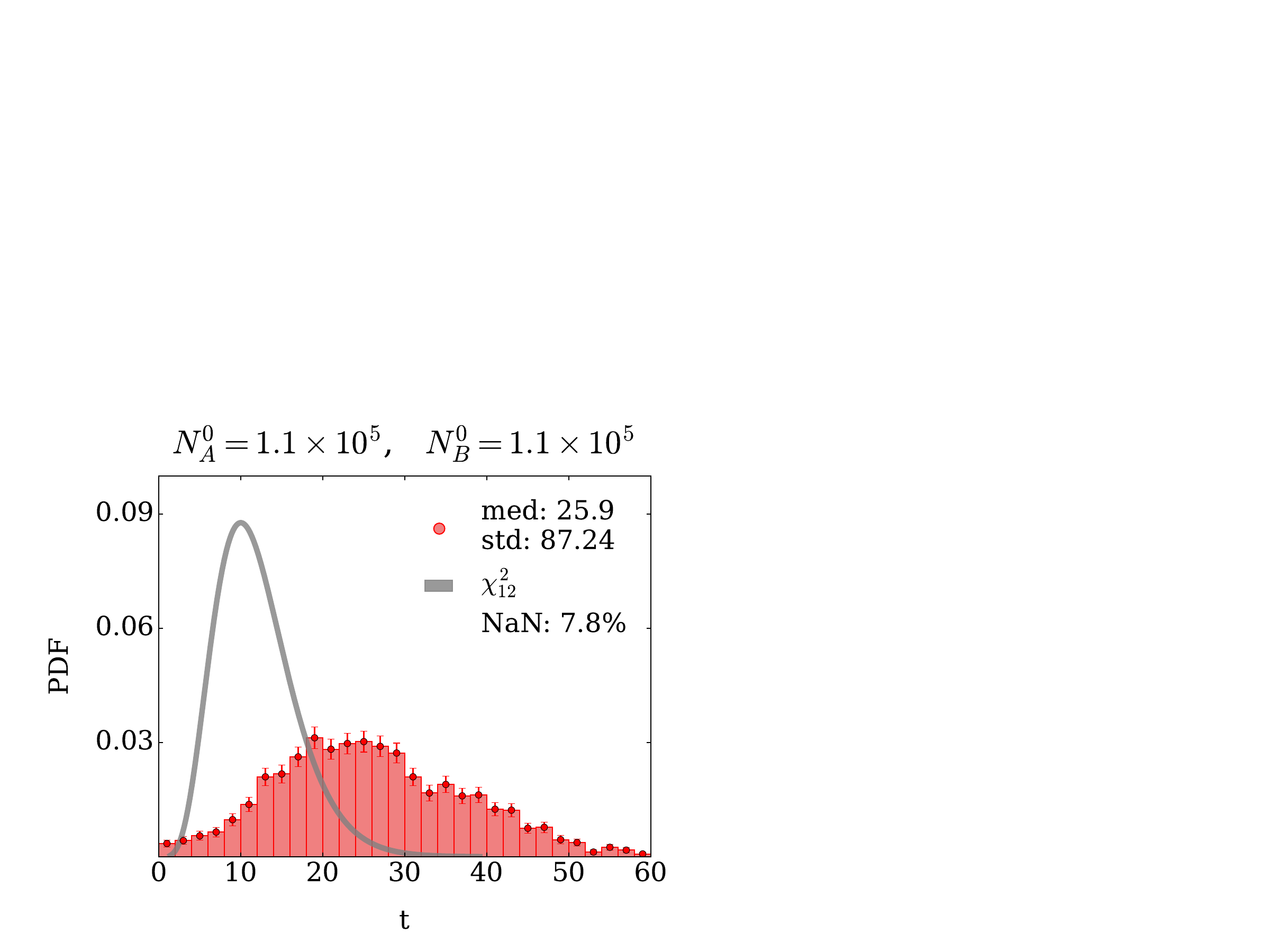}
\includetrimmedgraphics{0.038}{0.02}{0.47}{0.4}{0.274}{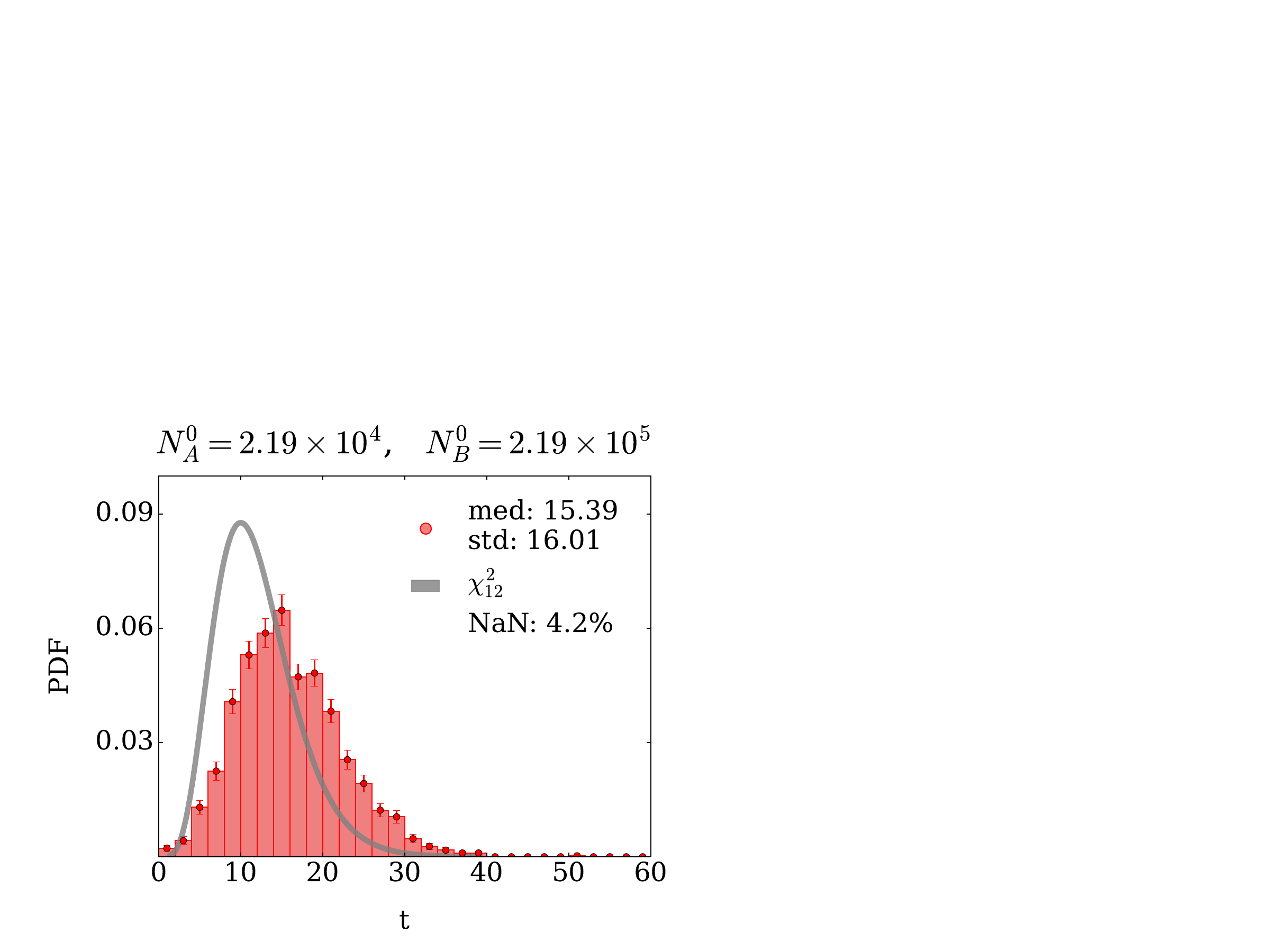}
\includetrimmedgraphics{0.038}{0.02}{0.47}{0.4}{0.274}{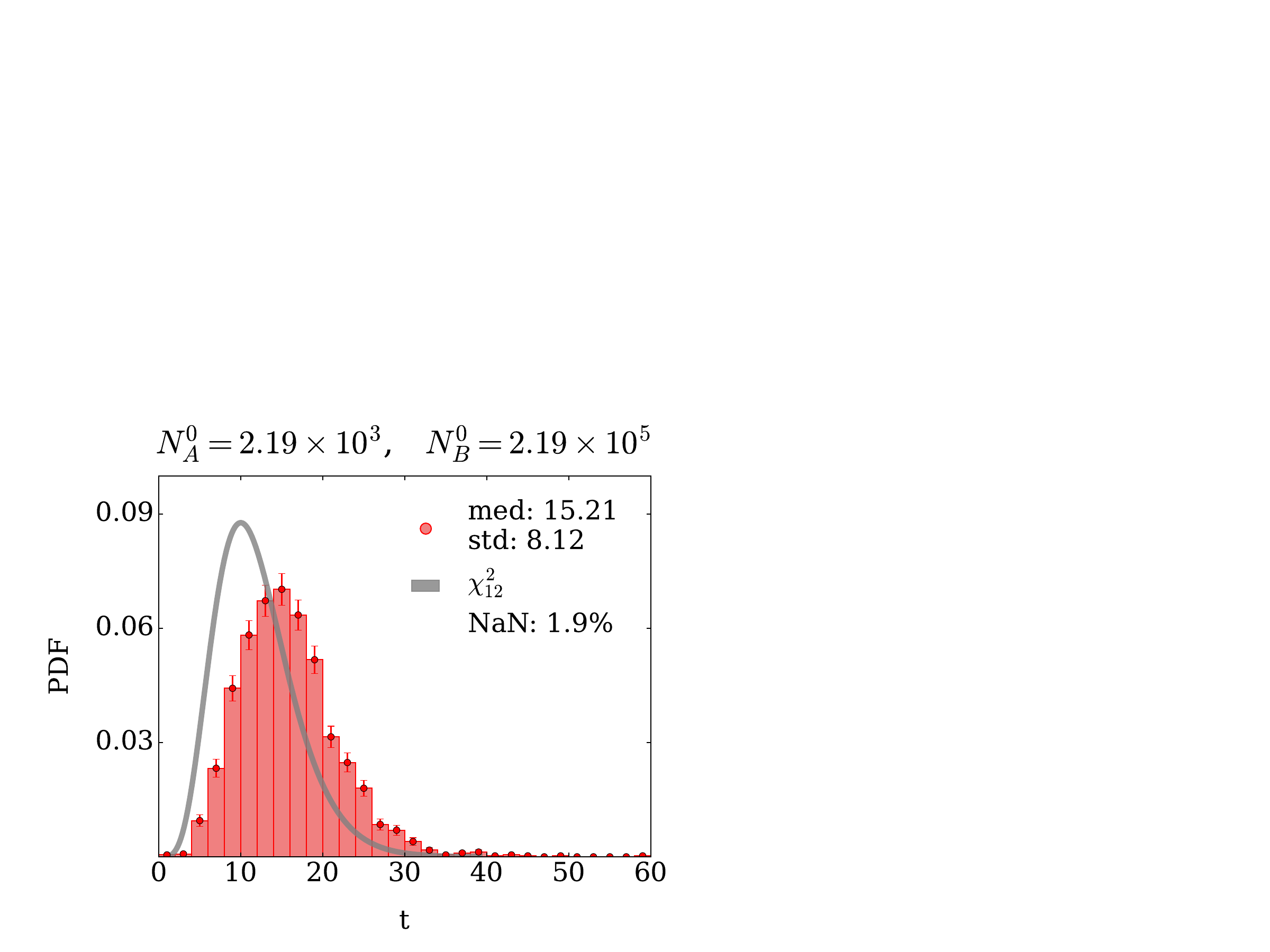}
 \caption{The distribution of the \ac{NPLM} test statistic $t_\mathbf{B}\leri{\mathbf{A}}$ under the null hypothesis. Left -- $N_\mathbf{B} = N_\mathbf{A} = N/2$, middle -- $N_\mathbf{A} = N/10$ and $N_\mathbf{B} = N$, right -- $N_\mathbf{A} = N/100$ and $N_\mathbf{B} = N$ with $N\approx 2.2\times 10^{5}$. Top - weight clipping of 9, bottom - no weight clipping (the fraction of toys resulted in infinite $t_\mathbf{B}\leri{\mathbf{A}}$ is given as well). Solid gray -- the expected $\chi^2_{12}$ distribution according to the Wilks-Wald theorem. Both sample $\mathbf{A}$ and sample $\mathbf{B}$ were drawn from the exponential distribution $b_0$ in Eq.~\eqref{eq:b0}.}
	\label{fig:t_A_null_dist}
\end{figure}

As discussed in~\cite{D_Agnolo_2019,DAgnolo:2019vbw,dAgnolo:2021aun}, the \ac{NPLM} method performs very well under two conditions.
The first, and the most relevant for our case, is the requirement of a large ratio between the sizes of the two samples. However, as we are interested in looking for small symmetry violations, we expect sample $\mathbf{A}$ and sample $\mathbf{B}$ to be of similar sizes. The implications of using the \ac{NPLM} method for different ratios of sample sizes are shown in Fig.~\ref{fig:t_A_null_dist}. The histograms shown in Fig.~\ref{fig:t_A_null_dist} represent the distribution of $t_{\mathbf{B}}\leri{\mathbf A}$ in Eq.~\eqref{eq:t_nref}, when the two samples were drawn from the same exponential distribution, as described in Sec.~\ref{subsec:modeling}. As shown in Fig.~\ref{fig:t_A_null_dist}, while the asymptotic property is maintained for 
a large $N_\mathbf{B}/N_\mathbf{A}$ ratio of 100, a significant deviation from it is found for samples of similar sizes.

The second limitation of the \ac{NPLM} method is the need for tightly constraining the weights of the \ac{NN} to achieve the asymptotic $\chi^2$ distribution. While \acp{NN} are very flexible, it is also well known that they are prone to overfit the data, easily producing sharp features tailored to a small number of data points. One method for regularizing the \ac{NN} is weight clipping -- limiting the weights $w$ to be smaller than some value $W$, and thus forbidding divergences of $f\leri{x}$ on scales $\Delta x\leq 1/W$~\cite{D_Agnolo_2019}. As shown in Fig.~\ref{fig:t_A_null_dist}, and in agreement with the results in~\cite{D_Agnolo_2019}, a weight clipping of 9 was required to match the asymptotic distribution for $N_\mathbf{B}/N_\mathbf{A}=100$, and it is expected to be more restricting for samples of similar sizes. 

Although this issue is generic to \ac{NN} models, especially with a large number of parameters, we note it is particularly severe for the \ac{NPLM} method. As noticed by the authors, the loss corresponding to Eq.~\eqref{eq:t_nref} is unbounded from below. Meaning, if $f\leri{x_\star}\rightarrow \infty$ for a point $x_\star\in  \leri{\mathbf{A}-\mathbf{B}\cap\mathbf{A}}$, the loss tends to negative infinity. Note that the derivative of the loss with respect to the \ac{NN} weights is also unbounded, which makes this ``slippery slope" almost irresistible to a gradient-descent-based algorithm. In order to avoid this divergence of the loss, a strict weight clipping must be implemented to ensure the network never accidentally identifies a point $x_\star$. Indeed, as can be seen in Fig.~\ref{fig:t_A_null_dist}, while a few percent of the trainings yielded an infinite loss when no weight clipping was applied, no divergences were reached with a weight clipping of 9. However, the tighter the weight clipping is, the less flexible the \ac{NN} becomes, impairing its ability to identify narrow signals.

The main reason for both these challenges is the treatment of the reference sample as an exact representation of the background distribution. Practically, this assumption is incorporated into the transition from Eq.~\eqref{eq:t_int} to Eq.~\eqref{eq:t_nref}, 
and it is advantageous because it makes it unnecessary to find a closed form for the null hypothesis distribution. It is equivalent to testing the null hypothesis 
\begin{align}
    \mathcal{H}_0 &: \quad \quad \quad \label{eq:NPLM_H_0}
    \begin{aligned} 
        n_\mathbf{A}\leri{x} &= \frac{N_\mathbf{A}}{N_\mathbf{B}}n_\mathbf{B}\leri{x} \,,\\ n_\mathbf{B}\leri{x} &= \tilde{n}_\mathbf{B}\leri{x} \,,
    \end{aligned}
    \end{align}
    against the alternative hypothesis
    \begin{align}
    \mathcal{H}_1 &: \quad \quad \quad  
    \begin{aligned} 
        n_\mathbf{A}\leri{x} &= \frac{N_\mathbf{A}}{N_\mathbf{B}}e^{f\leri{x}}n_\mathbf{B}\leri{x}\,, \\ n_\mathbf{B}\leri{x} &= \tilde{n}_\mathbf{B}\leri{x} \,,\label{eq:NPLM_H_1}
    \end{aligned}
\end{align}
where $\tilde{ \boldsymbol{\cdot}}\leri{x}$ denotes the \textbf{observed} value. The null and alternative hypotheses share the assumption that the \ac{NDF} from which sample $\mathbf{B}$ was drawn, $n_\mathbf{B}\leri{x}$, is the observed reference sample \ac{NDF}, $\tilde{n}_\mathbf{B}\leri{x}$. Although one expects this assumption to cancel in a likelihood ratio calculated over sample $\mathbf{B}$, it implicitly makes its way into the likelihood ratio calculated over sample $\mathbf{A}$. Given $\mathcal{H}_0$ and $\mathcal{H}_1$ above, the choice $n_\mathbf{B}\leri{x}=\tilde{n}_\mathbf{B}\leri{x}$ can be interpreted as fitting the common \ac{PDF} by maximizing its likelihood over the reference sample, where the parameter space being scanned includes an \textbf{exact} fit to it. This is the global \ac{MLE} for the reference sample in the absence of other constraints, but not the global \ac{MLE} for the symmetry hypothesis $p_\bold{A}=p_\bold{B}$.

Let us try to explain the two challenges of the \ac{NPLM} method in light of this observation. First, the choice of fitting the common \ac{PDF} from the reference sample alone is only justified if its statistical fluctuations are negligible compared to those of the observed sample. In the next section, we show that this approach follows the maximal likelihood principle only when the ratio $N_{\bold B}/N_{\bold A}$ is large. Therefore, a deviation from the asymptotic $\chi^2$ distribution is expected when the samples are balanced.

Second, the divergence of the \ac{NPLM} loss is indeed not a generic result of over-fitting, but an implication of the chosen null hypothesis.
Note that if a point $x_\star$ exists, for which $\tilde{n}_\mathbf{B}\leri{x_\star}=0$ but $\tilde{n}_\mathbf{A}\leri{x_\star}\neq0$, its likelihood under the \ac{NPLM} null hypothesis in Eq.~\eqref{eq:NPLM_H_0} is by definition zero, and thus the likelihood ratio diverges. Consequently, there is no finite solution to the equation $n_\mathbf{A}\leri{x_\star} \propto e^{f\leri{x_\star}}\tilde{n}_\mathbf{B}\leri{x_\star}$, and thus $f\leri{x_\star}$ diverges. \footnote{Essentially, $f\leri{x}$ is locally undefined on $x_\star$, and keeping it finite would imply that the likelihood of the alternative hypothesis is zero as well.}Had the \ac{SM} distribution been known analytically a priory, and it would have predicted events in $x_\star$ to be forbidden, their observation would completely rule out the \ac{SM} prediction. However, in the \ac{NPLM} case, the \ac{SM} distribution is not known, but is being deduced from data. The null hypothesis is model-agnostic, and is instead assuming a symmetry implying the equality of the underlying \acp{PDF} of the two samples. Of course, even had $\mathbf{A}$ and $\mathbf{B}$ been drawn from the same distribution, there could be points $x_\star$ randomly occurring in one sample but not in the other. It is thus clear that the \ac{NPLM} null hypothesis is not equivalent to the $p_{\mathbf{A}}= p_{\mathbf{B}}$ hypothesis. 

While the weight clipping scheme makes an attempt to correct these issues, it is doing so in an inconsistent and computationally intensive way. In the next section, we present the ``symmetrized formalism", generalizing the \ac{NPLM} method. As we will show, the symmetrized formalism introduces a natural and mathematically sound solution to both the challenges of the \ac{NPLM} method.

\subsection{The Symmetrized Formalism}\label{subsec:The Symmetrized Formalism}

The \ac{NPLM} formalism explored in~\cite{D_Agnolo_2019,DAgnolo:2019vbw,dAgnolo:2021aun} is aimed at testing the compatibility of an observed dataset $\mathbf{A}$ with a much larger reference dataset $\mathbf{B}$, representing the background distribution. However, the reference dataset $\mathbf{B}$, being of finite size, is nothing but a finite sample from its underlying distribution. Therefore, even had the underlying distributions of $\mathbf{A}$ and $\mathbf{B}$ been identical, both samples would fluctuate between repeated toy-experiments. Thus, we propose to construct a likelihood test to simultaneously fit both sample $\mathbf{A}$ and sample $\mathbf{B}$.
While this is a natural choice when the data samples are of similar sizes, and thus should be treated on equal footing\footnote{If the two samples are associated with different systematic uncertainties, these should be included in their likelihoods. While we do not discuss this case here, we intend to address it in future work. See e.g.~\cite{dAgnolo:2021aun} for addressing systematics within the \ac{NPLM} framework.}, it is never an invalid one, regardless of the size of the samples. 

The corresponding test statistic is then
\begin{align}
t=2\log\leri{\frac{{\text{max}_{\mu, \nu}}\leri{\mathcal{L}\leri{\mathcal{H}_1|\mathbf{A},\mathbf{B}}}}{{\text{max}}_{\nu}\leri{\mathcal{L}\leri{\mathcal{H}_0|\mathbf{A},\mathbf{B}}}}} =2\log\leri{\frac{{\text{max}_{\mu, \nu}}\leri{\mathcal{L}\leri{\mathcal{H}_1|\mathbf{A}}\mathcal{L}\leri{\mathcal{H}_1|\mathbf{B}}}}{{\text{max}}_{\nu}\leri{\mathcal{L}\leri{\mathcal{H}_0|\mathbf{A}}\mathcal{L}\leri{\mathcal{H}_0|\mathbf{B}}}}} \label{eq:t_symm}\,,
\end{align}
with the extended likelihoods as in Eq.~\eqref{eq:ext_likelihood}.
Since the null hypothesis $\mathcal{H}_0$ is that $\mathbf{A}$ and $\mathbf{B}$ were drawn from the same \ac{PDF}, where the alternative hypothesis allows their \acp{PDF} to be independent, we may heuristically re-express $t$ as
\begin{align}
t =2\log\leri{\frac{{\text{max}_{p_\mathbf{A}, p_\mathbf{B} }}\leri{\mathcal{L}\leri{N_\mathbf{A}, p_\mathbf{A}\leri{x}|\mathbf{A}}\mathcal{L}\leri{N_\mathbf{B}, p_\mathbf{B}\leri{x}|\mathbf{B}}}}{{\text{max}}_{p_0}\leri{\mathcal{L}\leri{N_\mathbf{A}, p_0\leri{x}|\mathbf{A}}\mathcal{L}\leri{N_\mathbf{B}, p_0\leri{x}|\mathbf{B}}}}}\,.
\end{align} 
It is then clear that if one sample is much larger than the other, it will dominate the maximization of the null hypothesis likelihood $\mathcal{L}\leri{\mathcal{H}_0|\mathbf{A,B}}$. If $N_\mathbf{B}\gg N_\mathbf{A}$, one would obtain $\hat{p}_0\leri{x}\approx \hat{p}_\mathbf{B}\leri{x}$, and get back the one sample likelihood
\begin{align}
t_{N_\mathbf{B}\gg N_\mathbf{A}}\rightarrow 2\log\leri{\frac{{\text{max}_{p_\mathbf{A} }}\leri{\mathcal{L}\leri{N_\mathbf{A}, p_\mathbf{A}\leri{x}|\mathbf{A}}}}{\mathcal{L}\leri{N_\mathbf{A}, \hat{p}_\mathbf{B}\leri{x}|\mathbf{A}}}}\,,
\end{align}
in agreement with the \ac{NPLM} test in~\cite{D_Agnolo_2019}, presented in the previous section. In the balanced case, $N_\mathbf{B}\approx N_\mathbf{A}$, the \ac{NPLM} assumption is therefore inappropriate. The ``symmetrized" test in Eq.~\eqref{eq:t_symm}, on the other hand, is generic, and can be used regardless of the ratio $N_\mathbf{A}/N_{\mathbf{B}}$. 

Let us now choose a specific parameterization of $\mathcal{H}_0$ and $\mathcal{H}_1$. While this is not the only possible choice, we follow the \ac{NPLM} spirit and assign
\begin{align}
\mathcal{H}_0 &: \quad \quad \quad 
\begin{aligned} \label{eq:H_0}
n_\mathbf{A}\leri{x} &= \frac{\tilde{N}_\mathbf{A}}{\int n_{\mathcal{R}}\leri{x}dx}e^{h\leri{x}}n_{\mathcal{R}}\leri{x} \\ n_\mathbf{B}\leri{x} &= \frac{\tilde{N}_\mathbf{B}}{\int n_{\mathcal{R}}\leri{x}dx}e^{h\leri{x}+r}n_{\mathcal{R}}\leri{x} \,, 
\end{aligned} \\ \nonumber\\
\mathcal{H}_1 &: \quad \quad \quad  
\begin{aligned} \label{eq:H_1}
n_\mathbf{A}\leri{x} &= \frac{\tilde{N}_\mathbf{A}}{\int n_{\mathcal{R}}\leri{x}dx}e^{f\leri{x}}n_\mathcal{R}\leri{x} \\ n_\mathbf{B}\leri{x} &= \frac{\tilde{N}_\mathbf{B}}{\int n_{\mathcal{R}}\leri{x}dx}e^{g\leri{x}}n_\mathcal{R}\leri{x} \,,
\end{aligned}
\end{align}
where $n_\mathcal{R}\leri{x}$ is some reference \ac{NDF} of our choosing, $r$ is a constant controlling the ratio between the expected number of events in $\mathbf{B}$ and $\mathbf{A}$, and $f\leri{x},g\leri{x}$ and $h\leri{x}$ will again be outputs of \acp{NN}. Then, the test statistic is given by
\begin{align}
t=2\log\leri{\frac{\underaccent{f\leri{x},g\leri{x}}{\text{max}}\leri{\mathcal{L}\leri{\mathcal{H}_1|\mathbf{A},\mathbf{B}}}}{\underaccent{h\leri{x},r}{\text{max}}\leri{\mathcal{L}\leri{\mathcal{H}_0|\mathbf{A},\mathbf{B}}}}}\,.
\end{align}Note that if we choose $n_\mathcal{R}=\tilde{n}_\mathbf{B}$, i.e. the observed distribution of sample $\mathbf{B}$, in the $N_\mathbf{B}\gg N_\mathbf{A}$ limit we get back the test statistic $t_{\mathbf{B}}\leri{\mathbf{A}}$ in Eq.~\eqref{eq:t_nref}.

Instead of using the observed distribution of one of the samples as our reference distribution, a natural choice would be to choose the observed distribution of the combined sample, $n_\mathcal{R}\leri{x} = \leri{\tilde{n}_\mathbf{A}\leri{x}+\tilde{n}_\mathbf{B}\leri{x}}$. This choice has a few motivations, as we now show. Under the symmetric hypothesis, both samples are drawn from the same distribution. Meaning, we can add them together to yield a sample of the null hypothesis \ac{PDF} of size $\tilde{N}_\mathbf{B}+\tilde{N}_\mathbf{A}$. As a result, the maximization of the null likelihood is analytic, and does not require additional numerical training. Explicitly, 
\begin{align} -2\log\leri{\underaccent{h\leri{x},r}{\text{max}}\leri{\mathcal{L}\leri{\mathcal{H}_0|\mathbf{A},\mathbf{B}}}}=2\sum_{x\in\mathbf{A},\mathbf{B}}\left[\frac{1}{\tilde{N}_{\mathbf{A}}+\tilde{N}_{\mathbf{B}}}\leri{\tilde{N}_\mathbf{A}e^{\hat{h}\leri{x}}+\tilde{N}_\mathbf{B}\leri{e^{\hat{h}\leri{x}+\hat{r}}-\hat{r}}}-\hat{h}\leri{x}\right] \,,
\end{align}
can be maximized for every $x$ in the sum, which must be its global maximum, by a constant $\hat{h}\leri{x}=0$ and $\hat{r}=0$. This corresponds to the likelihood of randomly assigning the observed data points $x\in\mathbf{A\cup B}$ into sample $\mathbf{A}$ or sample $\mathbf{B}$ with constant probabilities $p_\mathbf{A} = \tilde{N}_\mathbf{A}/\leri{\tilde{N}_\mathbf{B}+\tilde{N}_\mathbf{A}}$ and $p_\mathbf{B}=1-p_\mathbf{A}$, respectively (see also Sec.~\ref{subsubsec:cross_entropy}).

The resulting test statistic for this choice is then 
\begin{align}
t&=t_{\mathbf{A}+\mathbf{B}}\leri{\mathbf{A+B}}\equiv\nonumber\\
&\equiv-2\cdot\underaccent{f\leri{x},g\leri{x}}{\text{min}}\left[-\frac{1}{\tilde{N}_{\mathbf{A}}+\tilde{N}_{\mathbf{B}}}\sum_{x\in\mathbf{A},\mathbf{B}}\leri{\tilde{N}_{\mathbf{A}}\leri{e^{f\leri{x}}-1}+\tilde{N}_{\mathbf{B}}\leri{e^{g\leri{x}}-1}}+\sum_{x\in\mathbf{A}}f\leri{x}+\sum_{x\in\mathbf{B}}g\leri{x}\right] \,,\label{eq:t_0}
\end{align}
which can be separated into two independent tests, similar to $t_{\mathbf{B}}\leri{\mathbf{A}}$ in~Eq.~\eqref{eq:t_nref}. One of $\mathbf{A}$ vs. the ``reference sample" $\mathbf{A}+\mathbf{B}$
\begin{align}
t_{\mathbf{A}+\mathbf{B}}\leri{\mathbf{A}}=-2\cdot\underaccent{f\leri{x}}{\text{min}}\left[-\frac{1}{\tilde{N}_{\mathbf{A}}+\tilde{N}_{\mathbf{B}}}\sum_{x\in\mathbf{A},\mathbf{B}}\tilde{N}_{\mathbf{A}}\leri{e^{f\leri{x}}-1}+\sum_{x\in\mathbf{A}}f\leri{x}\right] \,,\label{eq:t_A+B_A}
\end{align}
and one of $\mathbf{B}$ vs. $\mathbf{A}+\mathbf{B}$
\begin{align}
t_{\mathbf{A}+\mathbf{B}}\leri{\mathbf{B}}=-2\cdot\underaccent{g\leri{x}}{\text{min}}\left[-\frac{1}{\tilde{N}_{\mathbf{A}}+\tilde{N}_{\mathbf{B}}}\sum_{x\in\mathbf{A},\mathbf{B}}\tilde{N}_{\mathbf{B}}\leri{e^{g\leri{x}}-1}+\sum_{x\in\mathbf{B}}g\leri{x}\right] \,.\label{eq:t_A+B_B}
\end{align}

This highlights another advantage of choosing the reference distribution to be the observed combined sample. Recall that the loss in~Eq.~\eqref{eq:t_nref} diverged to negative infinities if $f\rightarrow \infty$ for points $x_\star$ that are only included in the observed sample, but not in the reference sample. As explained in Sec.~\ref{subsec:NPLM_challenges}, this is a result of a null hypothesis that has a zero likelihood. Conversely, using the ``symmetric" parameterization, such divergences never appear, since there are no points that belong to either $\mathbf{A}$ or $\mathbf{B}$, but not to $\mathbf{A}\cup\mathbf{B}$. Instead, perfectly fitting the points $x_\star$ would result in a finite (and constant) $f\leri{x_\star}$, and a negatively infinite $g\leri{x_\star}$ -- none of which cause the loss to diverge. This is a reflection of the fact that the symmetric hypothesis always assumes non-zero probability for events that occurred in either sample. As the loss is now bounded from below, we expect over-fitting to be significantly less prominent and remove the weight clipping entirely. The symmetric parameterization is particularly helpful for reliably detecting asymmetric \ac{BSM} signals that lie in a region with low statistics.

\section{Methods}\label{sec:methods}
\subsection{Signal and background models}\label{subsec:modeling}
To demonstrate the performance of the symmetrized formalism, we tested it on two toy models. In the first toy model, the distributions used to generate the data were known analytically. Following~\cite{D_Agnolo_2019}, we chose the symmetric component of the two samples to be drawn from an exponential distribution, denoted as $b_{0}\leri{x}$. In the asymmetric case, three different signals were added only to sample $\mathbf{A}$, according to the three \ac{NP} signals introduced in~\cite{D_Agnolo_2019} -- two resonance signals, localized at the tail $\leri{S_1}$ and the bulk $\leri{S_3}$ of the distribution, and one non-local signal $\leri{S_2}$. The explicit expressions of the \acp{PDF} are - 
\begin{align}
b_{0}\leri{x}&= \exp\leri{-x}\,,\label{eq:b0}\\
    S_1\leri{x}&=\frac{1}{\sqrt{2\pi}\sigma} \exp\leri{-\frac{(x-\bar{x})^2}{2\sigma^2}}\,, \quad \quad \bar{x}=6.4\,,\sigma=0.16\,,\label{eq:S1} \\ 
    S_2\leri{x}&=\frac12 x^2\exp\leri{-x}\,, \label{eq:S2}\\ 
    S_3\leri{x}&= \frac{1}{\sqrt{2\pi}\sigma}\exp\leri{-\frac{(x-\bar{x})^2}{2\sigma^2}}\,, \quad \quad \bar{x}=1.6\,,\sigma=0.16\,.\label{eq:S3}
\end{align}
 
In the second model, the background and signal distributions were not known, but instead generated from \ac{MC} simulations of $\sqrt{s}=13\,$TeV $pp$ collisions. This case is chosen to emulate a search for \ac{LFUV}, where one compares two samples related by a $\mu\leftrightarrow e$ symmetry. Similarly to~\cite{Birman:2022xzu}, we used all the processes that give rise to events with one muon and one electron of opposite charges in the final state. Following~\cite{PhysRevD.90.015025}, we further selected only $e^\pm\mu^\mp$ events in which the electron has a larger transverse momentum ($p_T$) than the muon, and obtained $\sim2.2 \times 10^{5}$ events at an integrated luminosity of $20\,\rm{fb}^{-1}$. The background production channels included Drell-Yan, di-boson, $Wt$, $t \overline{t}$ and \ac{SM} Higgs $(H \rightarrow WW/\tau\tau)$. 
The sample was generated using MadGraph 2.6.4~\cite{Alwall:2014hca} and Pythia 8.2~\cite{Sjostrand:2014zea}, and the response of the
ATLAS detector was emulated using Delphes 3~\cite{deFavereau:2013fsa}. The events were selected such that $p^{e}_T\geq p^{\mu}_T\geq15\,\text{GeV}$~, $|\eta_e|\,,|\eta_\mu|\leq 2.3$ and the isolation threshold for the electron (muon) was 0.15 (0.25). 

To look for \ac{LFUV}, one could compare the sample with $p_T^e>p_T^\mu$ (denoted as $e\mu$), to a sample in which the muon is the leading lepton, $p_T^\mu>p_T^e$ (denoted as $\mu e$), and look for differences between them~\cite{ATLAS:2023mvd}. While in a realistic analysis, one should also account for known detection effects inducing small systematic asymmetries between these samples, we neglect these effects here. To this end, in the symmetric case, samples $\mathbf{A}$ and $\mathbf{B}$ were generated by random selection from the full $e\mu$ sample via bootstrapping with replacement -- i.e. assigning a uniform probability for choosing any event in the $e\mu$ sample, where events can be repeated~\cite{10.1214/aos/1176344552}.
In the asymmetric case, an additional signal was added only to sample $\mathbf{A}$, corresponding to an \ac{LFUV} Higgs decay $H\rightarrow \tau^\mp e^\pm \rightarrow \mu^\mp e^\pm 2\nu$ where the Higgs is produced by vector boson fusion and gluon-gluon fusion. The signal events were also sampled by bootstrapping with replacement from a signal template of 456 events, corresponding to a branching ratio of 1\%. The variable $x$ was chosen as the collinear mass~\cite{ATLAS:2016joj} of the $e\mu$ pair, in units of $0.1\,$TeV. The resulting collinear mass distributions of the signal and background are given in Appendix~\ref{app:datasets}. Here we simply show a proof of concept of our method, and characterize its sensitivity to physical asymmetries. A full study of the performance accounting for other production channels and kinematic variables as well as the asymmetric detector effects should follow in the future. We further comment on these in Sec.~\ref{subsec:asymmetric_results} and in the discussion in Sec.~\ref{sec:AAOQ}.

All trainings lasted $5\times 10^{5}$ epochs, for a \ac{NN} with one hidden layer of 4 neurons. The optimizer used was Adam~\cite{kingma2017adam}, with a learning rate of $10^{-3}$. For both models, the total number of background events and signal events were drawn from a Poisson distribution, and varied between toy experiments. Our computations were based on the \ac{NPLM} package code~\cite{Grosso_New_Physics_Learning_2021}, which outputs $t_\mathbf{B}\leri{\mathbf{A}}$ as in Eq.~\eqref{eq:t_nref}. The code was re-wrapped and modified to calculate any $t_{\mathbf{Y}_r}\leri{\mathbf{Y}_o}$, where $\mathbf{Y}_i$ is some sample -- a set of individual events. 
For the symmetric case, in which both samples were drawn from the same distribution, we ran roughly 2000 toys to obtain the background-only distribution of the statistic for each benchmark luminosity and/or expected relative sizes of the samples. For the asymmetric case, we ran roughly 100 toys for each expected signal strength and used the median to estimate the sensitivity of the search.

\subsection{Parameters of interest and asymptotic distribution}\label{subsec:dof}

Since we chose a \ac{NN} with four neurons and one hidden layer, each function described by it (${f\leri{x},g\leri{x},h\leri{x}}$ in Eq.~\eqref{eq:H_0} and \eqref{eq:H_1} as well as ${f\leri{x}}$ in Eq.~\eqref{eq:NPLM_H_1}) should have 13 free parameters. Both in the $t_{\mathbf{A}}\leri{\mathbf{B}}$ case and in the $t_{\mathbf{A}+\mathbf{B}}\leri{\mathbf{A+B}}$ case, the null hypothesis has 12 fewer free parameters than the alternative hypothesis. For $t_{\mathbf{B}}\leri{\mathbf{A}}$ the null hypothesis has one free parameter -- the total number of events $N_\mathbf{A}$, which we set to be the observed one $\tilde{N}_\mathbf{A}$. In the alternative hypothesis, all 13 parameters in the function $f$ are free. For $t_{\mathbf{A}+\mathbf{B}}\leri{\mathbf{A+B}}$, the null hypothesis has 12 constraints, setting $f\leri{x}=g\leri{x}+r$, where $r$ is a free constant (this is again just a free total number of events, which we set to be the observed one). Therefore, we expect the distribution of $t$ in the symmetric case -- in which both $\mathbf{A}$ and $\mathbf{B}$ were drawn from the same distribution, to follow a $\chi^2_{12}$ distribution at the limit of high enough statistics.

We note that in the original \ac{NPLM} papers~\cite{D_Agnolo_2019,DAgnolo:2019vbw,dAgnolo:2021aun}, the number of expected events in sample $\mathbf{A}$ under the null hypothesis was fixed, rather than a free parameter, and the number of events in the reference sample $\mathbf{B}$ was constant, rather than Poisson-distributed. In Appendix~\ref{appendix:moreresults}, we show the background-only distribution for $t_\mathbf{B}\leri{\mathbf{A}}$ using the exact calculation in~\cite{D_Agnolo_2019}, implemented in~\cite{Grosso_New_Physics_Learning_2021}. As shown in Appendix~\ref{appendix:moreresults}, although the number of \ac{DOF} expected for the asymptotic $\chi^2$ is larger by one, we find that this implementation faces the same challenges described in Sec.~\ref{subsec:NPLM_challenges}.

\subsection{Weight clipping}\label{subsec:weightclipping}

In principle, weight clipping could enhance the sensitivity of the test when wide signals are considered, as it smooths out local statistical fluctuations by limiting the gradients of the fitting functions. However, unlike in the \ac{NPLM} procedure~\cite{D_Agnolo_2019}, in this work no weight clipping optimization was performed, although it may improve the agreement with the asymptotic distribution for both the symmetrized and \ac{NPLM} methods. This choice is in line with the \ac{DDP} approach, which should allow rapid scanning of a large number of distributions and final states and thus can not rely on time-consuming optimizations. 

Since the symmetrized formalism is not subject to artificial divergences, no weight clipping was applied in our studies. For our implementations of the \ac{NPLM} tests (Fig.~\ref{fig:t_A_null_dist}), we applied a weight clipping of $9$, as was recommended in~\cite{D_Agnolo_2019}, and no weight clipping, to illustrate the challenges of this procedure.

\subsection{Ideal significance calculation}\label{subsec:Zid}
For the asymmetric case, we are interested in comparing the sensitivity of the symmetrized formalism to that of an ideal analysis. The latter uses a profile-likelihood test statistic $q_0$ where the (symmetric) background \ac{NDF} and the shape of the signal are known. At the asymptotic limit of infinite data, the distribution of $q_0$ under the null hypothesis is $\chi^2_1$, as the number of signal events is the only parameter of interest. Therefore, the ideal significance is given by 
\begin{align}
Z_{\rm id} = \sqrt{q_0}&\equiv\sqrt{2\leri{-N_s+\sum_x \leri{N_b(x)+N_s(x)}\log\frac{N_s(x)+N_b(x)}{N_b(x)}}}\,,\label{eq:Zid_binned}
\end{align}
where $N_s\leri{x}$ ($N_b\leri{x}$) is the number of signal (background) events at some point $x$, and $N_s$ ($N_b$) is the total number of signal (background) events. For comparison, we present the ideal symmetrized test score in Appendix~\ref{appendix:ideal_t_AB}. 

The calculation of $Z_{\rm id}$ was carried out as follows. When the \acp{PDF} of the background and signal were analytically known, $Z_{\rm id}$ was calculated from the expected number densities, i.e.
\begin{align}
    Z^{S_{i}}_{\rm id} =\sqrt{2\leri{-N_s+\int^{\infty}_0 dx \leri{N_b b_0\leri{x}+N_s S_i\leri{x}}\log\leri{1+\frac{N_s S_i\leri{x}}{N_b b_0\leri{x}}}}}\,,
\end{align}
where $b_0$ and $S_i$ are given in Eqs.~\eqref{eq:b0}-\eqref{eq:S3}, and $N_b$ and $N_s$ are the expected total number of background events and signal events in sample $\mathbf{A}$, respectively. When the samples are resampled from a known template, but the \acp{PDF} are not analytically known, one would expect an infinite ideal significance due to singularities resulting from the uniqueness of the data points. In this case, $Z_{\rm id}$ can be calculated via binning the data, however the result might depend on the binning scheme. We thus do not use this metric for characterizing the \ac{LFUV} performance, and present the \ac{BR} corresponding to a $2-\sigma$ significance instead.

\subsection{Empiric background-only distribution generation by permutation}\label{subsec:permutations}
A study of the null hypothesis distribution could also be done through a permutation test -- i.e. generating toy data by joining together two observed samples $\mathbf{A}$ and $\mathbf{B}$, and randomly repeatedly re-dividing the combined sample into two new samples $\mathbf{A}'$ and $\mathbf{B}'$, of the same sizes as $\mathbf{A}$ and $\mathbf{B}$. Within the null hypothesis, the events in $\mathbf{A}$ and in $\mathbf{B}$ are believed to have been sampled from a single \ac{PDF}\,. Thus, every relabeling of those events, such that the total number of events in each new sample $N_{\mathbf{A}'}\,,N_{\mathbf{B}'}$ is fixed to be the observed one, has the same probability, regardless of the underlying \ac{PDF}. While this strategy does not account for the fluctuations of the combined sample, it is the natural choice when the common (``background-only") component of the distribution is unknown, as it solely relies on assuming a symmetry relating the two samples. As such, permutation tests can be used to empirically generate the null hypothesis distribution of $t_{\mathbf{A}+\mathbf{B}}$, preserving some of its unique features that could cause deviations from the $\chi^2$ distribution, while not requiring a detailed \ac{MC} simulations of the \ac{SM} background.
 
We are then interested in testing how reliable the permutation procedure is as a proxy for the true background-only distribution for the purpose of detecting asymmetries. To show this concretely, we calculated the distribution of the symmetrized test $t_{\mathbf{A}+\mathbf{B}}$ as repeatedly sampled from the background \ac{PDF} $b_0$\,, as well as the distribution of $t_{\mathbf{A}+\mathbf{B}}$ resulting from relabelings (permutations) of representative toy datasets. We compared both the overall shapes of these distributions, as well as the corresponding resulting significances of asymmetric deviations in the original datasets. 

The permutation procedure was performed on samples that were expected to be of equal sizes under the null hypothesis, at two different scales. In the first case, the expected number of background events in the samples was $N_\mathbf{A}=N_{\mathbf{B}}\approx 5.5\times 10^{4}$. For this case, the original null hypothesis distribution was in good agreement with the asymptotic $\chi^2$. In the second case, the two samples were smaller by an order of magnitude, $N_\mathbf{A}=N_{\mathbf{B}}\approx 5.5\times 10^{3}$, and thus the observed distribution of $t_{\mathbf{A+B}}$ deviated from the asymptotic distribution when both samples were drawn from $b_0$. Examining these two cases allows us to test how well the permutation procedure reproduces the features of the original distribution. 

We studied two scenarios of representative datasets -- with and without signal. The former datasets were sampled from $b_0$, and originally resulted in the median significance of the background-only case (generated from $b_0$). The latter datasets correspond to the asymmetric $\sim 2\sigma$ $S_3$ signal injected to sample $\mathbf{A}$. Examining these two scenarios is useful for demonstrating the robustness of the permutation method.

\section{Results}\label{sec:results}
\subsection{The symmetric case}

\begin{figure}[H]
\centering
 \includetrimmedgraphics{0.038}{0.02}{0.47}{0.4}{0.274}{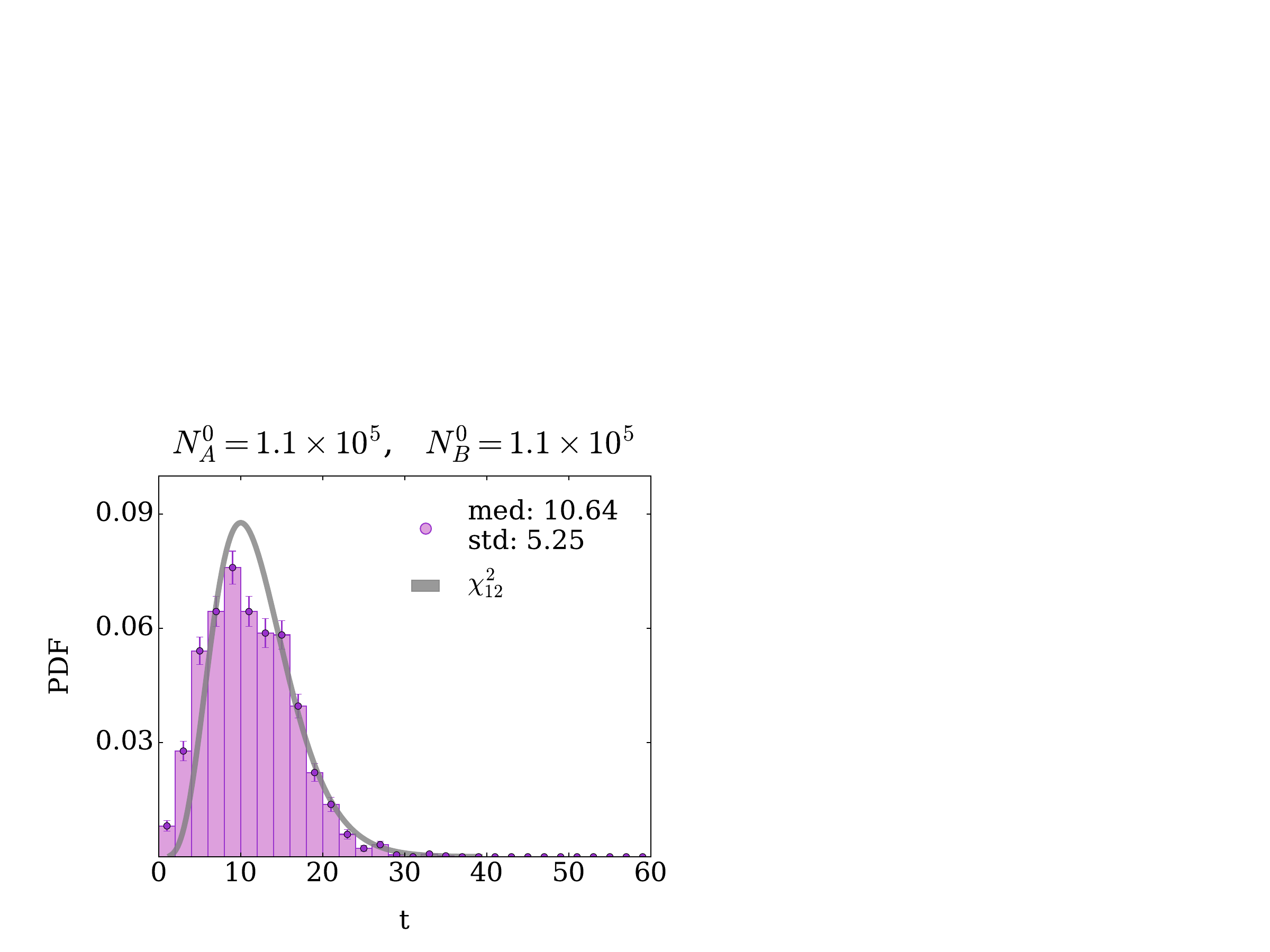}
\includetrimmedgraphics{0.038}{0.02}{0.47}{0.4}{0.274}{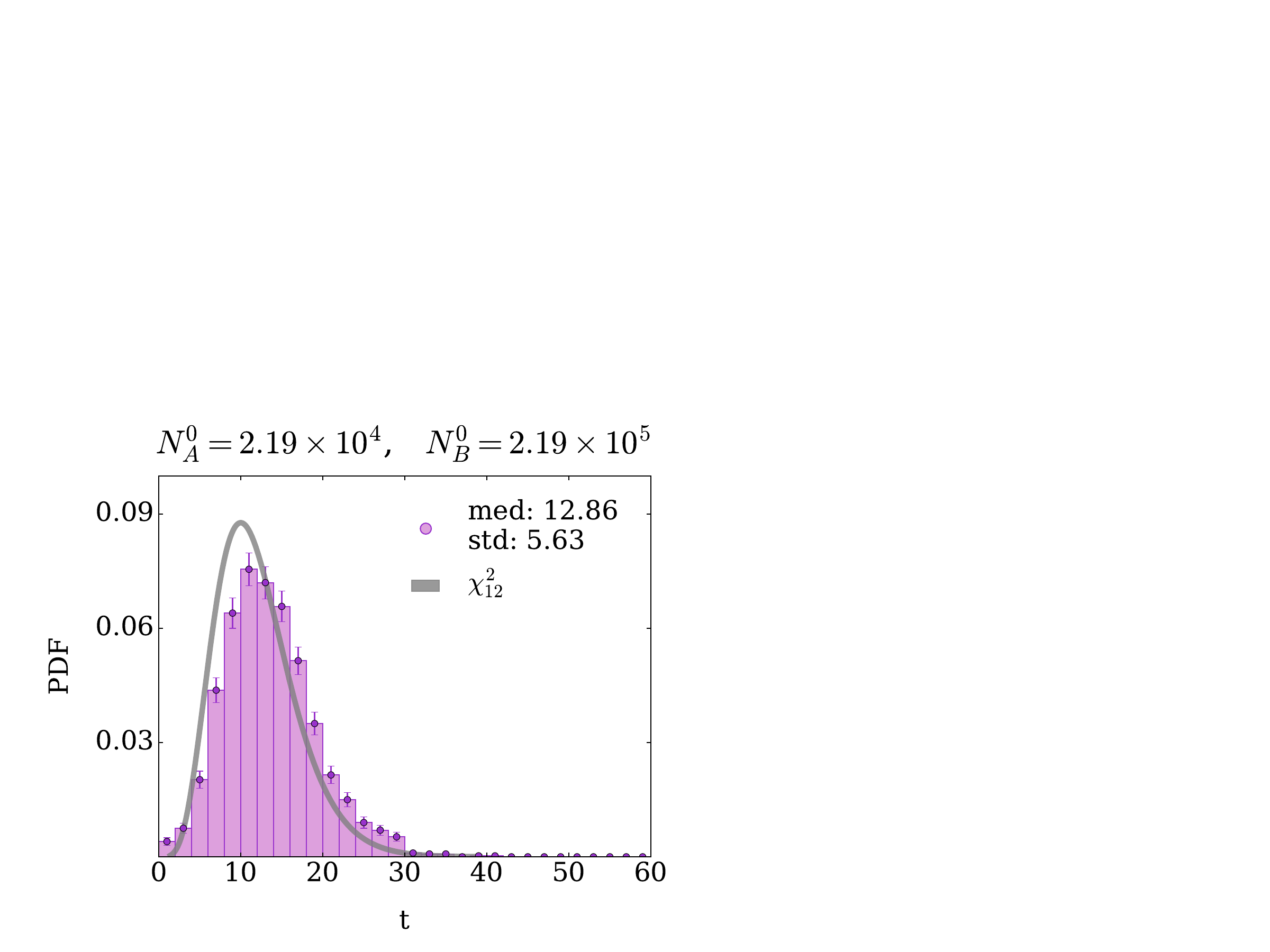}
\includetrimmedgraphics{0.038}{0.02}{0.47}{0.4}{0.274}{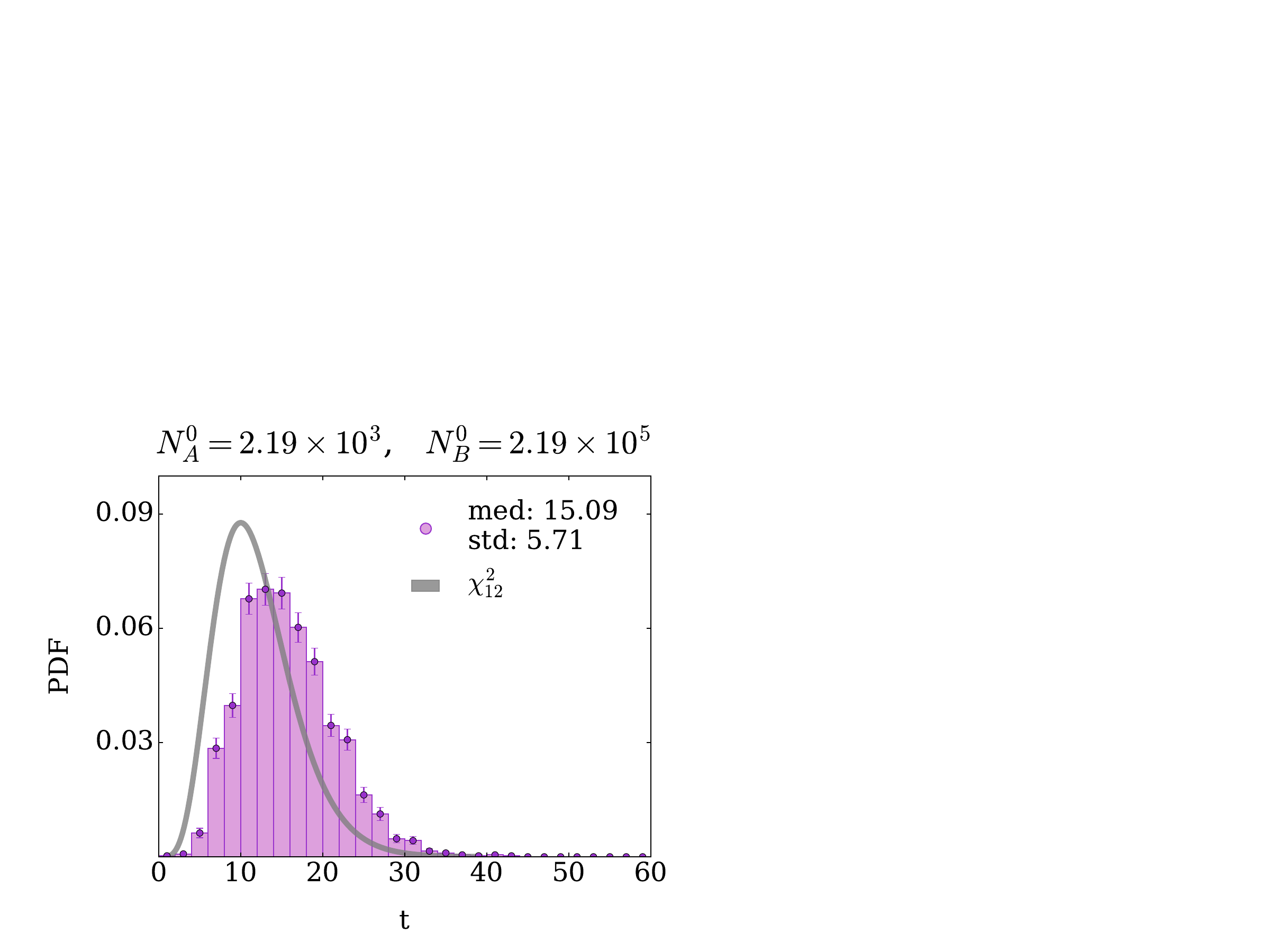}\\
\includetrimmedgraphics{0.038}{0.02}{0.47}{0.4}{0.274}{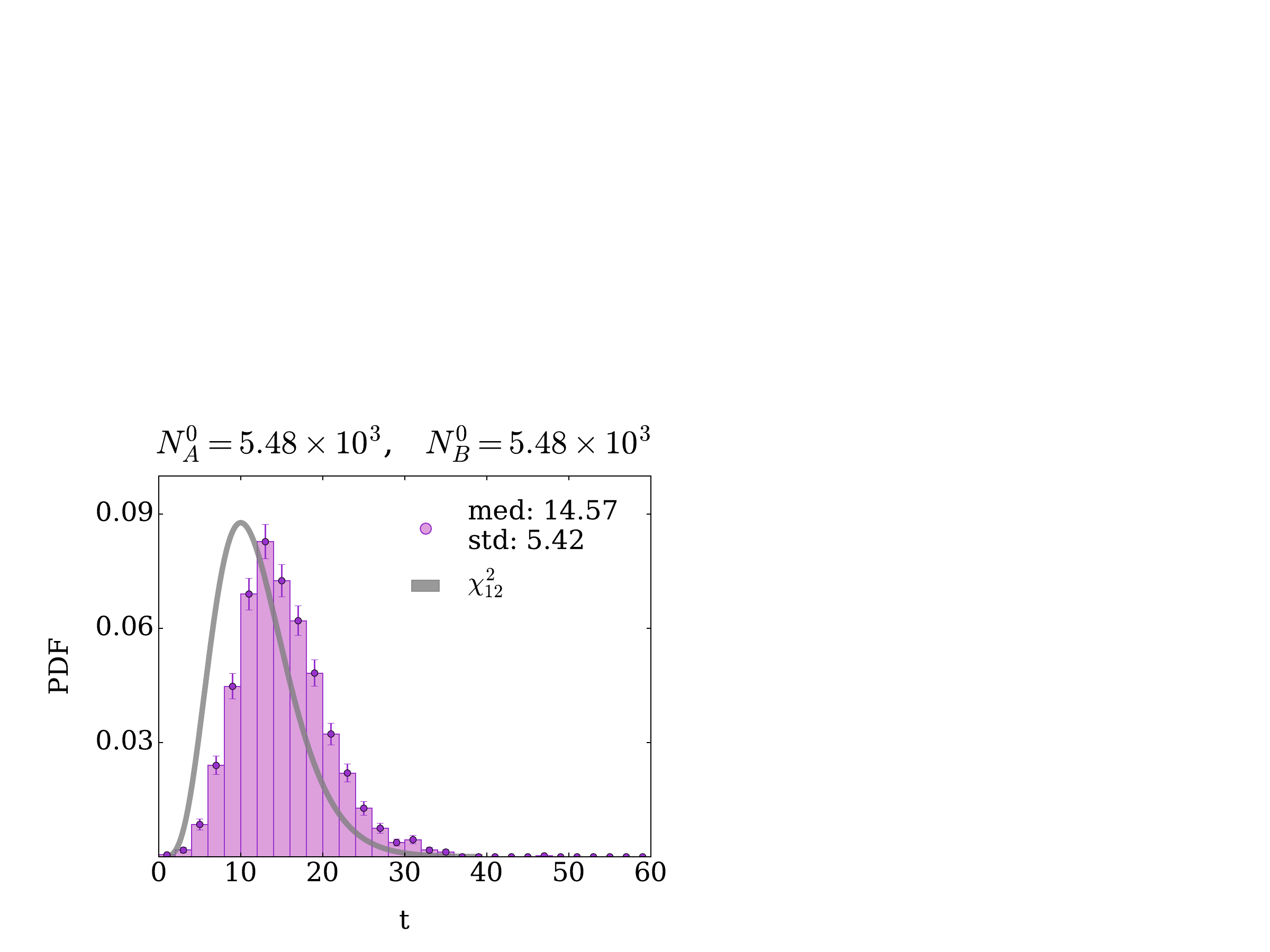}
\includetrimmedgraphics{0.038}{0.02}{0.47}{0.4}{0.274}{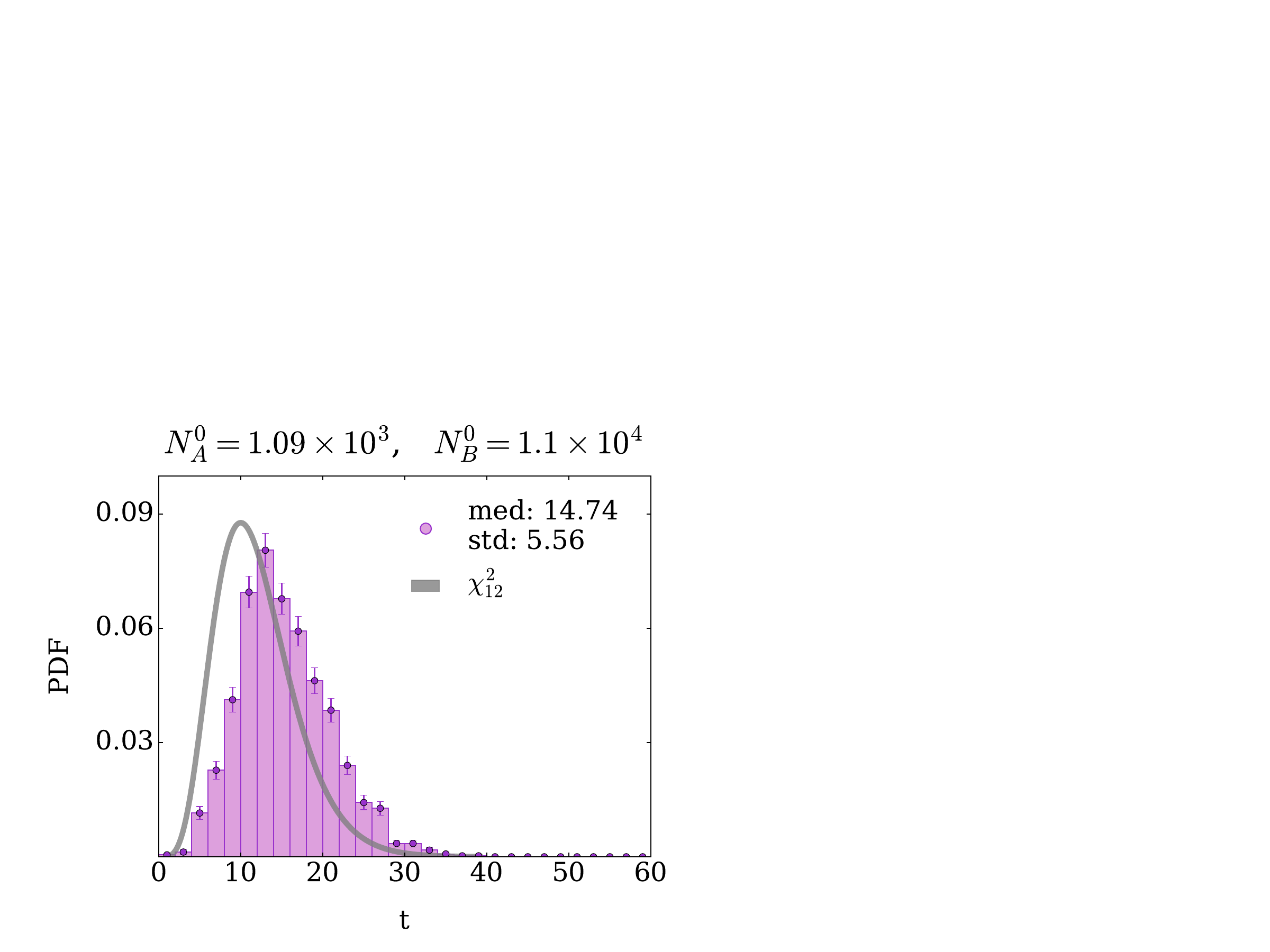}
\includetrimmedgraphics{0.038}{0.02}{0.47}{0.4}{0.274}{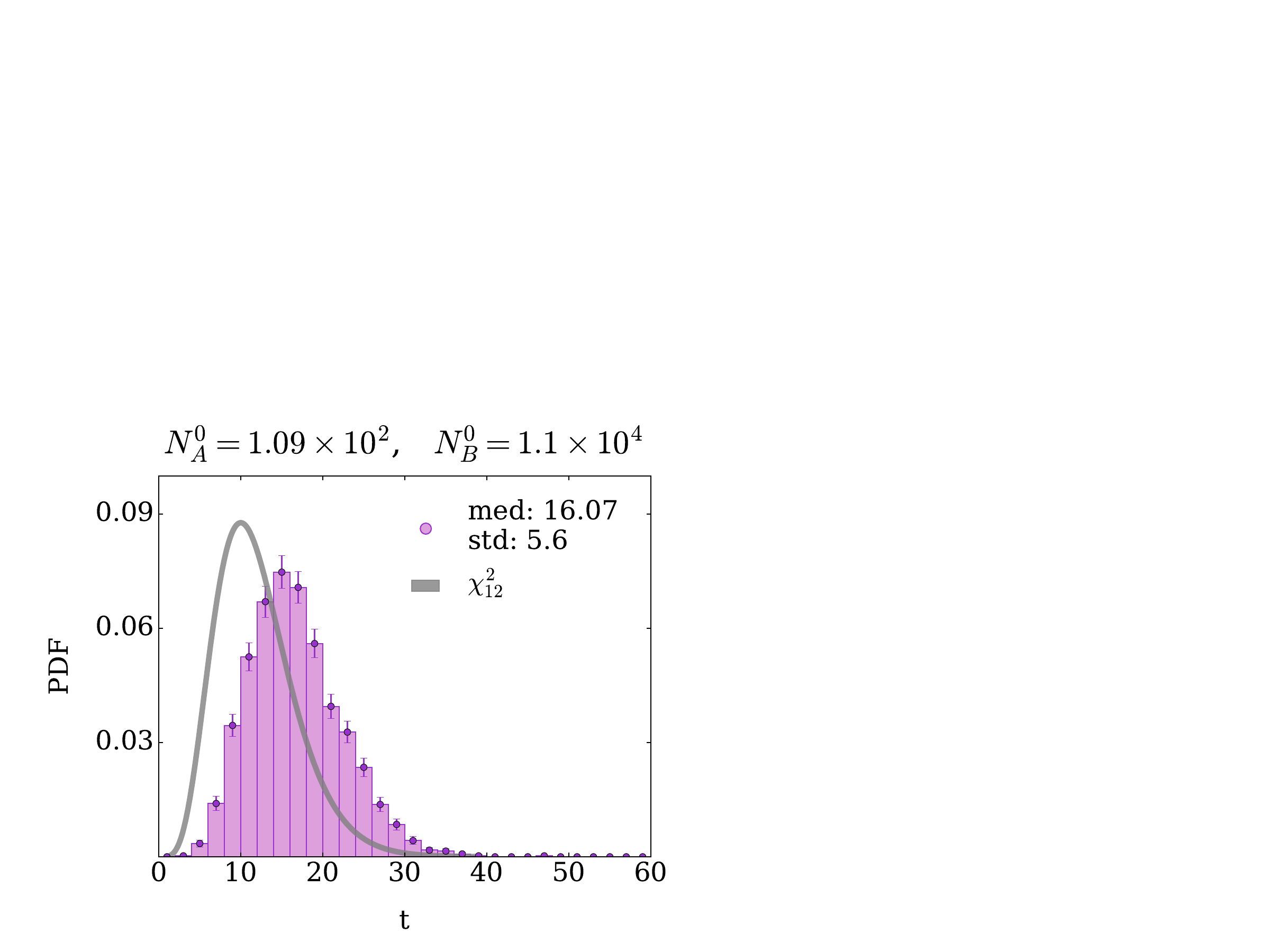}
	\caption{The distribution of the symmetrized test statistic $t_\mathbf{A+B}\leri{\mathbf{A+B}}$ under the null hypothesis. Left -- $N_\mathbf{B} = N_\mathbf{A} = N/2$, middle -- $N_\mathbf{A} = N/10$ and $N_\mathbf{B} = N$, right -- $N_\mathbf{A} = N/100$ and $N_\mathbf{B} = N$, all with no weight clipping. Top -- $N\approx 2.2\times 10^{5}$\,, bottom $N\approx10^{4}$. Solid gray -- the expected $\chi^2_{12}$ distribution according to the Wilks-Wald theorem. Both sample $\mathbf{A}$ and sample $\mathbf{B}$ were drawn from the exponential distribution $b_0$ in Eq.~\eqref{eq:b0}.}
	\label{fig:sym_null_exps}
\end{figure}

\begin{figure}[H]
\centering
 \includetrimmedgraphics{0.038}{0.02}{0.47}{0.4}{0.274}{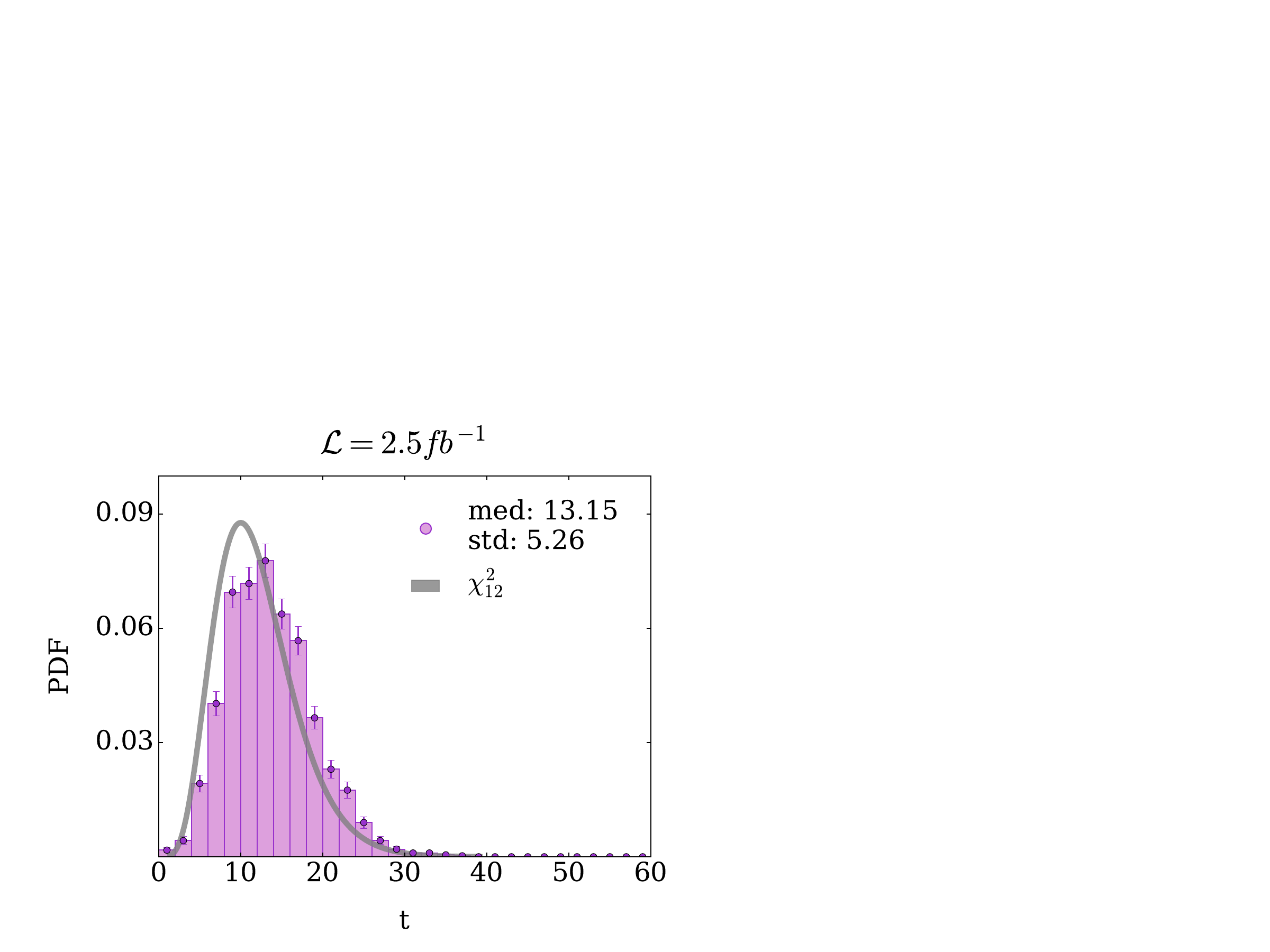}
 \includetrimmedgraphics{0.038}{0.02}{0.47}{0.4}{0.274}{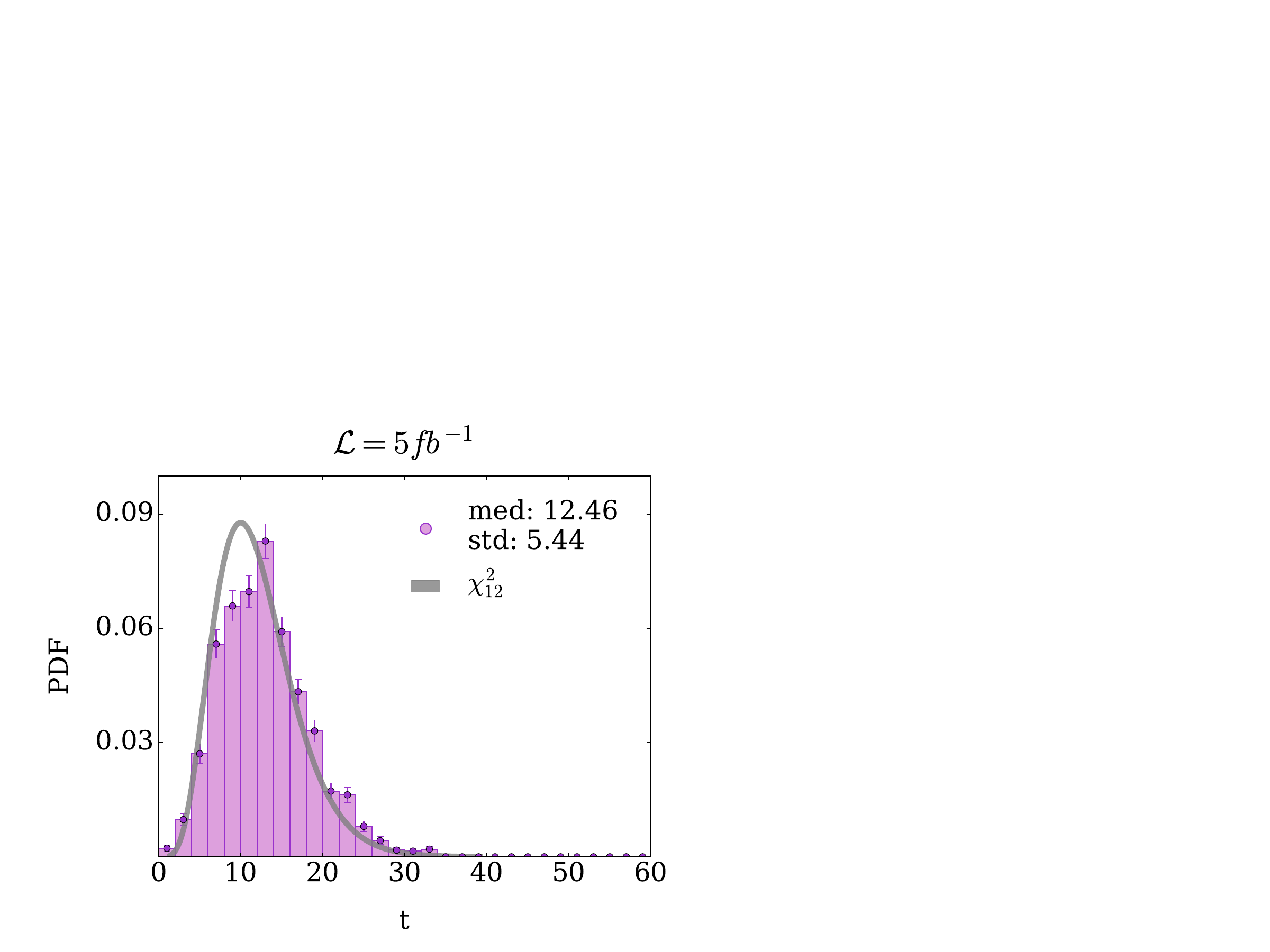}\\
 \includetrimmedgraphics{0.038}{0.02}{0.47}{0.4}{0.274}{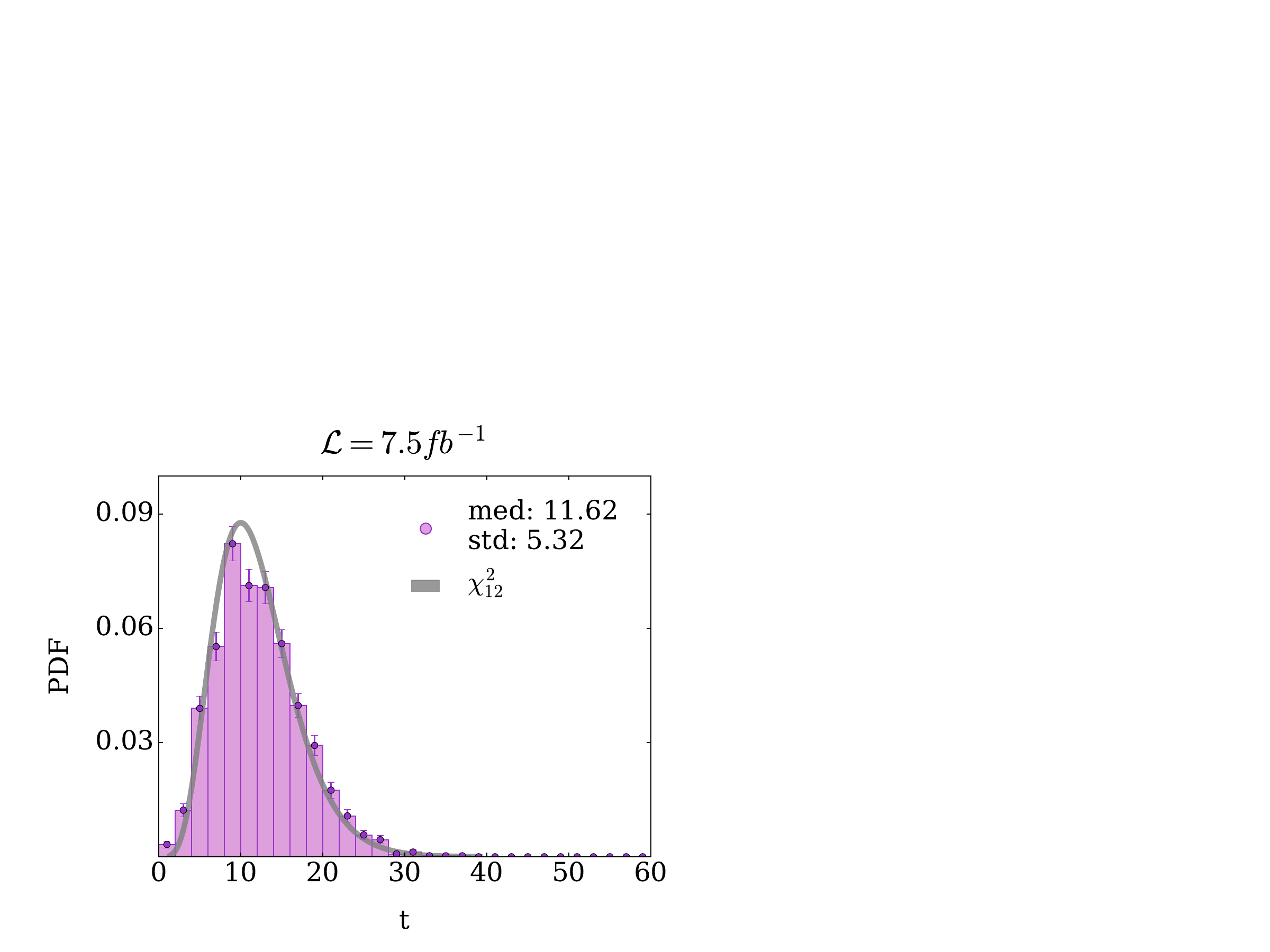}
 \includetrimmedgraphics{0.038}{0.02}{0.47}{0.4}{0.274}{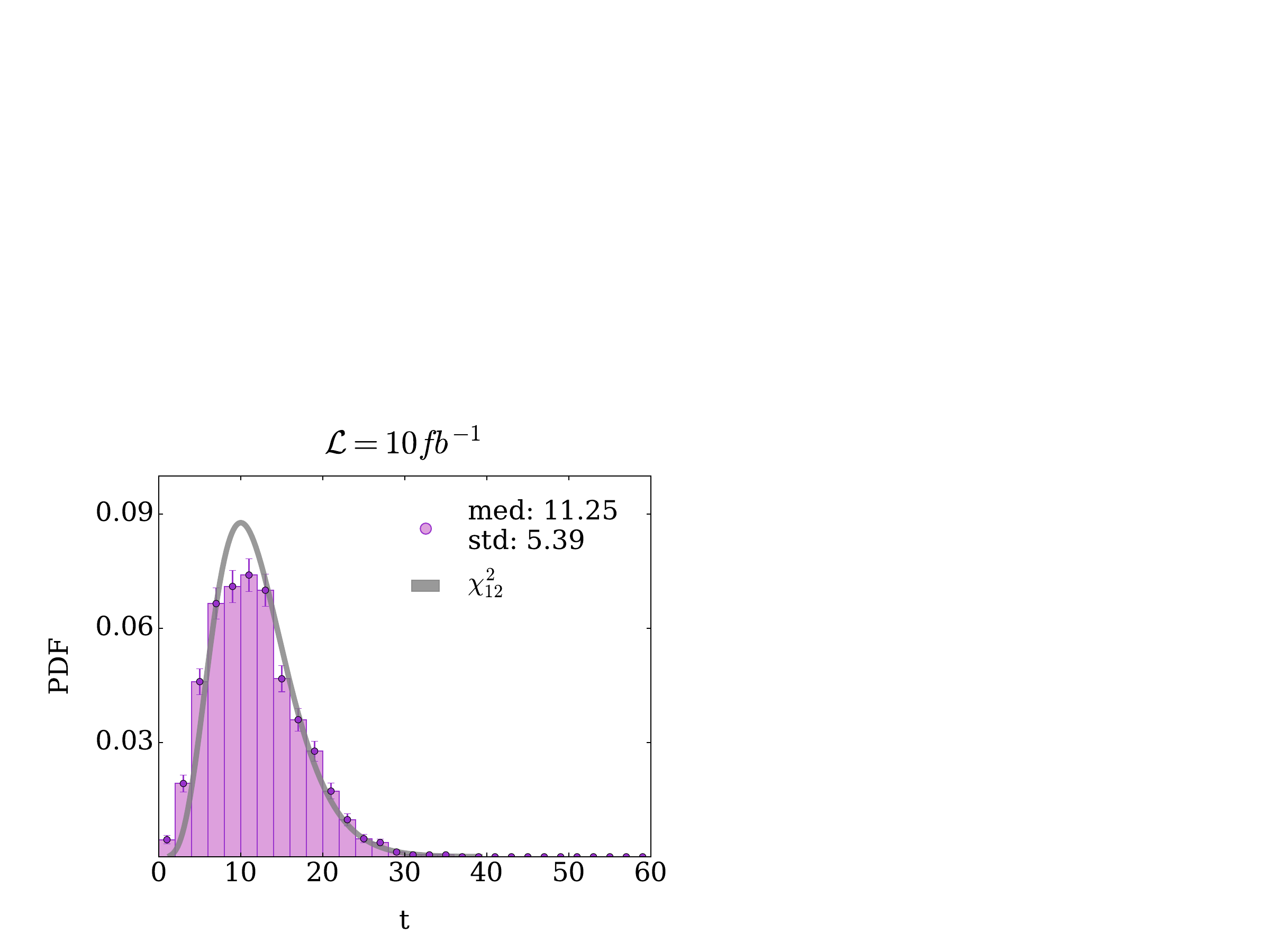}
 
	\caption{The distribution of the symmetrized test statistic $t_\mathbf{A+B}\leri{\mathbf{A+B}}$ under the null hypothesis for $N_\mathbf{B} = N_\mathbf{A}$, all with no weight clipping. Both sample $\mathbf{A}$ and sample $\mathbf{B}$ were drawn from an $e\mu$ sample as described in Sec.~\ref{subsec:modeling}. The integrated luminosities mentioned are the ones at which the \ac{SM} $e\mu$ yield is equal to the size of each sample. Solid gray -- the expected $\chi^2_{12}$ distribution according to the Wilks-Wald theorem.}
	\label{fig:sym_null_em}
\end{figure}

The distributions of $t_{\mathbf{A}+\mathbf{B}}\leri{\mathbf{A+B}}$ under the null hypothesis are presented in Fig.~\ref{fig:sym_null_exps} for the $b_0$ exponential background, and in Fig.~\ref{fig:sym_null_em} for the $e\mu$ background. As expected, the symmetrized formalism yields a relatively good agreement with the asymptotic behavior of the background-only distribution for any ratio between the sample sizes $N_\mathbf{A}/N_\mathbf{B}$. This is a result of the fact that the symmetrized formalism treats both samples equally (democratically), and thus is invariant to interchanging their labels. 

In particular, the symmetrized formalism accounts for the fluctuations of both samples, whereas the fluctuations of the reference sample are neglected in $t_{\mathbf{B}}\leri{\mathbf{A}}$. In turn, the \ac{NPLM} test requires a stricter weight clipping to reproduce the asymptotic $\chi^2_{12}$ distribution the larger $N_\mathbf{A}/N_\mathbf{B}$ is. As can be seen in the plots, the null hypothesis distribution of the symmetrized test tends to lower values of $t$ when a larger number of events is maintained in each of the samples. While there could still be some residual sensitivity to the ratio between the sizes of the samples, possibly favoring samples of similar sizes, it is significantly less prominent than in the \ac{NPLM} case. It appears that the deviation from a $\chi^2$ distribution is mostly affected by the size of the smallest of the two samples (rather than the ratio between them).

In addition, while a few percent of the trainings of $t_{\mathbf{B}}\leri{\mathbf{A}}$ yielded an infinitely negative loss when the weight clipping was relaxed, as predicted by the authors of~\cite{D_Agnolo_2019}, we find that in all trainings $t_{\mathbf{A}+\mathbf{B}}\leri{\mathbf{A+B}}$ remained finite, regardless of the weight clipping. This is expected, as $t_{\mathbf{A}+\mathbf{B}}\leri{\mathbf{A+B}}$ may never diverge. As explained in Sec.~\ref{subsec:The Symmetrized Formalism}, this is both due to the non-singular (and thus generic) parameterization used in the calculation of $t_{\mathbf{A}+\mathbf{B}}\leri{\mathbf{A+B}}$, and due to the fact that the null hypothesis in the symmetrized formalism always has a non-zero likelihood, as it is truly model agnostic.

\subsection{The asymmetric case}\label{subsec:asymmetric_results}

\begin{figure}[H]
    \centering
    \includetrimmedgraphics{0.09}{0.27}{0.08}{0.07}{0.33}{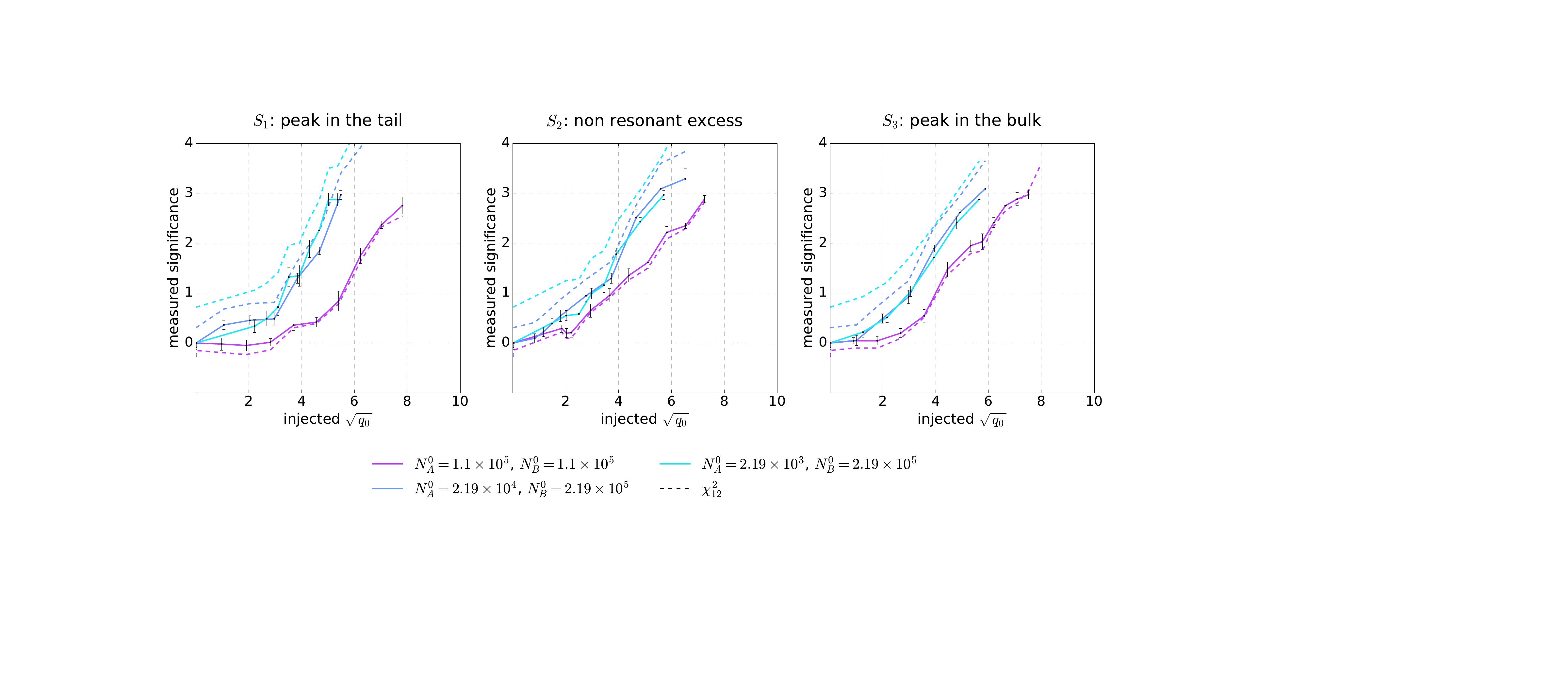}
    \caption{Significance measured with the symmetrized test as a function of the injected ideal $\sqrt{q_0}$ test for analytically known background and signal shapes. Solid curves represent different ratios between the datasets. Dashed curves are the corresponding $\chi^2$ approximations. Left: Resonant signal in the tail. Middle: non-resonant excess along the distribution. Right: Resonant signal in the bulk. }
    \label{fig:performance_exp}
\end{figure}

The sensitivity to the $S_1\,, S_2\,,$ and $S_3$ signals injected on top of the exponential background $b_0$, are presented in Fig.~\ref{fig:performance_exp}. The median significance is shown, for different $N_\mathbf{A}/N_\mathbf{B}$, as a function of the median significance of the ideal test $Z_{id}$ described in Sec. \ref{subsec:Zid}. Due to limited statistics in the high values of the background-only \ac{PDF}, the errors $\varepsilon$ on the median values are assumed to be symmetric and calculated as $\varepsilon = Z(t_{\text{med}})-Z(t_{\text{med}}-\sigma/\sqrt{N_{\text{toys}}})$, where $t_{\text{med}}$ is the median and $\sigma$ is the observed standard deviation of the test statistic over the $N_{\text{toys}}$ toy datasets. We find the performance in all these cases to be similar when compared to the corresponding ideal significance; the measured significance increases with the injected one at a larger slope for a larger ratio between the number of background events in samples $\mathbf{A}$ and $\mathbf{B}$.

\begin{figure}
    \centering
    \includetrimmedgraphics{0.038}{0.02}{0.47}{0.48}{0.4}{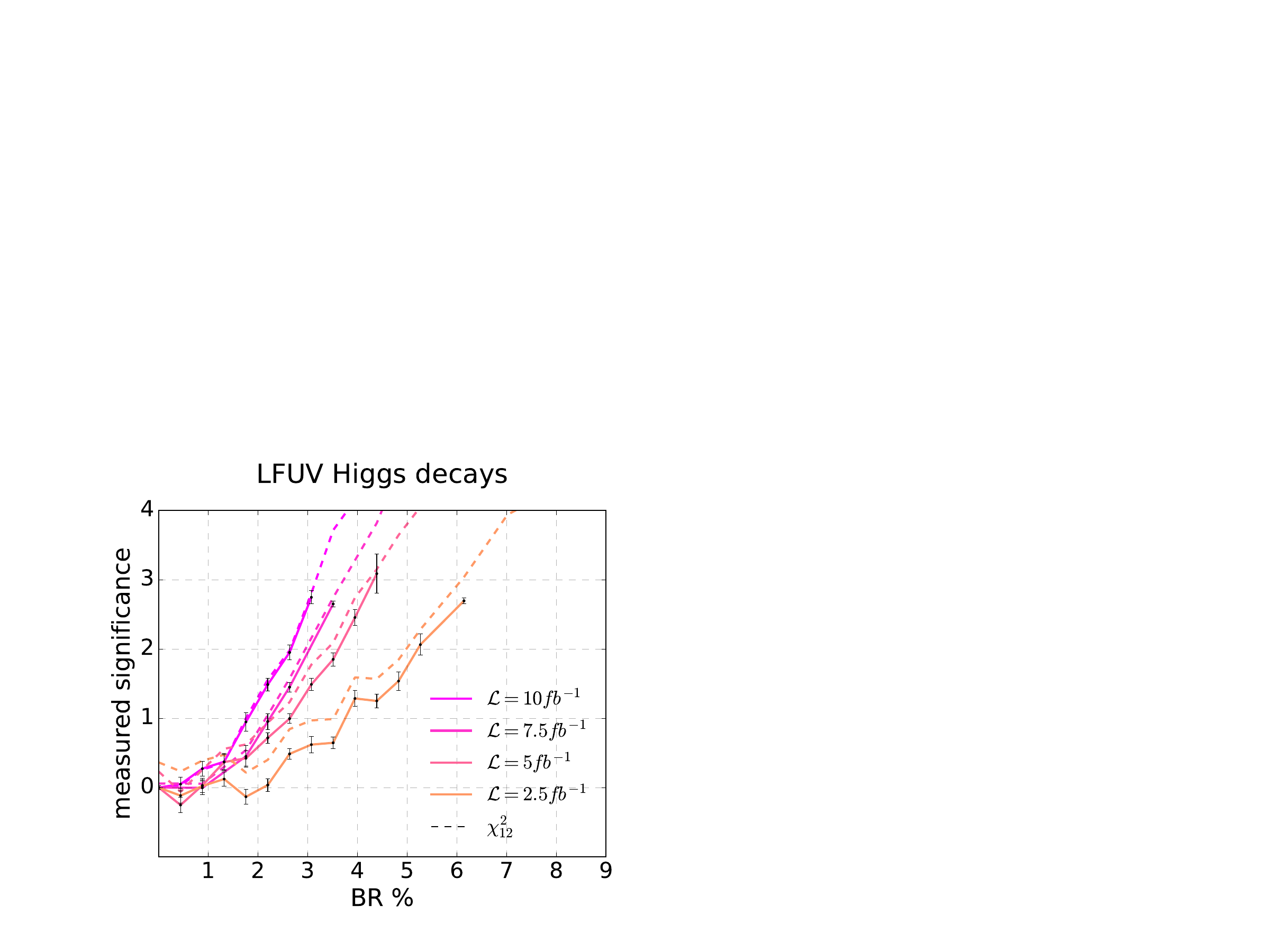}
    \caption{Sensitivity to Higgs \ac{LFUV} signals for different luminosities (solid curves). The measured significance of the symmetrized test is shown as a function of the Higgs \ac{LFUV} \ac{BR}. Dashed curves are the corresponding $\chi^2$ approximations.} 
    \label{fig:performance_em}
\end{figure}

In Fig.~\ref{fig:performance_em} we present the sensitivity of the symmetrized formalism to Higgs \ac{LFUV} decays $H\rightarrow \tau e \rightarrow \mu e 2\nu$ as a function of the \ac{BR} for different luminosities. As can be seen in the plot, the $2\sigma$ sensitivity expectation follows the expected scaling with $\sqrt{\mathcal{L}}$\,, and reaches a \ac{BR} of $\sim 2.7\%$ for $\mathcal{L}=10\,\text{fb}^{-1}$. We note that the current best bounds on the \ac{BR} of $H\rightarrow \tau e$ is at the $0.2\%$ level, measured at an integrated luminosity of $\mathcal{L}=137-138\,\text{fb}^{-1}$~\cite{CMS:2021rsq,ATLAS:2023mvd}. Extrapolating our results to $\mathcal{L}=140\,\text{fb}^{-1}$ yields an expected $2 \sigma$ sensitivity to a BR of $\sim 0.7\%$. 
These two sensitivities should not be directly compared: first, the traditional search employs significant background rejection exploiting many kinematic variables in a single multivariate discriminator. In our case, the signal to background separation is based solely on the collinear mass, and the extension of the method for more than one variable, as done in \cite{DAgnolo:2019vbw} for the \ac{NPLM}, is left for future work. Second, in the traditional analysis, the signal shape and background shape are known while in our approach both are unknown. On the other hand, in our analysis the two samples were drawn from the same distribution, while in practice detector and phase-space effects will cause some asymmetry between $e\mu$ and $\mu e$ samples. While these are also to be addressed in future work, we refer the reader to e.g.~\cite{ATLAS:2023mvd} where efficiency corrections were applied to restore the symmetry between the samples. We discuss further possible solutions in our discussion in Sec.~\ref{sec:Conclusions}.

\subsection{Permutation tests}

\begin{figure}
\centering
      \includetrimmedgraphics{0.038}{0.02}{0.47}{0.4}{0.265}{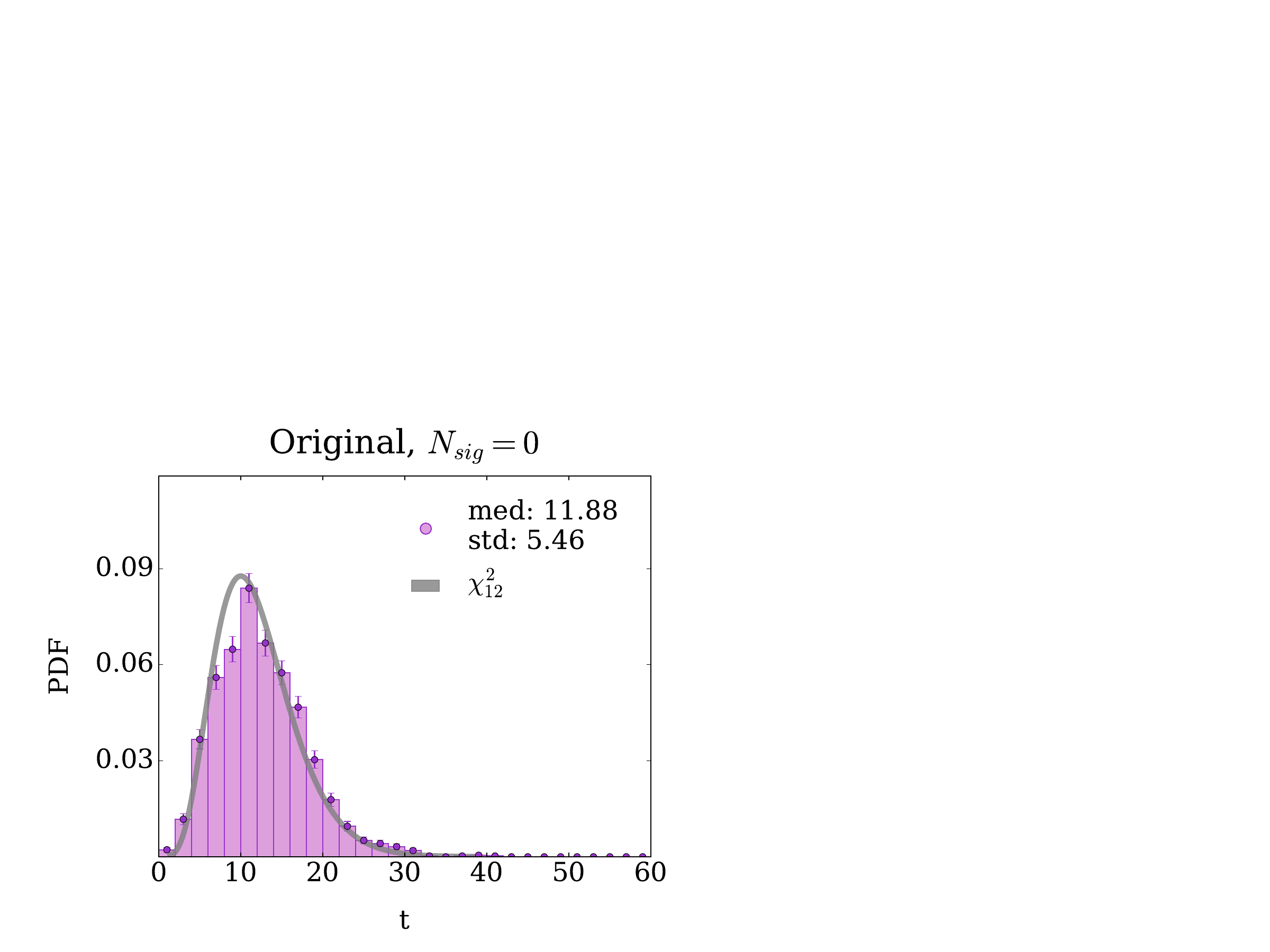}
    \includetrimmedgraphics{0.038}{0.02}{0.47}{0.4}{0.265}{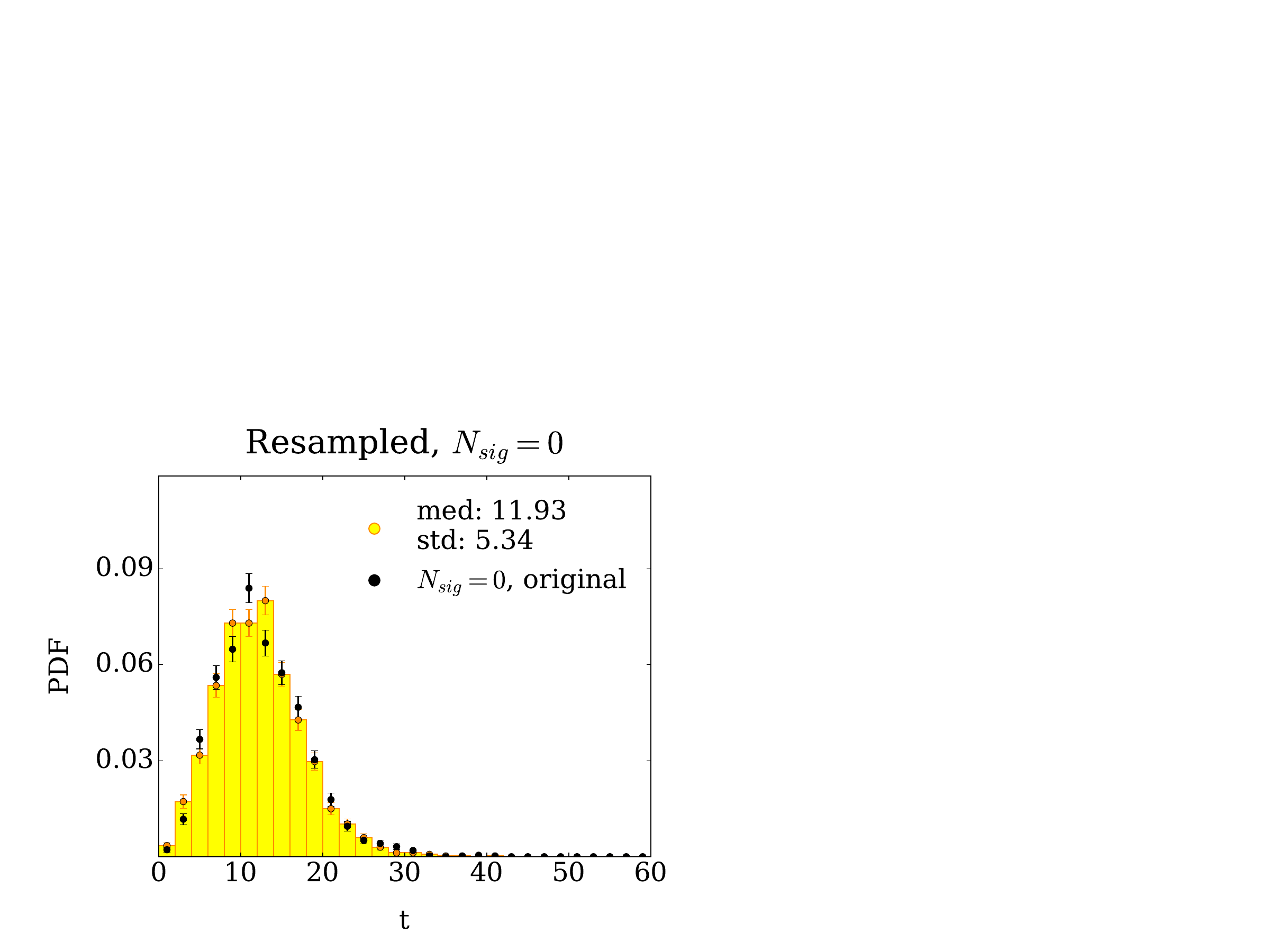}
    \includetrimmedgraphics{0.038}{0.02}{0.47}{0.4}{0.265}{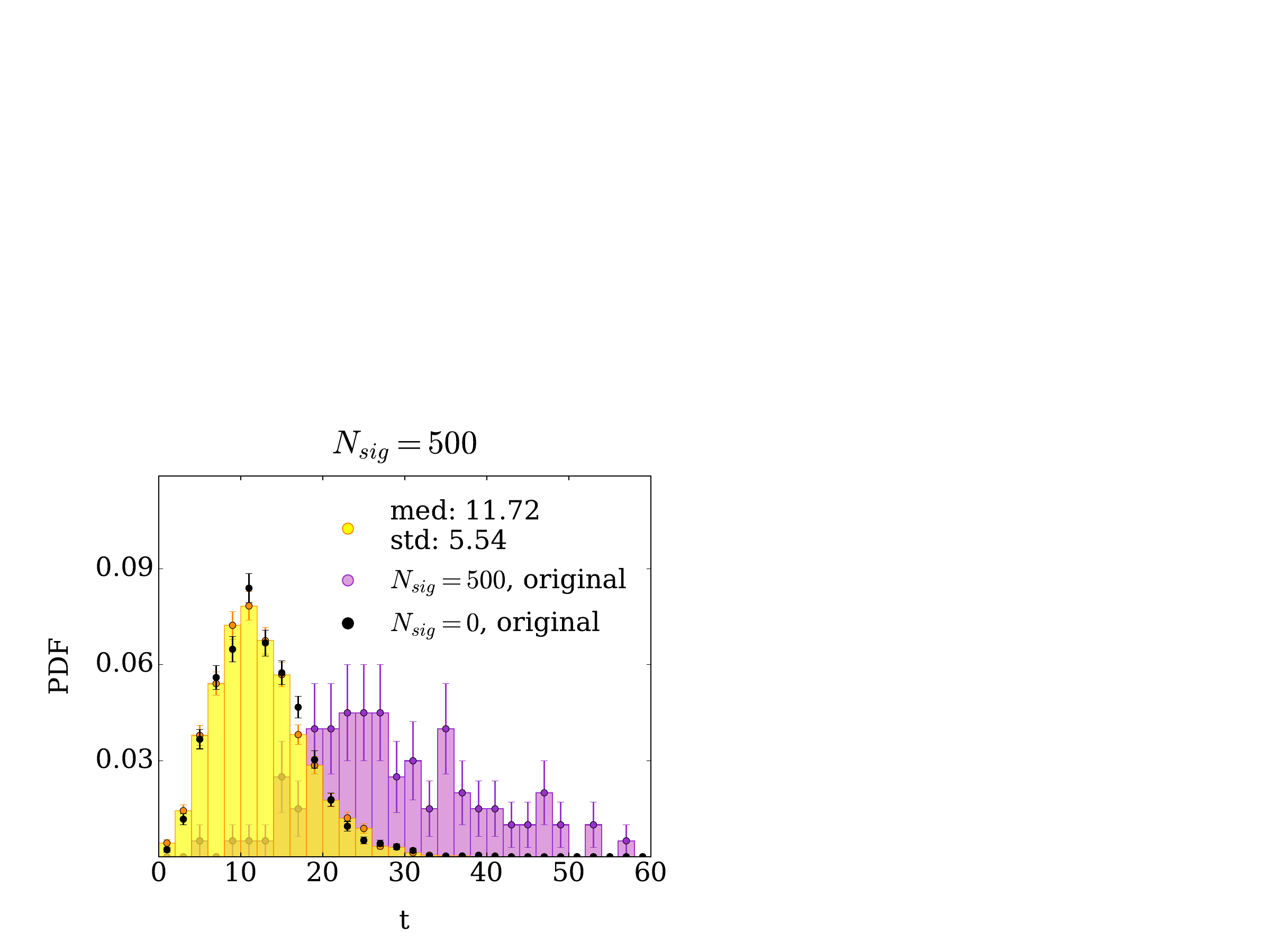}
    \\
    \includetrimmedgraphics{0.038}{0.02}{0.47}{0.4}{0.265}{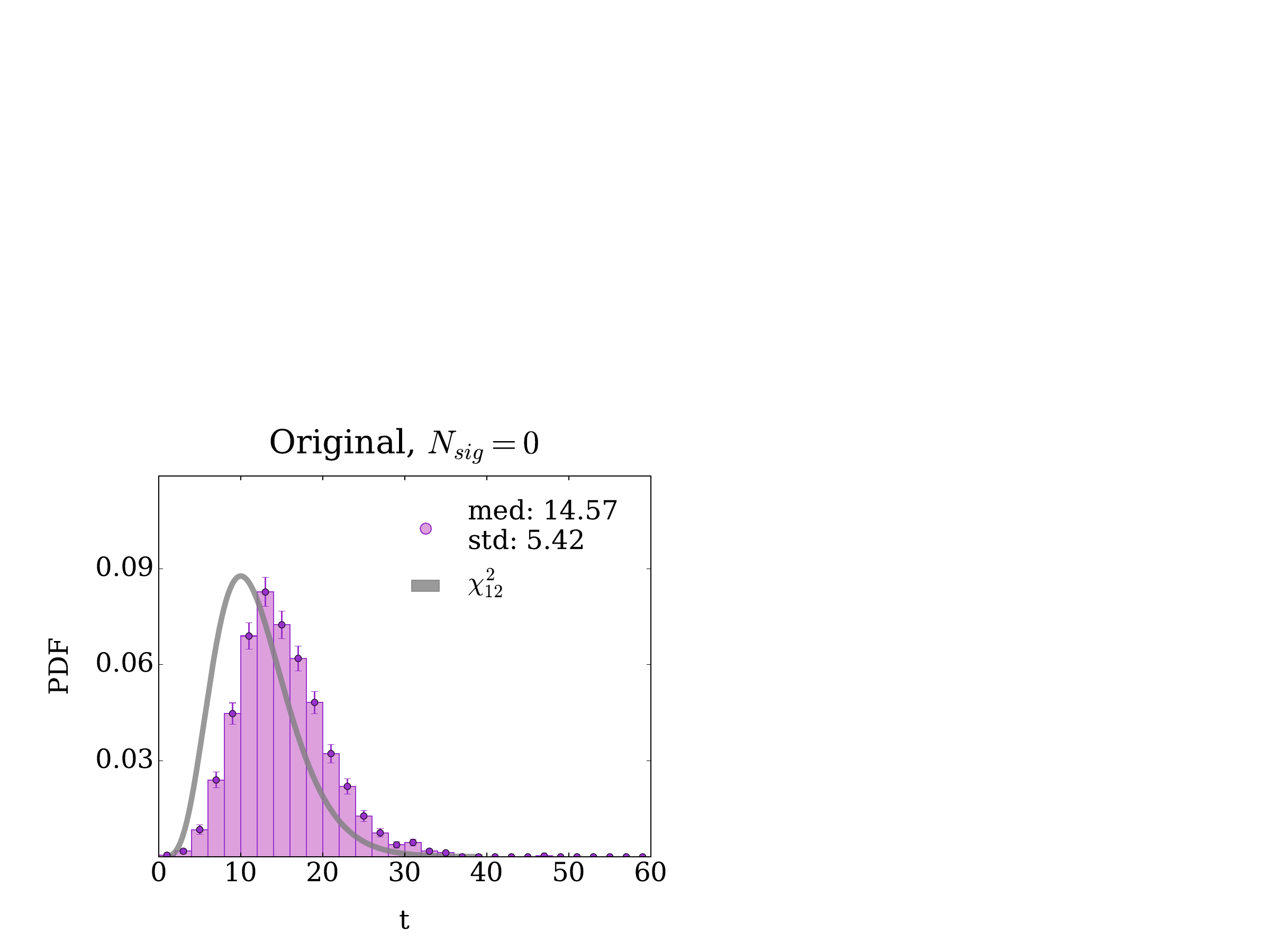}
    \includetrimmedgraphics{0.038}{0.02}{0.47}{0.4}{0.265}{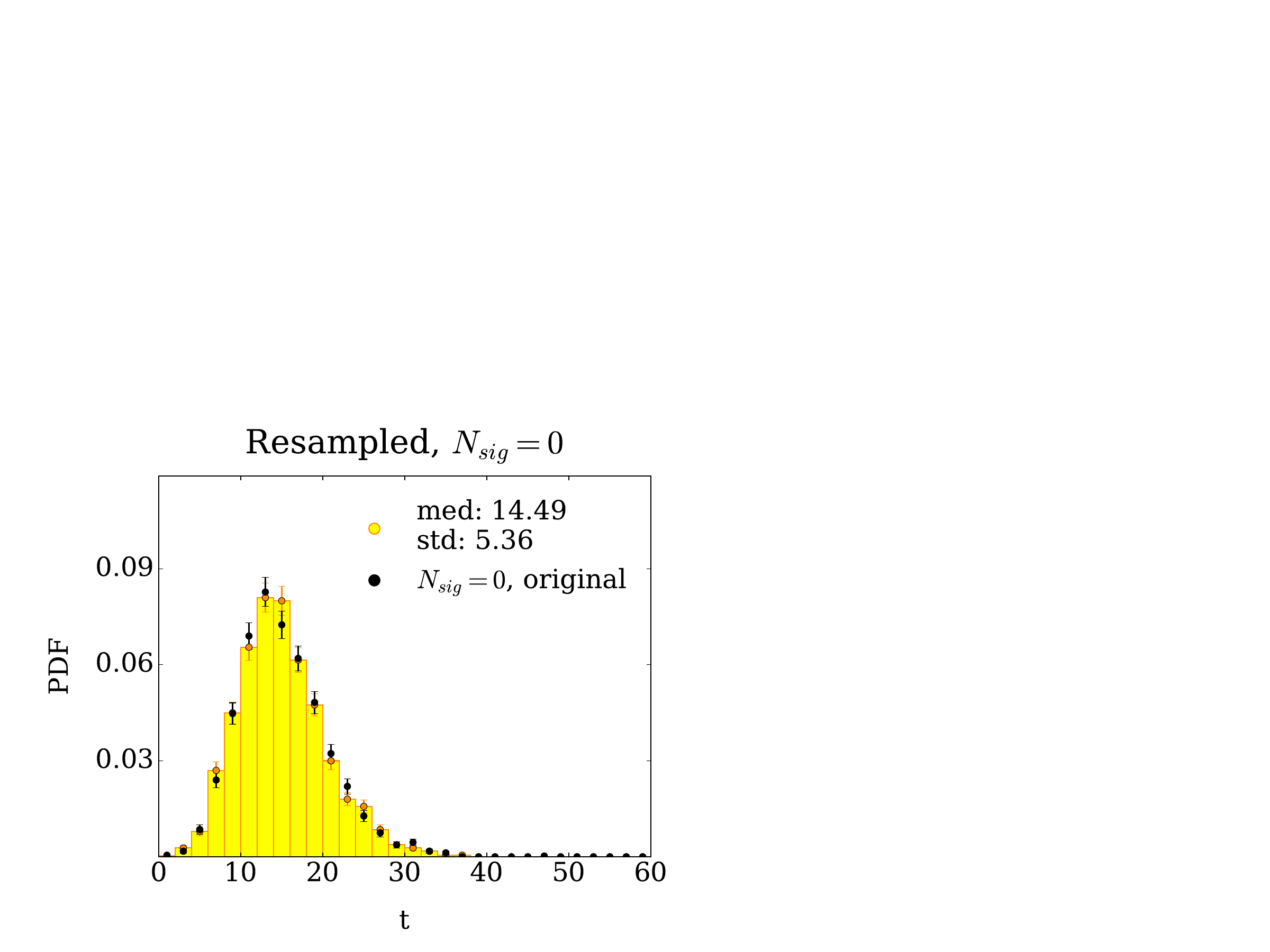}
    \includetrimmedgraphics{0.038}{0.02}{0.47}{0.4}{0.265}{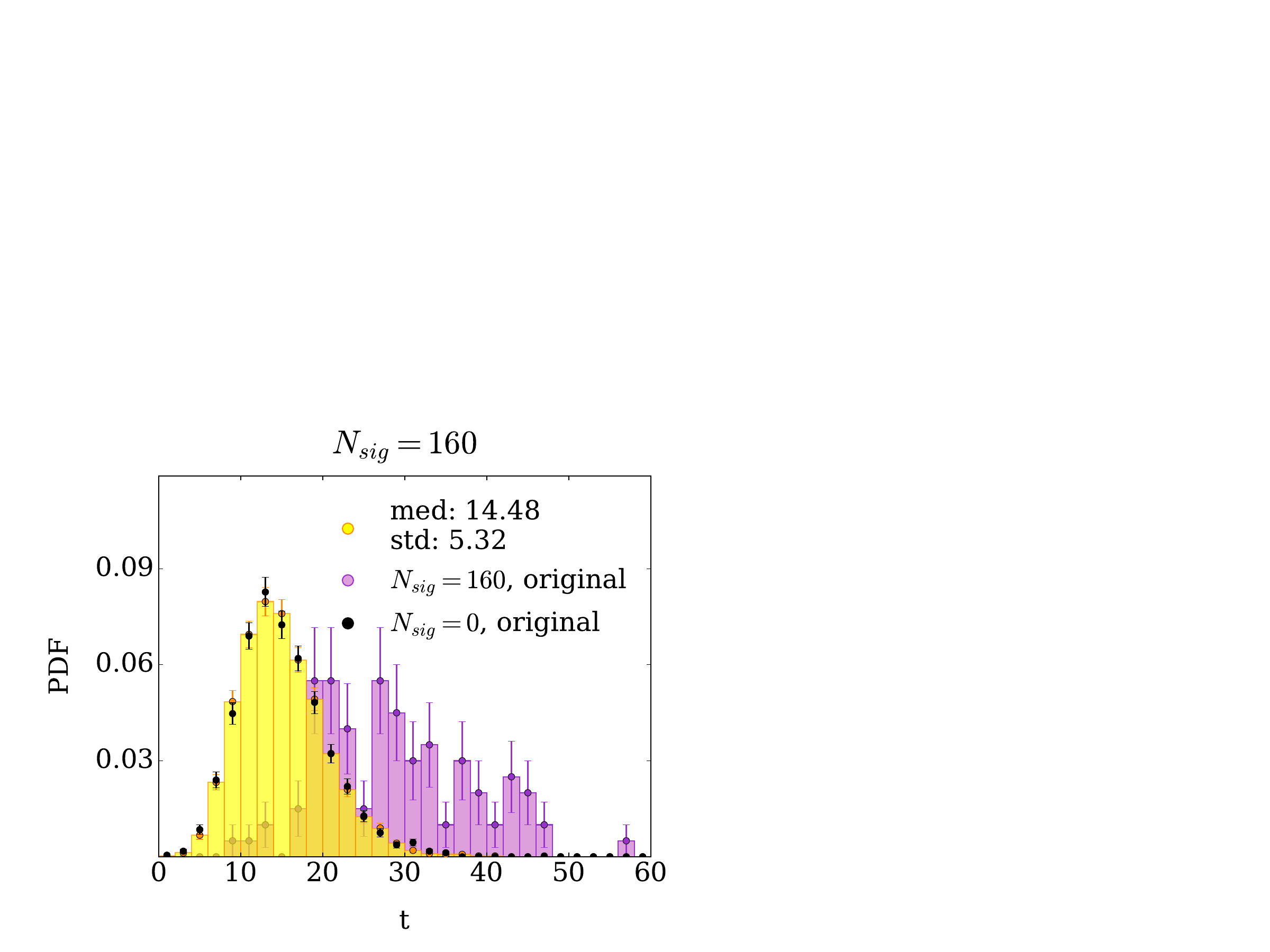}
    \caption{\small \acp{PDF} of the symmetrized test score $t_{\mathbf{A+B}}$ for $N_\mathbf{A} = N_\mathbf{B} \approx 5.5\times 10^{4}$ ($5.5\times 10^{3}$) in the first (second) row. Purple -- $t_{\mathbf{A+B}}$ distribution over datasets generated from the $b_0$ background and expected $N_{\rm sig} = 0\,, 500$ ($N_{\rm sig} =0\,, 160$) $S_3$ signal events. Yellow -- $t_{\mathbf{A+B}}$ distribution over permutations of a representative dataset, corresponding to the median $t_{\mathbf{A+B}}$ for each signal benchmark. The black points in the middle and the right columns  illustrate the $t_{\mathbf{A+B}}$ distribution of the $b_0$ background-generated datasets of the respective sizes (the left column distributions).}
    \label{fig:Permutations}
\end{figure}

The permutation procedure for modeling the background-only distribution from a given dataset is tested in Fig.~\ref{fig:Permutations}. Two distinguished scenarios are examined: the case in which the original background-only \ac{PDF} follows a $\chi^2$ distribution (top row, $N_\mathbf{A} = N_\mathbf{B}\approx 5.5\times 10^{4}$), and the case in which it diverges from it (bottom row, $N_\mathbf{A} = N_\mathbf{B} \approx 5.5\times 10^{3}$). The null hypothesis distribution is given as a reference in the left column for both scenarios. The middle and right columns present the $t_{\mathbf{A}+\mathbf{B}}$ distribution over permutations of a representative dataset, plotted in yellow. In the middle column, the representative dataset contained only background events sampled from $b_0$, corresponding to a $0\sigma$ significance. In the right column, the representative dataset in the top (bottom) row contained an additional expected 500 (160) signal events that were injected into sample $\mathbf{A}$, corresponding to $\sim 2\sigma$ significance. In purple we show the distribution of $t_{\mathbf{A}+\mathbf{B}}$ over datasets sampled from the true $b_0$ background with the additional $S_3$ signal of the respective expected size. 

As seen in the plots, for all these cases, the resulting distribution of $t_{\mathbf{A}+\mathbf{B}}$ from the permuted samples (yellow) was in good agreement with the background-only distribution generated from events sampled from $b_0$ only. In addition, the significance calculated for the unpermuted datasets by comparing their $t_{\mathbf{A}+\mathbf{B}}$ score to the corresponding permuted distributions matched quite closely the significances calculated from the unpermuted background-only samples. For $N_\mathbf{A} = N_\mathbf{B} \approx 5.5\times 10^{4}$, the measured significances relative to the permuted samples were $0.01\sigma$ and $2.17\sigma$\,, respectively, and for $N_\mathbf{A} = N_\mathbf{B} \approx 5.5\times 10^{3}$, the measured significances relative to the permuted samples were $0\sigma$ and $2.09\sigma$\,, respectively. This indicates that using permutations of observed samples to generate the \ac{PDF} of the background-only case is a valid option, which is less consuming than running a full background-only simulation, while still conserving the main particular features of data. Note that the resampled distribution and the fully-known background-only distribution are in good agreement although the fluctuations of the combined $\mathbf{A}+\mathbf{B}$ sample are not captured by permuting a single toy dataset. We validated that this agreement is maintained almost independently of the degree of asymmetry in the permuted dataset. This reflects the robustness of the symmetrized formalism as a test of the symmetry hypothesis -- i.e. of the two datasets being sampled from the same distribution, which is also the only assumption made when generating the null hypothesis distribution by assigning all permuted datasets an equal probability.

\section{Conclusions}\label{sec:Conclusions}
\acresetall

\subsection{Summary}

In this work, we introduced the symmetrized formalism for testing whether two samples, $\mathbf{A}$ and  $\mathbf{B}$, are sampled from the same underlying distribution (null hypothesis, $\mathcal{H}_0$) or from two different distributions (alternative hypothesis, $\mathcal{H}_1$). While our main motivation was to study violations of symmetries of the \ac{SM} as a means for discovering \ac{NP}, our method may be applied to test any assumption predicting two datasets to distribute according to a common, unknown, source. 

Our symmetrized formalism is a generalization of the \ac{NPLM} method, first presented in~\cite{D_Agnolo_2019}. In the \ac{NPLM} formalism, one deduces the distribution of the background model from sample $\mathbf{B}$, which is assumed to be a control sample much larger than sample $\mathbf{A}$. Then, one tests its agreement with the observed distribution of sample $\mathbf{A}$, using a \ac{NN} to parameterize the ratio between the observed \ac{NDF} of sample $\mathbf{B}$, $\tilde{n}_{\mathbf{B}}$, and the \ac{NDF} from which sample $\mathbf{A}$ is sampled, $n_\mathbf{A}$. A maximum log-likelihood ratio test calculated over the observed sample $\mathbf{A}$, $t_{\mathbf{B}}\leri{\mathbf{A}}$, is employed to fit the \ac{NN} parameters and calculate statistical significances. We discussed two major challenges of the \ac{NPLM} method. First, it requires a large ratio between the size of sample $\mathbf{B}$ and the size of sample $\mathbf{A}$ to yield robust and predictable results. This is disadvantageous when searching for small symmetry violations, inducing small differences between samples of similar sizes. Second, the method requires restricting the \ac{NN} weights to control the shape of the null hypothesis \ac{PDF} and avoid divergences of the log-likelihood ratio. 

The symmetrized test, on the other hand, is constructed as the log-likelihood ratio between the null hypothesis assuming $p_\mathbf{A}=p_{\mathbf{B}}$, with $p$ being a \ac{PDF}, and the alternative hypothesis assuming $p_\mathbf{A}\neq p_{\mathbf{B}}$, calculated both on sample $\mathbf{A}$ and on sample $\mathbf{B}$. The background model distribution is deduced from the combined $\mathbf{A+B}$ sample, and taken as the observed one (although other choices of fitting functions are possible). The ratios between the true distribution from which each sample was drawn and the observed distribution of the combined sample are parameterized by two independent \acp{NN}, one for sample $\mathbf{A}$ and one for sample $\mathbf{B}$, maximizing the two log-likelihoods ratios. The symmetrized test score $t_{\mathbf{A+B}}$ is obtained by summing together the two individual tests. 

We have shown that the symmetrized test is less sensitive to the relative sizes of the samples, and only mildly dependent on their absolute sizes. In addition, it avoids artificial divergences when calculating the significance by which the samples are inconsistent with the null hypothesis by construction. Finally, it eliminates the need for fine-tuning the weight clipping parameter of the \ac{NN}, without degrading the search sensitivity. In the context of the \ac{DDP}, the symmetrized formalism has an additional advantage; the \ac{PDF} of the test statistic in the symmetric scenario approaches the $\chi^2_n$ distribution. Thus, it avoids the need to perform time-consuming optimizations for each search in order to reach sufficient sensitivity to asymmetries. The ability to use permutations to generate the distribution of the symmetric scenario from observed data was demonstrated as well. 

The performance of the method was tested with two types of samples. The first type was generated from an analytic exponential decaying function, with various signal shapes injected along the distribution. The second type emulated a search for asymmetries in physical processes containing one electron and one muon of opposite signs in the final state, and a \ac{LFUV} signal expected in processes such as the decay of a Higgs boson into an electron and a tau. 
In all cases, a signal that could be discovered at a $\mathrm{4\sigma-6\sigma}$ significance in an ideal analysis (with an exact knowledge of the background and the signal shape), could be identified at $\mathrm{2\sigma}$ by our method, which requires no prior knowledge of neither the background nor the signal shapes.

\subsection{Immediate and future applications}\label{sec:applications}

The symmetrized formalism may be readily used to test for differences between any two datasets. An expected ``symmetry" relating these datasets does not have to be a \ac{SM} symmetry, but any assumption that two datasets were sampled from the same, unknown, distribution. This includes the standard procedure of comparing a measurement to an auxiliary background measurement, or comparing data to simulation. In particular, since our method is a generalization of the \ac{NPLM} procedure, and alleviates some of its challenges, it may also be used for all cases that would have been addressed by the \ac{NPLM}. In the context of searching for violations of \ac{SM} symmetries, one can apply our method as is to analyses in which the symmetry should not be significantly violated by detector effects. Some examples include searching for CP violation in $e^+e^-$ data, and searching for asymmetries between $e^+\mu^-$ and $e^-\mu^+$ samples in $pp$ collisions~\cite{ATLAS:2021tar}. 

We note that such realistic applications may benefit from performing a multi-dimensional analysis, using more than one kinematical variable. Although here we demonstrated the symmetrized formalism in one-dimensional analyses, it could be easily generalized to the multi-dimensional case by changing the input layer of the \acp{NN}. However, a more detailed study of the performance using different choices of kinematical variables should be conducted for specific physics cases. Additionally, the extent to which the asymptotic approximation holds when multiple variables are used, as well as the gain in sensitivity, should be tested.

Following studies could also expand the scope of the current work to account for known or expected asymmetries, one example being potential systematic discrepancies between the samples. In particular, in searches for \ac{LFUV}, detector effects related to the different trigger, reconstruction, identification and isolation efficiency values of the different leptons, as well as the different probability that other objects are misidentified as these leptons, should be addressed. These can be viewed as additional nuisance parameters, which can be restricted by simulation or other measurements. As mentioned, a method for correcting these effects has been introduced and used in~\cite{ATLAS:2023mvd}. It has shown that an efficiency ratio factor can be applied to each measured event and cancel asymmetries induced by the different efficiency values, while the contribution from misidentified leptons can be treated similarly to any analysis using leptons. One could either apply this method directly to reweigh the samples, or use it to model the expected yield for electrons and muons. Another possibility, inline with the approach we have taken here, is to use the symmetrized formalism to test for differences between the auxiliary measurements of the asymmetry within the \ac{SM}, and the asymmetry found in data. 

\section{Alternative approaches and open questions}
\label{sec:AAOQ}

\subsection{Fitting the symmetric and asymmetric components and the cross-entropy loss}\label{subsubsec:cross_entropy}

In this work, we have been interested in the \textbf{difference} between $p_\mathbf{A}\leri{x}$ and $p_\mathbf{B}\leri{x}$, or in the deviation of $f\leri{x}-g\leri{x}$ from constant. Rather than choosing $N_\mathbf{A}$ and $N_{\mathbf{B}}$ as our functions of interest, another possible choice of variables, which seems rather natural in our case, is
\begin{align}
    N\leri{x} &\equiv N_{\mathbf{A}}\leri{x}+N_{\mathbf{B}}\leri{x} = 2e^{\frac{f'\leri{x}+g'\leri{x}}{2}}\cosh\leri{\frac{f'\leri{x}-g'\leri{x}}{2}}\tilde{N}\leri{x} = 2 e^{\Sigma\leri{x}}\cosh\leri{\delta\leri{x}}\tilde{N}\leri{x}\,,\\
    \Delta\leri{x}& \equiv N_{\mathbf{A}}\leri{x}-N_{\mathbf{B}}\leri{x} = 2e^{\frac{f'\leri{x}+g'\leri{x}}{2}}\sinh\leri{\frac{f'\leri{x}-g'\leri{x}}{2}}\tilde{N}\leri{x} = \tanh\leri{\delta\leri{x}}N\leri{x}\,,
\end{align}
where $f'\leri{x} = f\leri{x}+\log\leri{\tilde{N}_{\mathbf A}/\leri{\tilde{N}_{\mathbf A}+\tilde{N}_{\mathbf B}}}$\,, $g'\leri{x} = g\leri{x}+\log\leri{\tilde{N}_{\mathbf B}/\leri{\tilde{N}_{\mathbf A}+\tilde{N}_{\mathbf B}}}$, $\Sigma\leri{x} = \frac{f'\leri{x}+g'\leri{x}}{2}$ and $\delta\leri{x} = \frac{f'\leri{x}-g'\leri{x}}{2}$. The deviation of $\delta\leri{x}$ from a constant is our function of interest. The maximal log-likelihood of a hypothesis $\mathcal{H}$ can be written as
\begin{align}
&\text{max}\leri{2\log\leri{\mathcal{L}\leri{\mathcal{H}|\mathbf{A,B}}}}=-2\left[\hat{N}_{\mathbf{A}}-\sum_{x\in\mathbf{A}}\log\leri{\hat{N}_{\mathbf{A}}\leri{x|\mathcal{H}}}+\hat{N}_{\mathbf{B}}-\sum_{x\in\mathbf{B}}\log\leri{\hat{N}_{\mathbf{B}}\leri{x|\mathcal{H}}}\right] \nonumber \\ 
    &=-2\left[\sum_{x\in{\mathbf{A},\mathbf{B}}}\hat{N}\leri{x|\mathcal{H}}-\sum_{x\in{\mathbf{A},\mathbf{B}}}\log\leri{\hat{N}\leri{x|\mathcal{H}}}-\sum_{x\in\mathbf{A}}\log\leri{\sigma\leri{2\hat{\delta}\leri{x|\mathcal{H}}}}-\sum_{x\in\mathbf{B}}\log\leri{1-\sigma\leri{2\hat{\delta}\leri{x|\mathcal{H}}}}\right]\,.
\end{align}

In principle, one could fit both $\delta\leri{x}$ and $\Sigma\leri{x}$ with two separate \ac{NN} functions (or one bigger \ac{NN} with a two-dimensional output), however, the training can no longer be generically split into two independent trainings.

Another possibility is to consider $N\leri{x}$ to be independent of $\delta\leri{x}$. In that case, we get that our test statistic for $\delta$ is exactly the total cross-entropy. This makes sense, as the cross-entropy is just the likelihood of assigning either the label $\mathbf{A}$ or the label $\mathbf{B}$ to observed events, with a labeling probability $p^\delta_\mathbf{A} = \frac{n_{\mathbf{A}}\leri{x}}{n_{\mathbf{A}}\leri{x}+n_{\mathbf{B}}\leri{x}}=\frac{1}{1+e^{-2\delta\leri{x}}}$ and $p^\delta_\mathbf{B} = \frac{n_{\mathbf{B}}\leri{x}}{n_{\mathbf{A}}\leri{x}+n_{\mathbf{B}}\leri{x}}=\frac{1}{1+e^{2\delta\leri{x}}}$, respectively. The symmetric null hypothesis ($\delta\leri{x}=\mathrm{constant}$) would set the labeling probabilities, $p^\delta_\mathbf{A}$ and $p^\delta_\mathbf{B}$, to be independent of $x$, and just equal to the observed frequencies (or to a prior frequency set by the null hypothesis). Accordingly, the final test statistic would be
\begin{align}
 t^\delta_{\mathbf{A}+\mathbf{B}}\leri{\mathbf{A}+\mathbf{B}} = 2\sum_{x\in\mathbf{A}}\log\left(\frac{\leri{\tilde{N}_{\mathbf A}+\tilde{N}_{\mathbf B}}\sigma\leri{2\hat{\delta}\leri{x}}}{\tilde{N}_{\mathbf A}}\right)+2\sum_{x\in\mathbf{B}}\log\left(\frac{\leri{\tilde{N}_{\mathbf A}+\tilde{N}_{\mathbf B}}\leri{1-\sigma\leri{2\hat{\delta}\leri{x}}}}{\tilde{N}_{\mathbf B}}\right).\label{eq:t_delta_cross_entropy}
\end{align}

In this context, we note that other proposals using the cross-entropy as the loss, namely - most classifiers, are really performing a maximum likelihood test. In particular, the \ac{NN} test score presented in~\cite{Birman:2022xzu} for identifying symmetry violations would have been equivalent (up to a linear transformation) to $t^\delta_{\mathbf{A}+\mathbf{B}}\leri{\mathbf{A}+\mathbf{B}}$ in Eq.~\eqref{eq:t_delta_cross_entropy}, had the batch size been taken to be the full sample. The original \ac{NPLM} paper~\cite{D_Agnolo_2019} has examined the use of the cross-entropy for finding $n_\mathbf{A}\leri{x}/n_\mathbf{B}\leri{x}$, a proposal that was later implemented in two following papers~\cite{Letizia:2022xbe,Grosso:2023ltd}. 
The cross-entropy, unlike $t_{\mathbf B}\leri{\mathbf A}$, does not diverge, which follows from the same logic forbidding divergences of the symmetric test statistic with the symmetric null hypothesis. Had the cross-entropy score itself been used for statistical inference, it would have been equivalent to the symmetric test statistic in Eq.~\eqref{eq:t_delta_cross_entropy}. 
In~\cite{D_Agnolo_2019} and~\cite{Letizia:2022xbe}, the resulting $n_\mathbf{A}\leri{x}/n_\mathbf{B}\leri{x}$ found by minimizing the cross-entropy loss was substituted back into $t_{\mathbf B}\leri{\mathbf A}$ in Eq.~\eqref{eq:t_nref}, which is thus subject to the same conditions discussed in Sec.~\ref{subsec:NPLM_challenges}. On the other hand, in~\cite{Grosso:2023ltd}, the cross-entropy \acs{MLE} is substituted into a test statistic assuming the total number of events does not fluctuate, and is thus linearly related to the second term in the parentheses of Eq.~\eqref{eq:t_nref}. In this case, since the total number of events is not included in the likelihood, one should have added a Lagrange multiplier term setting the normalization of $p_{\mathbf A}\leri{x}\equiv e^{f\leri{x}}p_{\mathbf B}\leri{x}$ to make it an appropriate probability function, or should have restricted the form of $f\leri{x}$ to satisfy this constraint (see~\cite{Nachman:2021yvi}). However, the likelihood ratio is still unbounded, of course, and diverges if the cross-entropy yields $f\leri{x_\star}\rightarrow \infty$, as it could find points that are included in sample $\mathbf{A}$ but not in sample $\mathbf{B}$. This is the same issue explained in Sec.~\ref{subsec:NPLM_challenges}, stemming from the selected null hypothesis.

\subsection{Overfitting}
One of the reasons for choosing a likelihood-based test statistic is its well-known asymptotic distribution under the null hypothesis. While here we found a relatively good agreement with the predicted asymptotic distribution, one should note that this agreement is of course not perfect, and may depend on the hyper-parameters. Within the \ac{NPLM} framework, the deviation from the asymptotic distribution is treated as a result of overfitting -- i.e. perfectly fitting individual points, rather than the ensemble trends. Therefore, in~\cite{D_Agnolo_2019,DAgnolo:2019vbw,dAgnolo:2021aun}, the weights of the \ac{NN} have been restricted to a maximal absolute value (weight clipping), which was found by demanding some agreement of the background-only distribution on the test statistic with the asymptotic $\chi^2_n$ distribution. Here, using our symmetrized formalism, we have shown that the weight clipping could be removed, and the $\chi^2$ agreement was maintained for arbitrary ratios of sample sizes when the other hyper-parameters were similar to those used in~\cite{dAgnolo:2021aun}. On the other hand, small drifts from the asymptotic distribution were observed when increasing the number of epochs, and it is understandably more significant when the number of events is smaller. Although these deviations are negligible for setting the $2\sigma$ or $3\sigma$ sensitivity of the method, it would be useful to understand their origin, characterize their dependence on the model parameters, estimate their maximal severity and come up with potential ways to minimize them.

\begin{figure}
\centering
    \includetrimmedgraphics{0}{0}{0}{0}{0.3}{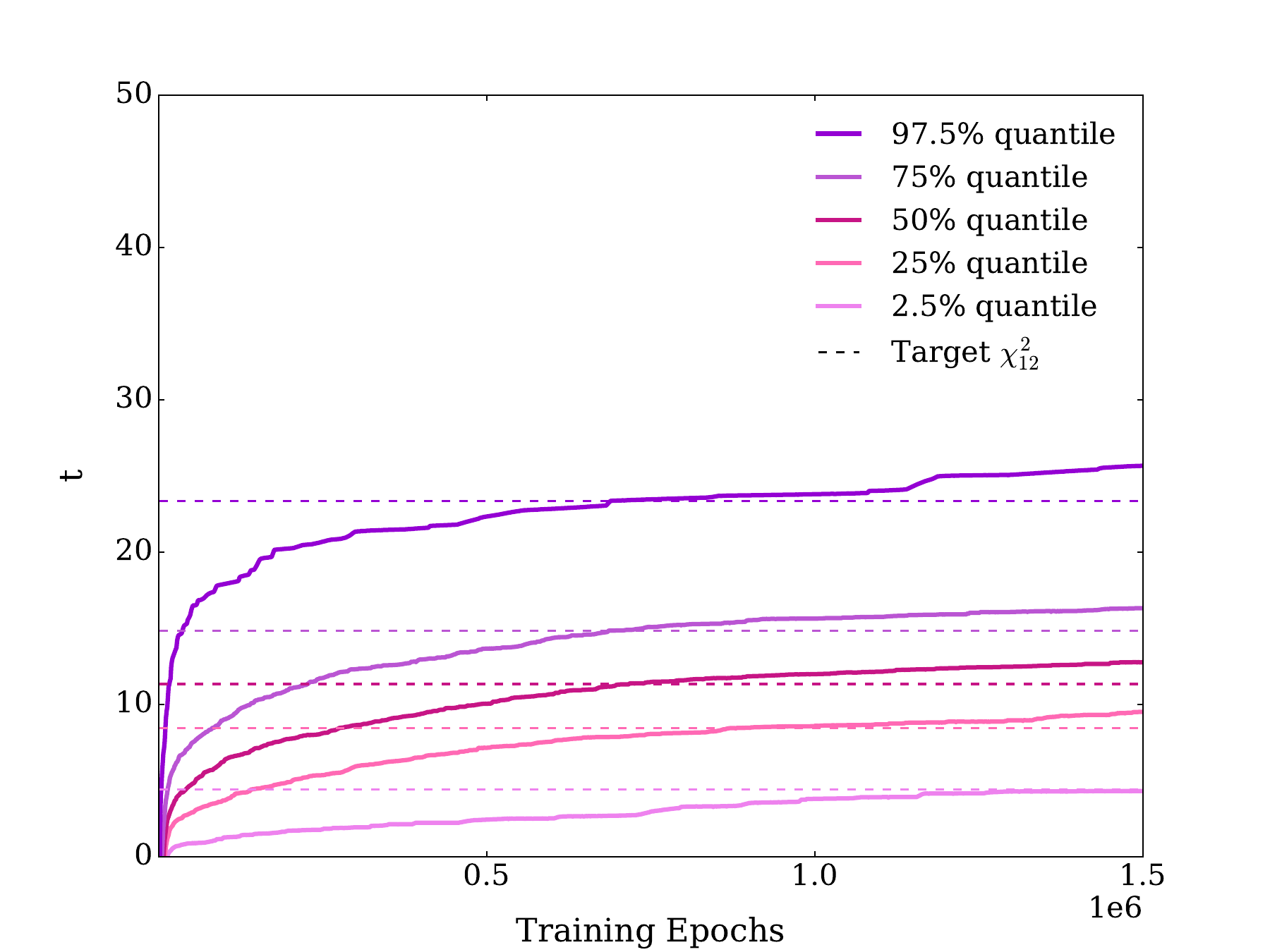}\\
    \caption{The percentiles of $t_{\mathbf{A+B}}\leri{\mathbf{A+B}}$ as a function of the number of epochs, for exponential distribution samples of sizes $N_\mathbf{A}=N_\mathbf{B}\approx 10^{5}$. The dashed lines represent the corresponding percentiles for a $\chi^2_{12}$ distribution.}
    \label{fig:Percentiles}
\end{figure}

While neither sample $\mathbf{A}$ nor sample $\mathbf{B}$ can be perfectly fit with the parameterization above and a limited number of variables, the choice of fitting functions given by a \ac{NN} with one hidden layer of $N_\text{neu}$ neurons may yield some degree of overfitting. An example of an ``overfit" solution, where the longest sequences of points appearing only in sample $\mathbf{A}$, $x^\star_{\mathbf{A}}$, and of points only appearing in sample $\mathbf{B}$, $x^\star_{\mathbf{B}}$ are isolated, is explained in appendix~\ref{appendix:overfit}. 
However, the gradient of the loss around these ``overfit" solutions is finite and small, while the parameters corresponding to it are quite far away from the initialized values. Therefore, gradient-descent-based methods are not guaranteed to find these solutions within a finite number of epochs. Indeed, as shown in Fig.~\ref{fig:Percentiles}, over a total of $1.5$~M epochs, these solutions were not found, but some degree of overfitting has been observed.

One way to try and reduce the degree of overfitting is by requiring that a deviation from the null hypothesis, i.e. from constant $f$ and $g$, would be correlated between the two samples. In the ``overfit" solutions, each fitting function is free to find $x^\star_{\mathbf{B}}$ and $x^\star_{\mathbf{A}}$ independently. If the two samples were drawn from the same distribution, the points $x^\star_{\mathbf{B}}$ and $x^\star_{\mathbf{A}}$ are by definition different from each other, and they might not even be close in $x$ space. However, if a true signal is responsible for finding a large number of say $x^\star_{\mathbf{A}}$ points, the points $x^\star_{\mathbf{A}}$ will also be identified in the fit for the $\mathbf{A}$ sample as having an excess of events, in addition to being identified as missing from the $\mathbf{B}$ sample. 

 In this regard, one may favor the parameterization introduced in Sec.~\ref{subsubsec:cross_entropy}, where the asymmetric part $\delta$ and the symmetric part $\Sigma$ are fitted. In the previous parameterization used in the main text, while effectively $\delta\leri{x}$ had the same number of degrees of freedom, the function describing it could have had two sinks and two peaks simultaneously, corresponding to the four non-overlapping regions of $x^\star_{\mathbf{B}}$ and $x^\star_{\mathbf{A}}$. Using this alternative parameterization, $\delta$ can only have two sinks, one sink and one peak, or two peaks, hopefully making a true signal easier to detect over statistical fluctuations.

 Another option would be to implement standard solutions from the \acs{ML} literature that were intended to tackle overfitting; e.g. early stopping (with or without cross-validation) and including penalty terms that effectively constrain the complexity of the model. While all these solutions could significantly reduce the degree of overfitting and could improve the sensitivity to asymmetries, there is no guarantee that they will yield the asymptotic $\chi^2$ distribution. Therefore, one should either characterize the obtained null hypothesis distribution using these tools, either empirically or analytically, or, as done in~\cite{dAgnolo:2021aun,Letizia:2022xbe}, tune the hyper-parameters controlling the impact of these effective constraints such that the asymptotic $\chi^2$ distribution is maintained, hopefully generically -- if possible. While we leave a detailed study of these options to future work, we note that using a validation set to fix a point for early stopping yielded a null hypothesis distribution that is much narrower than the expected $\chi^2_{12}$ for our case. This is to be expected, as within the null hypothesis, the two samples are indeed generated from the same distribution, and thus there is no real ``information" to be learned that would be significantly captured by a validation set, also containing a different set of samples that were drawn from the same distribution. In this case, another possibility would be to set a generic early stopping rule by defining a fixed number of epochs for the training, which could be chosen by some desired sensitivity to signals.

\begin{acknowledgments}
We thank Andrea Wulzer, Gaia Grosso, Raffaele Tito D'Agnolo and Benjamin Nachman for useful discussions and insightful comments on the manuscript. We also thank Mattias Birman for his collaboration in the early stages of this work.
This work is supported by the Sir Charles Clore Prize, grants from the Israel Science Foundation (grant number 2871/19), the German Israeli Foundation (grant number I-1506-303.7/2019) and the YedaSela (YeS) Center for Basic Research. IS acknowledges the support of the Weizmann Institute of Science and the Ariane de Rothschild Women Doctoral Program during principal parts of this work. IS is currently supported by the Office of High Energy Physics of the U.S. Department of Energy under contract DE-AC02-05CH11231 and by the CHE/PBC Fellowship for Outstanding Women Postdoctoral Fellows. 
\end{acknowledgments}

\appendix{}

\section{Signal and background distributions for \acs{LFUV} search}

\label{app:datasets}

\begin{figure}[h]
    \centering
\includetrimmedgraphics{0.038}{0.02}{0.47}{0.4}{0.35}{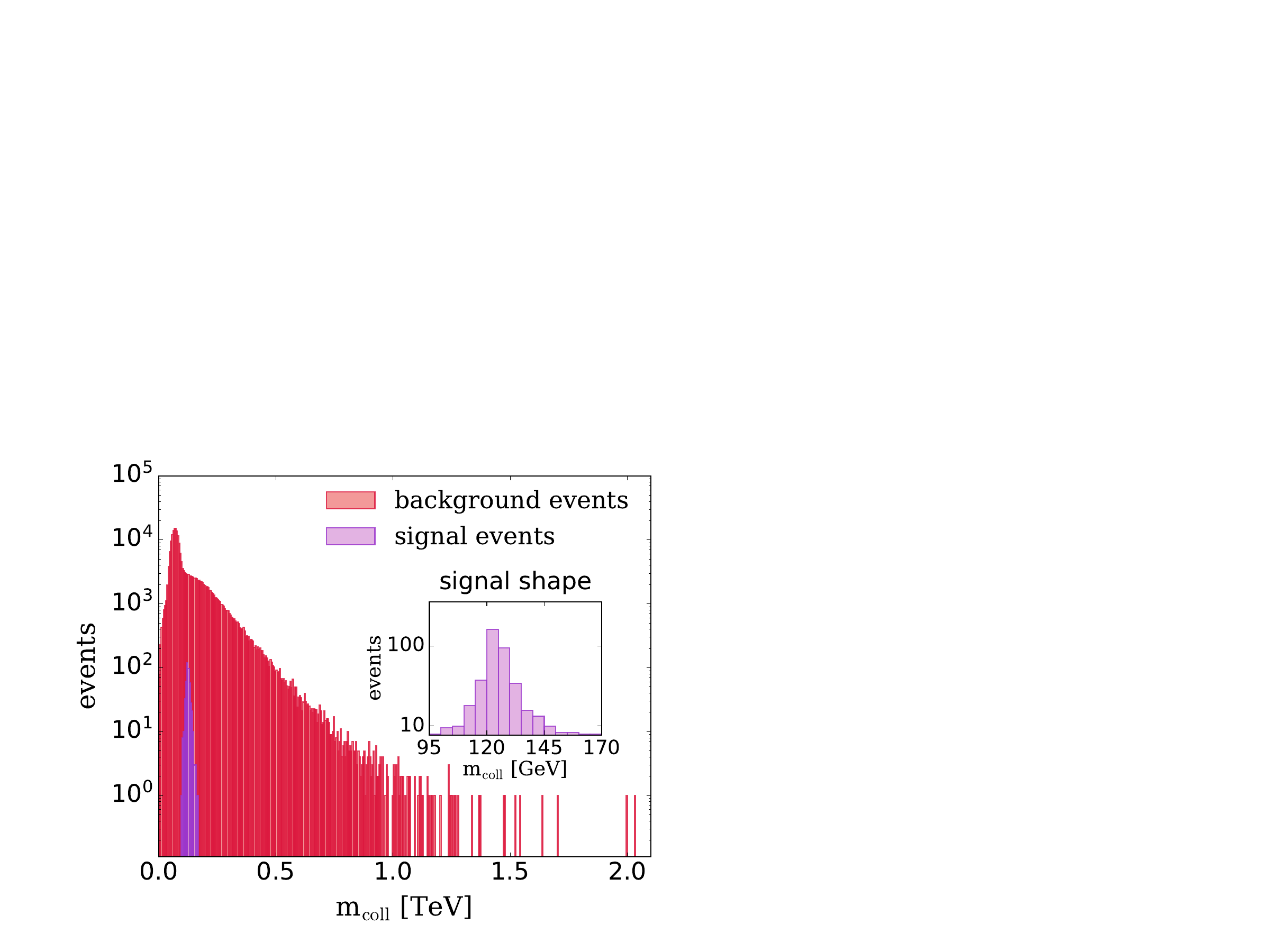}
    \caption{Collinear mass distributions of simulated Higgs \ac{LFUV} and background events.}
    \label{fig:data distribution}
\end{figure}

Fig. \ref{fig:data distribution} shows the collinear mass distributions of the signal ($H\rightarrow\tau e\rightarrow \mu e 2\nu$) and background samples used to study the sensitivity of the symmetrized method to Higgs \ac{LFUV} signal.

\section{\acs{NPLM} with known number of background events}

\label{appendix:moreresults}

\begin{figure}[H]
	\centering
 \includetrimmedgraphics{0.038}{0.02}{0.47}{0.4}{0.274}{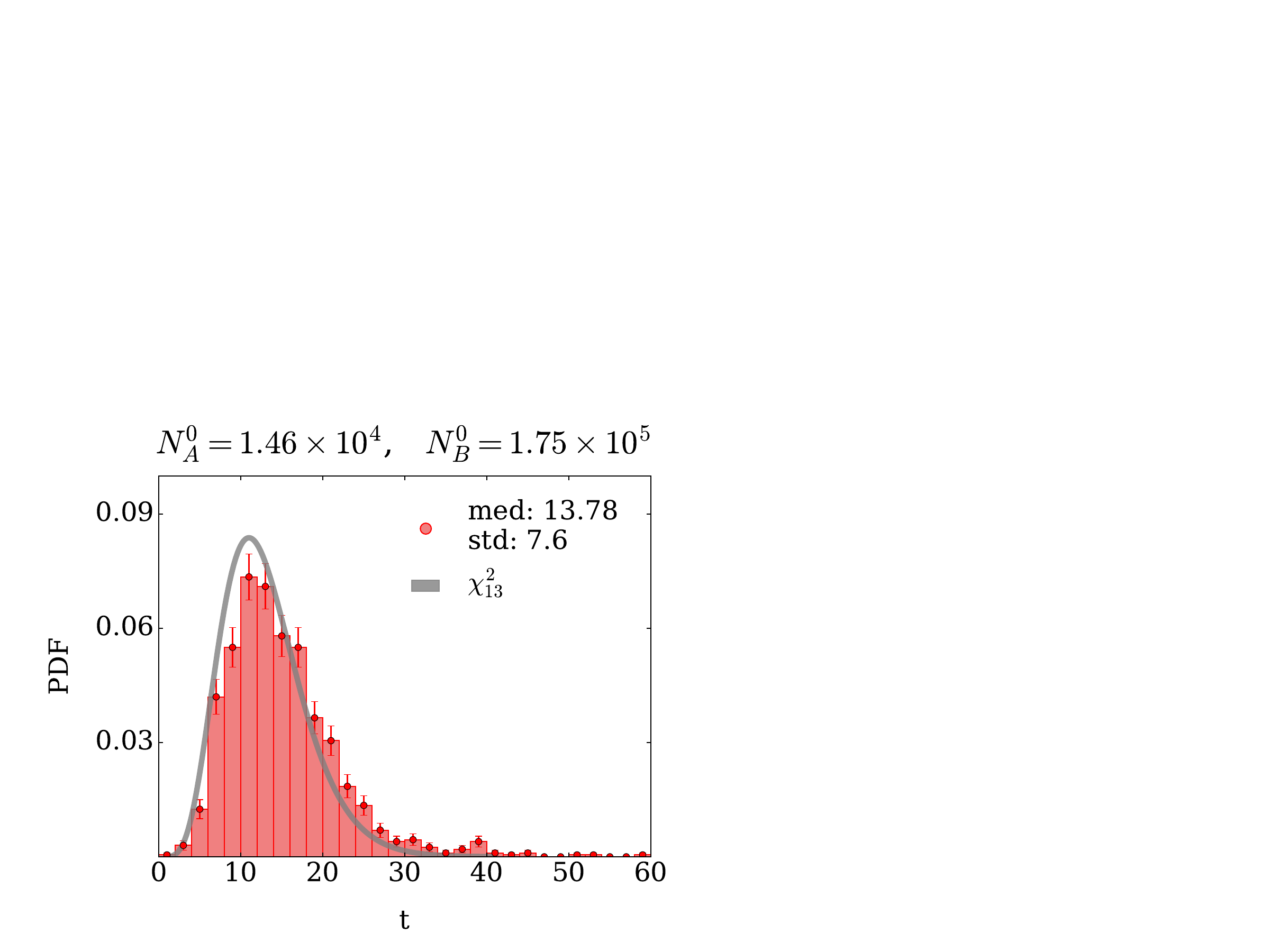}
\includetrimmedgraphics{0.038}{0.02}{0.47}{0.4}{0.274}{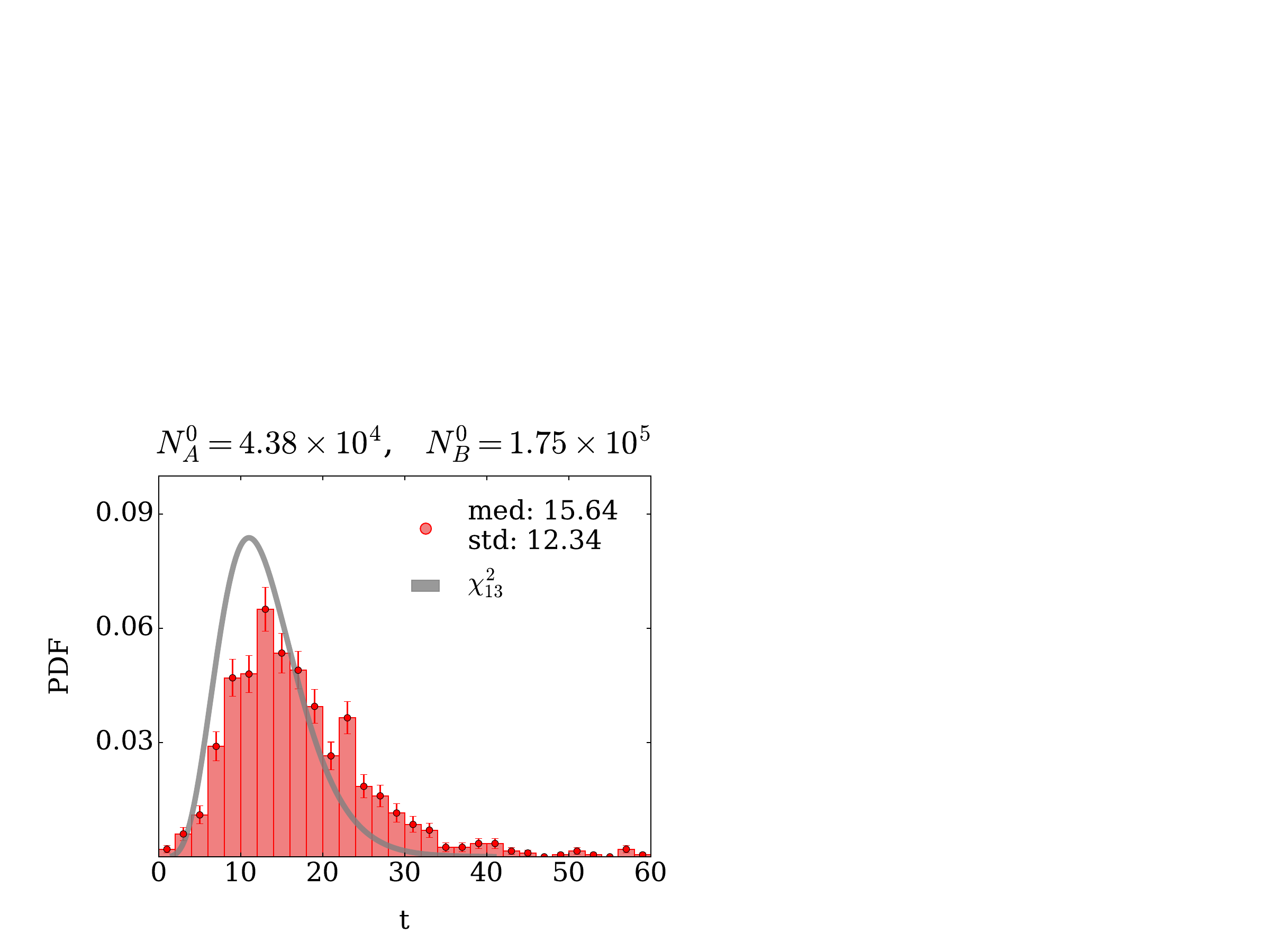}
\includetrimmedgraphics{0.038}{0.02}{0.47}{0.4}{0.274}{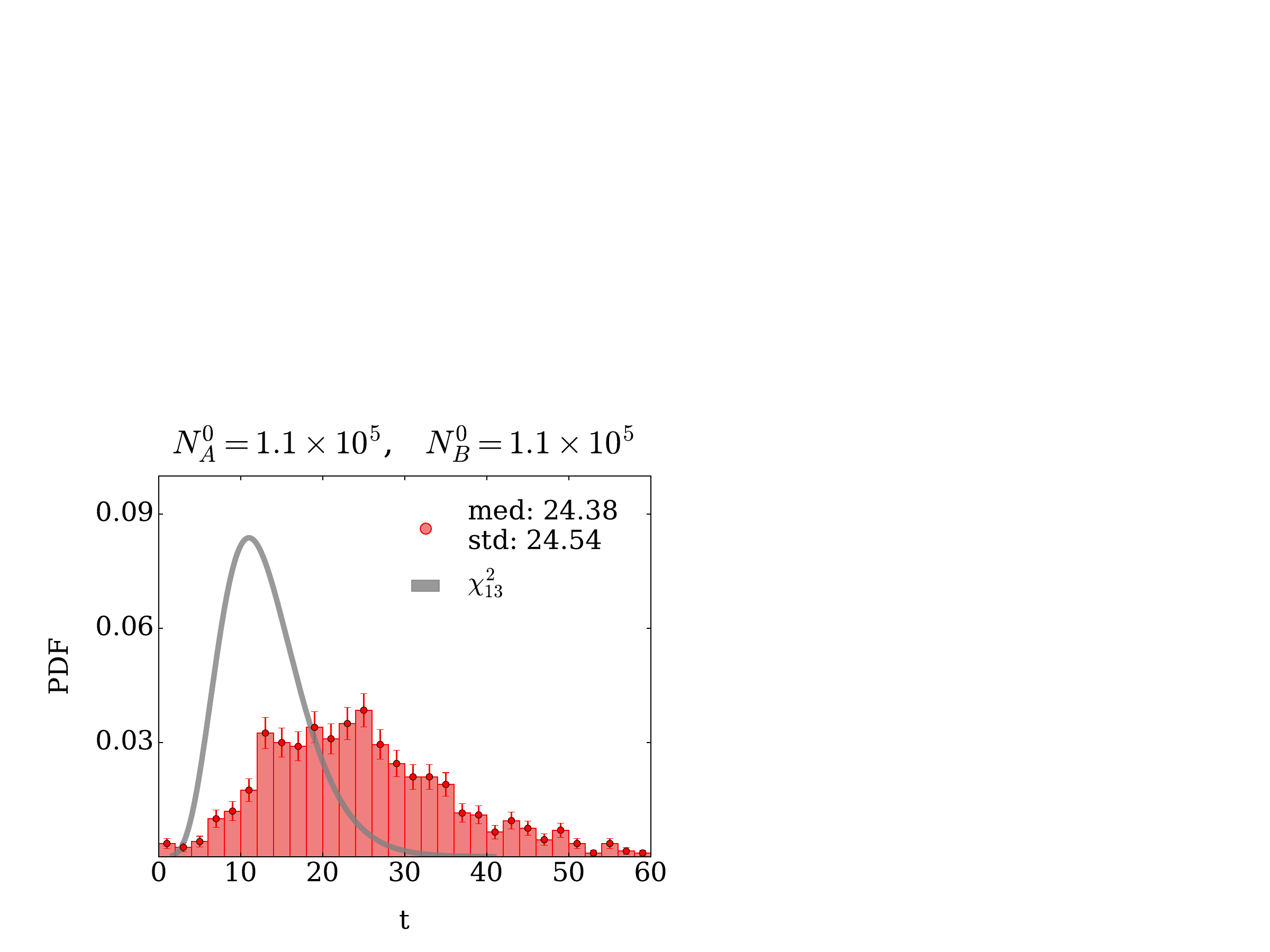}
	\caption{The distribution of $t_{\mathbf{B}}\leri{\mathbf{A}}$ under the null hypothesis. Left -- $N_\mathbf{B} = 1/2 N$ and $N_\mathbf{A} = 1/2 N$, middle -- $N_\mathbf{B} = 4/5 N$ and $N_\mathbf{A} = 1/5 N$, right -- $N_\mathbf{B} = 4/5 N$ and $N_\mathbf{A} = 1/15 N$ with $N\approx 2.2\times 10^{5}$. All are calculated with weight clipping of 9. Solid gray -- the expected $\chi^2_{13}$ distribution according to the Wilks-Wald theorem.}
	\label{fig:t_A_null_dist_13}
\end{figure}

In Fig.~\ref{fig:t_A_null_dist_13} we show the null hypothesis distributions of the \ac{NPLM} test statistic $t_\mathbf{B}\leri{\mathbf{A}}$ for the implementation in~\cite{D_Agnolo_2019}. Here, the total number of events in sample $\mathbf{B}$, $\tilde{N}_\mathbf{B}$, was fixed\footnote{This is consistent with the implementation in~\cite{D_Agnolo_2019}, but in any case the total number of events in sample $\mathbf{B}$ does not play much of a role in $t_{\mathbf{B}}\leri{\mathbf A}$. This is because the sum over sample $\mathbf{B}$ is normalized, and therefore statistical fluctuations of $\tilde{N}_\mathbf{B}$ would only enter through the sample having more or less statistics, but this is insignificant.}, and the total number of events in sample $\mathbf{A}$, $\tilde{N}_\mathbf{A}$, was Poisson distributed, while $N_\mathbf{A}$ was set to be the expected number of events (the Poisson mean). Therefore, the number of degrees of freedom of the asymptotic $\chi^2$ should be the same as the number of parameters in the \ac{NN} describing $f$, since the null hypothesis is fully constrained. 

\section{Ideal symmetrized significance}\label{appendix:ideal_t_AB}

Let us characterize our symmetrized test statistic in terms of the ideal score, obtained when the \ac{NN} is able to perfectly reconstruct the signal and the symmetric background. When the expected number of background events in both samples is known, the ideal $t_\mathbf{A+B}\leri{\mathbf{A+B}}$ score in Eq. \eqref{eq:t_0}, can be approximated by
\begin{align}
t_\mathbf{A+B}\leri{\mathbf{A+B}}^{\rm max}&\approx 2\Bigg[-N_s+\sum \leri{N_s(x)+N^\mathbf{A}_b(x)} \log\frac{\leri{N_s(x)+N^\mathbf{A}_b(x)}}{N^\mathbf{A}_b(x)}+\nonumber\\
&-\sum \leri{N_s(x)+N^\mathbf{A}_b(x)+N^\mathbf{B}_b(x)} \log\frac{N_b^\mathbf{A}\leri{N_s(x)+N^\mathbf{A}_b(x)+N^\mathbf{B}_b(x)}}{\leri{N_\mathbf{B}+N_\mathbf{A}}N^\mathbf{A}_b(x)}\Bigg]\,, 
\end{align}
which asymptotes $q_0$ from Eq.~\eqref{eq:Zid_binned} for $N_\mathbf{B}\gg N_\mathbf{A}$. In our case, we don't assume prior knowledge of the number of background events, and thus
\begin{align}
t_\mathbf{A+B}\leri{\mathbf{A+B}}^{\rm max}&\approx 2\Bigg[(N_s+N^\mathbf{A}_b)\log\leri{1-\frac{N_s}{N_s+N^\mathbf{A}_b}}+\sum \leri{N_s(x)+N^\mathbf{A}_b(x)} \log\frac{\leri{N_s(x)+N^\mathbf{A}_b(x)}}{N^\mathbf{A}_b(x)}\nonumber\\
&-\sum \leri{N_s(x)+N^\mathbf{A}_b(x)+N^\mathbf{B}_b(x)} \log\frac{N_b^\mathbf{A}\leri{N_s(x)+N^\mathbf{A}_b(x)+N^\mathbf{B}_b(x)}}{\leri{N_\mathbf{B}+N_\mathbf{A}}N^\mathbf{A}_b(x)}\Bigg]\,.
\end{align}

\section{Example of overfit solutions}
\label{appendix:overfit}
\begin{figure}[H]
    \centering
\includetrimmedgraphics{0.038}{0.02}{0.47}{0.4}{0.35}{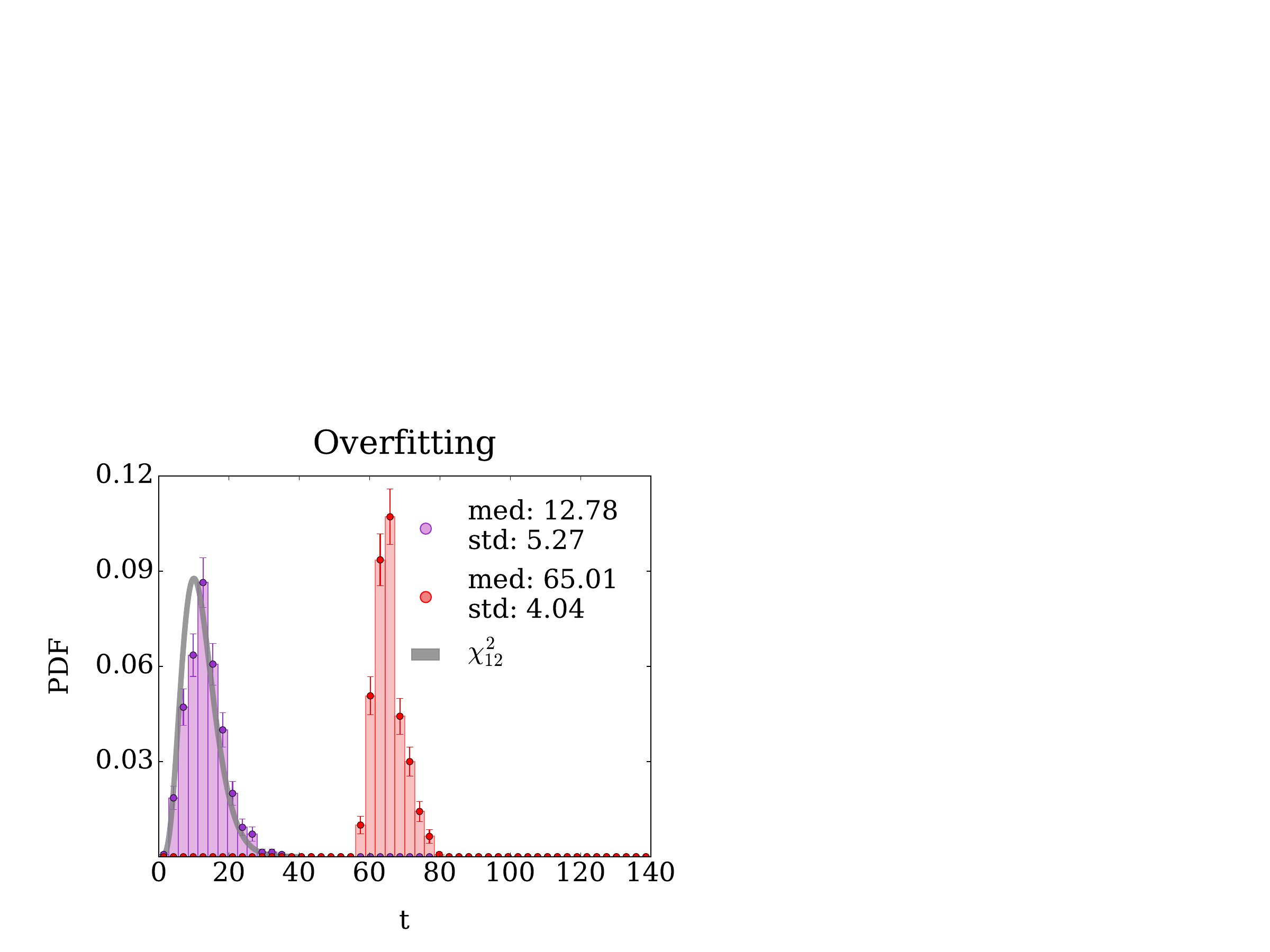}
    \caption{The background-only \ac{PDF} of $t_{\mathbf{A+B}}\leri{\mathbf{A}+\mathbf{B}}$ for an exponential distribution. Purple -- training for $1.5$~M epochs, red -- the ``overfit" solutions: setting $x_{\mathbf A}^{\star}$ ($x_{\mathbf B}^{\star}$) to capture the two longest sequences of points in $\mathbf{A}\setminus\mathbf{B}$ ($\mathbf{B}\setminus\mathbf{A}$). See text for more details.}
    \label{fig:overfit}
\end{figure}

Recall that the family of fitting functions can be written as
\begin{align}
    f\leri{x} = b_{\rm out}+\sum_{\alpha = 1}^{N_\text{neu}} w^\alpha_{\rm out} \sigma\leri{w_\alpha x +b_\alpha}\,,\label{eq:f2NN}
\end{align}
where $\sigma\leri{z} = 1/\leri{1+e^{-z}}$ is the logistic sigmoid function. For sufficiently large values of $w_\alpha$\,, the sigmoid approaches a step function, with a gradient (very) roughly set by $w^\alpha_{\rm out}w_\alpha$. A sum of two sigmoids, as shown in~\cite{D_Agnolo_2019} can produce a ``bump", which could be arbitrarily narrow, according to the weights. A sum of four sigmoids can produce at most two sinks or peaks. Let us observe the log-likelihoods of the alternative hypothesis in our parameterization
\begin{align}	t_{\mathbf{A}+\mathbf{B}}\leri{\mathbf{A}}&= -2\leri{\frac{1}{\tilde{N}_{\mathbf{B}}+\tilde{N}_{\mathbf{A}}}\sum_{x\in \mathbf{A},\mathbf{B}}\tilde{N}_\mathbf{A} \leri {e^{\hat{f}\leri{x}}-1}  -\sum_{x\in\mathbf{A}} \hat{f}\leri{x}}\,, \\
 t_{\mathbf{A}+\mathbf{B}}\leri{\mathbf{B}}&= -2\leri{\frac{1}{\tilde{N}_{\mathbf{B}}+\tilde{N}_{\mathbf{A}}}\sum_{x\in \mathbf{A},\mathbf{B}}\tilde{N}_\mathbf{B} \leri {e^{\hat{g}\leri{x}}-1}  -\sum_{x\in\mathbf{B}} \hat{g}\leri{x}}\,.~\label{eq:t_AD_A_D}
	\end{align}\,
If in $t_{\mathbf{A}+\mathbf{B}}\leri{\mathbf{A}}$ we set $e^{f\leri{x^\star_B}}\rightarrow 0$ for a total of $N^\star_\mathbf{B}$ points such that $x^\star_{\mathbf B}\in \leri{\mathbf{B}-\mathbf{B}\cap\mathbf{A}}$\,, and in $t_{\mathbf{A}+\mathbf{B}}\leri{\mathbf{B}}$ we set $e^{g\leri{x^\star_{\mathbf A}}}\rightarrow 0$ for a total of $N^\star_\mathbf{A}$ points such that $x^\star_{\mathbf{A}}\in \leri{\mathbf{A}-\mathbf{B}\cap\mathbf{A}}$\,, and keep $g$ and $f$ constant for all other points, we obtain
\begin{align}
    f^\star&=\log\leri{\frac{N_{\mathbf{A}}\leri{N_{\mathbf{A}}+N_{\mathbf{B}}}}{N_{\mathbf{A}}+N_{\mathbf{B}}-N^\star_\mathbf{B}}}\,,\\
    g^\star&=\log\leri{\frac{N_{\mathbf{B}}\leri{N_{\mathbf{A}}+N_{\mathbf{B}}}}{N_{\mathbf{A}}+N_{\mathbf{B}}-N^\star_\mathbf{A}}}\,,
\end{align}
and thus
\begin{align}	
t^\star_{\mathbf{A}+\mathbf{B}}\leri{\mathbf{A}}+t^\star_{\mathbf{A}+\mathbf{B}}\leri{\mathbf{B}}&= -2\leri{-N_{\mathbf A}\log\leri{1+\frac{N^\star_\mathbf{B}}{N_{\mathbf A}+N_{\mathbf B}-N^\star_\mathbf{B}}}-N_{\mathbf{B}}\log\leri{1+\frac{N^\star_\mathbf{A}}{N_{\mathbf{A}}+N_{\mathbf{B}}-N^\star_\mathbf{A}}}} \nonumber\\
&\approx 2\leri{\frac{N_{\mathbf A} N^\star_\mathbf{B}+N_{\mathbf B}N^\star_\mathbf{A}}{N_{\mathbf A}+N_{\mathbf B}} }\,,
 \end{align}
where in the last step we assumed $N^\star_\mathbf{B},N^\star_\mathbf{A}\ll N_\mathbf{B},N_\mathbf{A}$. For $N_\mathbf{B}\approx N_\mathbf{A}$, this yields a score that is approximately equal to the number of total points that may be removed using our fitting functions. For four sigmoids, we may then find the two longest sequences (``runs") of points of type $x^\star_{\mathbf{B}}$ and $x^\star_{\mathbf{A}}$, and remove them with a sufficiently negative value of $f$ and $g$. Note that the expected length of such sequence follows $\log_{N_{\mathbf{A}}/\leri{N_{\mathbf{A}}+N_{\mathbf{B}}}}\leri{N_{\mathbf{A}}}\approx \log_2{N_{\mathbf{A}}}-1$\,. In Fig.~\ref{fig:overfit}, we show the expected overfit score distribution for $N_{\mathbf{A}}=10^{5}$, centered at around $\sim70$\,, very far from the expected $12$ for a $\chi^2_{12}$ distribution.

\bibliographystyle{apsrev4-2}
\bibliography{references}

\begin{thebibliography}{24}%
\makeatletter
\providecommand \@ifxundefined [1]{%
 \@ifx{#1\undefined}
}%
\providecommand \@ifnum [1]{%
 \ifnum #1\expandafter \@firstoftwo
 \else \expandafter \@secondoftwo
 \fi
}%
\providecommand \@ifx [1]{%
 \ifx #1\expandafter \@firstoftwo
 \else \expandafter \@secondoftwo
 \fi
}%
\providecommand \natexlab [1]{#1}%
\providecommand \enquote  [1]{``#1''}%
\providecommand \bibnamefont  [1]{#1}%
\providecommand \bibfnamefont [1]{#1}%
\providecommand \citenamefont [1]{#1}%
\providecommand \href@noop [0]{\@secondoftwo}%
\providecommand \href [0]{\begingroup \@sanitize@url \@href}%
\providecommand \@href[1]{\@@startlink{#1}\@@href}%
\providecommand \@@href[1]{\endgroup#1\@@endlink}%
\providecommand \@sanitize@url [0]{\catcode `\\12\catcode `\$12\catcode
  `\&12\catcode `\#12\catcode `\^12\catcode `\_12\catcode `\%12\relax}%
\providecommand \@@startlink[1]{}%
\providecommand \@@endlink[0]{}%
\providecommand \url  [0]{\begingroup\@sanitize@url \@url }%
\providecommand \@url [1]{\endgroup\@href {#1}{\urlprefix }}%
\providecommand \urlprefix  [0]{URL }%
\providecommand \Eprint [0]{\href }%
\providecommand \doibase [0]{https://doi.org/}%
\providecommand \selectlanguage [0]{\@gobble}%
\providecommand \bibinfo  [0]{\@secondoftwo}%
\providecommand \bibfield  [0]{\@secondoftwo}%
\providecommand \translation [1]{[#1]}%
\providecommand \BibitemOpen [0]{}%
\providecommand \bibitemStop [0]{}%
\providecommand \bibitemNoStop [0]{.\EOS\space}%
\providecommand \EOS [0]{\spacefactor3000\relax}%
\providecommand \BibitemShut  [1]{\csname bibitem#1\endcsname}%
\let\auto@bib@innerbib\@empty
\bibitem [{\citenamefont {Volkovich}\ \emph {et~al.}(2022)\citenamefont
  {Volkovich}, \citenamefont {De~Vito~Halevy},\ and\ \citenamefont
  {Bressler}}]{Volkovich:2021txe}%
  \BibitemOpen
  \bibfield  {author} {\bibinfo {author} {\bibfnamefont {S.}~\bibnamefont
  {Volkovich}}, \bibinfo {author} {\bibfnamefont {F.}~\bibnamefont
  {De~Vito~Halevy}},\ and\ \bibinfo {author} {\bibfnamefont {S.}~\bibnamefont
  {Bressler}},\ }\href {https://doi.org/10.1140/epjc/s10052-022-10215-1}
  {\bibfield  {journal} {\bibinfo  {journal} {Eur. Phys. J. C}\ }\textbf
  {\bibinfo {volume} {82}},\ \bibinfo {pages} {265} (\bibinfo {year} {2022})},\
  \Eprint {https://arxiv.org/abs/2107.11573} {arXiv:2107.11573 [hep-ex]}
  \BibitemShut {NoStop}%
\bibitem [{\citenamefont {D'Agnolo}\ and\ \citenamefont
  {Wulzer}(2019)}]{D_Agnolo_2019}%
  \BibitemOpen
  \bibfield  {author} {\bibinfo {author} {\bibfnamefont {R.~T.}\ \bibnamefont
  {D'Agnolo}}\ and\ \bibinfo {author} {\bibfnamefont {A.}~\bibnamefont
  {Wulzer}},\ }\bibfield  {journal} {\bibinfo  {journal} {Physical Review D}\
  }\textbf {\bibinfo {volume} {99}},\ \href
  {https://doi.org/10.1103/physrevd.99.015014} {10.1103/physrevd.99.015014}
  (\bibinfo {year} {2019})\BibitemShut {NoStop}%
\bibitem [{\citenamefont {Birman}\ \emph {et~al.}(2022)\citenamefont {Birman},
  \citenamefont {Nachman}, \citenamefont {Sebbah}, \citenamefont {Sela},
  \citenamefont {Turetz},\ and\ \citenamefont {Bressler}}]{Birman:2022xzu}%
  \BibitemOpen
  \bibfield  {author} {\bibinfo {author} {\bibfnamefont {M.}~\bibnamefont
  {Birman}}, \bibinfo {author} {\bibfnamefont {B.}~\bibnamefont {Nachman}},
  \bibinfo {author} {\bibfnamefont {R.}~\bibnamefont {Sebbah}}, \bibinfo
  {author} {\bibfnamefont {G.}~\bibnamefont {Sela}}, \bibinfo {author}
  {\bibfnamefont {O.}~\bibnamefont {Turetz}},\ and\ \bibinfo {author}
  {\bibfnamefont {S.}~\bibnamefont {Bressler}},\ }\href
  {https://doi.org/10.1140/epjc/s10052-022-10454-2} {\bibfield  {journal}
  {\bibinfo  {journal} {Eur. Phys. J. C}\ }\textbf {\bibinfo {volume} {82}},\
  \bibinfo {pages} {508} (\bibinfo {year} {2022})},\ \Eprint
  {https://arxiv.org/abs/2203.07529} {arXiv:2203.07529 [hep-ph]} \BibitemShut
  {NoStop}%
\bibitem [{\citenamefont {Bressler}\ \emph {et~al.}(2014)\citenamefont
  {Bressler}, \citenamefont {Dery},\ and\ \citenamefont
  {Efrati}}]{PhysRevD.90.015025}%
  \BibitemOpen
  \bibfield  {author} {\bibinfo {author} {\bibfnamefont {S.}~\bibnamefont
  {Bressler}}, \bibinfo {author} {\bibfnamefont {A.}~\bibnamefont {Dery}},\
  and\ \bibinfo {author} {\bibfnamefont {A.}~\bibnamefont {Efrati}},\ }\href
  {https://doi.org/10.1103/PhysRevD.90.015025} {\bibfield  {journal} {\bibinfo
  {journal} {Phys. Rev. D}\ }\textbf {\bibinfo {volume} {90}},\ \bibinfo
  {pages} {015025} (\bibinfo {year} {2014})}\BibitemShut {NoStop}%
\bibitem [{\citenamefont {Lester}\ and\ \citenamefont
  {Brunt}(2017)}]{Lester_2017}%
  \BibitemOpen
  \bibfield  {author} {\bibinfo {author} {\bibfnamefont {C.~G.}\ \bibnamefont
  {Lester}}\ and\ \bibinfo {author} {\bibfnamefont {B.~H.}\ \bibnamefont
  {Brunt}},\ }\bibfield  {journal} {\bibinfo  {journal} {Journal of High Energy
  Physics}\ }\textbf {\bibinfo {volume} {2017}},\ \href
  {https://doi.org/10.1007/jhep03(2017)149} {10.1007/jhep03(2017)149} (\bibinfo
  {year} {2017})\BibitemShut {NoStop}%
\bibitem [{\citenamefont {Workman}\ and\ \citenamefont
  {Others}(2022)}]{Workman:2022ynf}%
  \BibitemOpen
  \bibfield  {author} {\bibinfo {author} {\bibfnamefont {R.~L.}\ \bibnamefont
  {Workman}}\ and\ \bibinfo {author} {\bibnamefont {Others}} (\bibinfo
  {collaboration} {Particle Data Group}),\ }\href
  {https://doi.org/10.1093/ptep/ptac097} {\bibfield  {journal} {\bibinfo
  {journal} {PTEP}\ }\textbf {\bibinfo {volume} {2022}},\ \bibinfo {pages}
  {083C01} (\bibinfo {year} {2022})}\BibitemShut {NoStop}%
\bibitem [{\citenamefont {Aad}\ \emph {et~al.}(2017)\citenamefont {Aad} \emph
  {et~al.}}]{ATLAS:2016joj}%
  \BibitemOpen
  \bibfield  {author} {\bibinfo {author} {\bibfnamefont {G.}~\bibnamefont
  {Aad}} \emph {et~al.} (\bibinfo {collaboration} {ATLAS}),\ }\href
  {https://doi.org/10.1140/epjc/s10052-017-4624-0} {\bibfield  {journal}
  {\bibinfo  {journal} {Eur. Phys. J. C}\ }\textbf {\bibinfo {volume} {77}},\
  \bibinfo {pages} {70} (\bibinfo {year} {2017})},\ \Eprint
  {https://arxiv.org/abs/1604.07730} {arXiv:1604.07730 [hep-ex]} \BibitemShut
  {NoStop}%
\bibitem [{\citenamefont {Aad}\ \emph {et~al.}(2022)\citenamefont {Aad} \emph
  {et~al.}}]{ATLAS:2021tar}%
  \BibitemOpen
  \bibfield  {author} {\bibinfo {author} {\bibfnamefont {G.}~\bibnamefont
  {Aad}} \emph {et~al.} (\bibinfo {collaboration} {ATLAS}),\ }\href
  {https://doi.org/10.1016/j.physletb.2022.137106} {\bibfield  {journal}
  {\bibinfo  {journal} {Phys. Lett. B}\ }\textbf {\bibinfo {volume} {830}},\
  \bibinfo {pages} {137106} (\bibinfo {year} {2022})},\ \Eprint
  {https://arxiv.org/abs/2112.08090} {arXiv:2112.08090 [hep-ex]} \BibitemShut
  {NoStop}%
\bibitem [{\citenamefont {d'Agnolo}\ \emph {et~al.}(2022)\citenamefont
  {d'Agnolo}, \citenamefont {Grosso}, \citenamefont {Pierini}, \citenamefont
  {Wulzer},\ and\ \citenamefont {Zanetti}}]{dAgnolo:2021aun}%
  \BibitemOpen
  \bibfield  {author} {\bibinfo {author} {\bibfnamefont {R.~T.}\ \bibnamefont
  {d'Agnolo}}, \bibinfo {author} {\bibfnamefont {G.}~\bibnamefont {Grosso}},
  \bibinfo {author} {\bibfnamefont {M.}~\bibnamefont {Pierini}}, \bibinfo
  {author} {\bibfnamefont {A.}~\bibnamefont {Wulzer}},\ and\ \bibinfo {author}
  {\bibfnamefont {M.}~\bibnamefont {Zanetti}},\ }\href
  {https://doi.org/10.1140/epjc/s10052-022-10226-y} {\bibfield  {journal}
  {\bibinfo  {journal} {Eur. Phys. J. C}\ }\textbf {\bibinfo {volume} {82}},\
  \bibinfo {pages} {275} (\bibinfo {year} {2022})},\ \Eprint
  {https://arxiv.org/abs/2111.13633} {arXiv:2111.13633 [hep-ph]} \BibitemShut
  {NoStop}%
\bibitem [{\citenamefont {Karagiorgi}\ \emph {et~al.}(2022)\citenamefont
  {Karagiorgi}, \citenamefont {Kasieczka}, \citenamefont {Kravitz},
  \citenamefont {Nachman},\ and\ \citenamefont {Shih}}]{Karagiorgi:2022qnh}%
  \BibitemOpen
  \bibfield  {author} {\bibinfo {author} {\bibfnamefont {G.}~\bibnamefont
  {Karagiorgi}}, \bibinfo {author} {\bibfnamefont {G.}~\bibnamefont
  {Kasieczka}}, \bibinfo {author} {\bibfnamefont {S.}~\bibnamefont {Kravitz}},
  \bibinfo {author} {\bibfnamefont {B.}~\bibnamefont {Nachman}},\ and\ \bibinfo
  {author} {\bibfnamefont {D.}~\bibnamefont {Shih}},\ }\href
  {https://doi.org/10.1038/s42254-022-00455-1} {\bibfield  {journal} {\bibinfo
  {journal} {Nature Rev. Phys.}\ }\textbf {\bibinfo {volume} {4}},\ \bibinfo
  {pages} {399} (\bibinfo {year} {2022})}\BibitemShut {NoStop}%
\bibitem [{\citenamefont {Wilks}(1938)}]{Wilks:1938dza}%
  \BibitemOpen
  \bibfield  {author} {\bibinfo {author} {\bibfnamefont {S.~S.}\ \bibnamefont
  {Wilks}},\ }\href {https://doi.org/10.1214/aoms/1177732360} {\bibfield
  {journal} {\bibinfo  {journal} {Annals Math. Statist.}\ }\textbf {\bibinfo
  {volume} {9}},\ \bibinfo {pages} {60} (\bibinfo {year} {1938})}\BibitemShut
  {NoStop}%
\bibitem [{\citenamefont {Wald}(1943)}]{10.2307/1990256}%
  \BibitemOpen
  \bibfield  {author} {\bibinfo {author} {\bibfnamefont {A.}~\bibnamefont
  {Wald}},\ }\href {http://www.jstor.org/stable/1990256} {\bibfield  {journal}
  {\bibinfo  {journal} {Transactions of the American Mathematical Society}\
  }\textbf {\bibinfo {volume} {54}},\ \bibinfo {pages} {426} (\bibinfo {year}
  {1943})}\BibitemShut {NoStop}%
\bibitem [{\citenamefont {D'Agnolo}\ \emph {et~al.}(2021)\citenamefont
  {D'Agnolo}, \citenamefont {Grosso}, \citenamefont {Pierini}, \citenamefont
  {Wulzer},\ and\ \citenamefont {Zanetti}}]{DAgnolo:2019vbw}%
  \BibitemOpen
  \bibfield  {author} {\bibinfo {author} {\bibfnamefont {R.~T.}\ \bibnamefont
  {D'Agnolo}}, \bibinfo {author} {\bibfnamefont {G.}~\bibnamefont {Grosso}},
  \bibinfo {author} {\bibfnamefont {M.}~\bibnamefont {Pierini}}, \bibinfo
  {author} {\bibfnamefont {A.}~\bibnamefont {Wulzer}},\ and\ \bibinfo {author}
  {\bibfnamefont {M.}~\bibnamefont {Zanetti}},\ }\href
  {https://doi.org/10.1140/epjc/s10052-021-08853-y} {\bibfield  {journal}
  {\bibinfo  {journal} {Eur. Phys. J. C}\ }\textbf {\bibinfo {volume} {81}},\
  \bibinfo {pages} {89} (\bibinfo {year} {2021})},\ \Eprint
  {https://arxiv.org/abs/1912.12155} {arXiv:1912.12155 [hep-ph]} \BibitemShut
  {NoStop}%
\bibitem [{\citenamefont {Letizia}\ \emph {et~al.}(2022)\citenamefont
  {Letizia}, \citenamefont {Losapio}, \citenamefont {Rando}, \citenamefont
  {Grosso}, \citenamefont {Wulzer}, \citenamefont {Pierini}, \citenamefont
  {Zanetti},\ and\ \citenamefont {Rosasco}}]{Letizia:2022xbe}%
  \BibitemOpen
  \bibfield  {author} {\bibinfo {author} {\bibfnamefont {M.}~\bibnamefont
  {Letizia}}, \bibinfo {author} {\bibfnamefont {G.}~\bibnamefont {Losapio}},
  \bibinfo {author} {\bibfnamefont {M.}~\bibnamefont {Rando}}, \bibinfo
  {author} {\bibfnamefont {G.}~\bibnamefont {Grosso}}, \bibinfo {author}
  {\bibfnamefont {A.}~\bibnamefont {Wulzer}}, \bibinfo {author} {\bibfnamefont
  {M.}~\bibnamefont {Pierini}}, \bibinfo {author} {\bibfnamefont
  {M.}~\bibnamefont {Zanetti}},\ and\ \bibinfo {author} {\bibfnamefont
  {L.}~\bibnamefont {Rosasco}},\ }\href
  {https://doi.org/10.1140/epjc/s10052-022-10830-y} {\bibfield  {journal}
  {\bibinfo  {journal} {Eur. Phys. J. C}\ }\textbf {\bibinfo {volume} {82}},\
  \bibinfo {pages} {879} (\bibinfo {year} {2022})},\ \Eprint
  {https://arxiv.org/abs/2204.02317} {arXiv:2204.02317 [hep-ph]} \BibitemShut
  {NoStop}%
\bibitem [{\citenamefont {Nachman}\ and\ \citenamefont
  {Thaler}(2021)}]{Nachman:2021yvi}%
  \BibitemOpen
  \bibfield  {author} {\bibinfo {author} {\bibfnamefont {B.}~\bibnamefont
  {Nachman}}\ and\ \bibinfo {author} {\bibfnamefont {J.}~\bibnamefont
  {Thaler}},\ }\href {https://doi.org/10.1103/PhysRevD.103.116013} {\bibfield
  {journal} {\bibinfo  {journal} {Phys. Rev. D}\ }\textbf {\bibinfo {volume}
  {103}},\ \bibinfo {pages} {116013} (\bibinfo {year} {2021})},\ \Eprint
  {https://arxiv.org/abs/2101.07263} {arXiv:2101.07263 [physics.data-an]}
  \BibitemShut {NoStop}%
\bibitem [{\citenamefont {Grosso}\ \emph {et~al.}(2023)\citenamefont {Grosso},
  \citenamefont {Lai}, \citenamefont {Letizia}, \citenamefont {Pazzini},
  \citenamefont {Rando}, \citenamefont {Rosasco}, \citenamefont {Wulzer},\ and\
  \citenamefont {Zanetti}}]{Grosso:2023ltd}%
  \BibitemOpen
  \bibfield  {author} {\bibinfo {author} {\bibfnamefont {G.}~\bibnamefont
  {Grosso}}, \bibinfo {author} {\bibfnamefont {N.}~\bibnamefont {Lai}},
  \bibinfo {author} {\bibfnamefont {M.}~\bibnamefont {Letizia}}, \bibinfo
  {author} {\bibfnamefont {J.}~\bibnamefont {Pazzini}}, \bibinfo {author}
  {\bibfnamefont {M.}~\bibnamefont {Rando}}, \bibinfo {author} {\bibfnamefont
  {L.}~\bibnamefont {Rosasco}}, \bibinfo {author} {\bibfnamefont
  {A.}~\bibnamefont {Wulzer}},\ and\ \bibinfo {author} {\bibfnamefont
  {M.}~\bibnamefont {Zanetti}},\ }\href
  {https://doi.org/10.1088/2632-2153/acebb7} {\bibfield  {journal} {\bibinfo
  {journal} {Mach. Learn. Sci. Tech.}\ }\textbf {\bibinfo {volume} {4}},\
  \bibinfo {pages} {035029} (\bibinfo {year} {2023})},\ \Eprint
  {https://arxiv.org/abs/2303.05413} {arXiv:2303.05413 [hep-ex]} \BibitemShut
  {NoStop}%
\bibitem [{\citenamefont {Alwall}\ \emph {et~al.}(2014)\citenamefont {Alwall},
  \citenamefont {Frederix}, \citenamefont {Frixione}, \citenamefont {Hirschi},
  \citenamefont {Maltoni}, \citenamefont {Mattelaer}, \citenamefont {Shao},
  \citenamefont {Stelzer}, \citenamefont {Torrielli},\ and\ \citenamefont
  {Zaro}}]{Alwall:2014hca}%
  \BibitemOpen
  \bibfield  {author} {\bibinfo {author} {\bibfnamefont {J.}~\bibnamefont
  {Alwall}}, \bibinfo {author} {\bibfnamefont {R.}~\bibnamefont {Frederix}},
  \bibinfo {author} {\bibfnamefont {S.}~\bibnamefont {Frixione}}, \bibinfo
  {author} {\bibfnamefont {V.}~\bibnamefont {Hirschi}}, \bibinfo {author}
  {\bibfnamefont {F.}~\bibnamefont {Maltoni}}, \bibinfo {author} {\bibfnamefont
  {O.}~\bibnamefont {Mattelaer}}, \bibinfo {author} {\bibfnamefont {H.~S.}\
  \bibnamefont {Shao}}, \bibinfo {author} {\bibfnamefont {T.}~\bibnamefont
  {Stelzer}}, \bibinfo {author} {\bibfnamefont {P.}~\bibnamefont {Torrielli}},\
  and\ \bibinfo {author} {\bibfnamefont {M.}~\bibnamefont {Zaro}},\ }\href
  {https://doi.org/10.1007/JHEP07(2014)079} {\bibfield  {journal} {\bibinfo
  {journal} {JHEP}\ }\textbf {\bibinfo {volume} {07}},\ \bibinfo {pages}
  {079}},\ \Eprint {https://arxiv.org/abs/1405.0301} {arXiv:1405.0301 [hep-ph]}
  \BibitemShut {NoStop}%
\bibitem [{\citenamefont {Sj\"ostrand}\ \emph {et~al.}(2015)\citenamefont
  {Sj\"ostrand}, \citenamefont {Ask}, \citenamefont {Christiansen},
  \citenamefont {Corke}, \citenamefont {Desai}, \citenamefont {Ilten},
  \citenamefont {Mrenna}, \citenamefont {Prestel}, \citenamefont {Rasmussen},\
  and\ \citenamefont {Skands}}]{Sjostrand:2014zea}%
  \BibitemOpen
  \bibfield  {author} {\bibinfo {author} {\bibfnamefont {T.}~\bibnamefont
  {Sj\"ostrand}}, \bibinfo {author} {\bibfnamefont {S.}~\bibnamefont {Ask}},
  \bibinfo {author} {\bibfnamefont {J.~R.}\ \bibnamefont {Christiansen}},
  \bibinfo {author} {\bibfnamefont {R.}~\bibnamefont {Corke}}, \bibinfo
  {author} {\bibfnamefont {N.}~\bibnamefont {Desai}}, \bibinfo {author}
  {\bibfnamefont {P.}~\bibnamefont {Ilten}}, \bibinfo {author} {\bibfnamefont
  {S.}~\bibnamefont {Mrenna}}, \bibinfo {author} {\bibfnamefont
  {S.}~\bibnamefont {Prestel}}, \bibinfo {author} {\bibfnamefont {C.~O.}\
  \bibnamefont {Rasmussen}},\ and\ \bibinfo {author} {\bibfnamefont {P.~Z.}\
  \bibnamefont {Skands}},\ }\href {https://doi.org/10.1016/j.cpc.2015.01.024}
  {\bibfield  {journal} {\bibinfo  {journal} {Comput. Phys. Commun.}\ }\textbf
  {\bibinfo {volume} {191}},\ \bibinfo {pages} {159} (\bibinfo {year}
  {2015})},\ \Eprint {https://arxiv.org/abs/1410.3012} {arXiv:1410.3012
  [hep-ph]} \BibitemShut {NoStop}%
\bibitem [{\citenamefont {de~Favereau}\ \emph {et~al.}(2014)\citenamefont
  {de~Favereau}, \citenamefont {Delaere}, \citenamefont {Demin}, \citenamefont
  {Giammanco}, \citenamefont {Lema\^\i{}tre}, \citenamefont {Mertens},\ and\
  \citenamefont {Selvaggi}}]{deFavereau:2013fsa}%
  \BibitemOpen
  \bibfield  {author} {\bibinfo {author} {\bibfnamefont {J.}~\bibnamefont
  {de~Favereau}}, \bibinfo {author} {\bibfnamefont {C.}~\bibnamefont
  {Delaere}}, \bibinfo {author} {\bibfnamefont {P.}~\bibnamefont {Demin}},
  \bibinfo {author} {\bibfnamefont {A.}~\bibnamefont {Giammanco}}, \bibinfo
  {author} {\bibfnamefont {V.}~\bibnamefont {Lema\^\i{}tre}}, \bibinfo {author}
  {\bibfnamefont {A.}~\bibnamefont {Mertens}},\ and\ \bibinfo {author}
  {\bibfnamefont {M.}~\bibnamefont {Selvaggi}} (\bibinfo {collaboration}
  {DELPHES 3}),\ }\href {https://doi.org/10.1007/JHEP02(2014)057} {\bibfield
  {journal} {\bibinfo  {journal} {JHEP}\ }\textbf {\bibinfo {volume} {02}},\
  \bibinfo {pages} {057}},\ \Eprint {https://arxiv.org/abs/1307.6346}
  {arXiv:1307.6346 [hep-ex]} \BibitemShut {NoStop}%
\bibitem [{\citenamefont {Aad}\ \emph {et~al.}(2023)\citenamefont {Aad} \emph
  {et~al.}}]{ATLAS:2023mvd}%
  \BibitemOpen
  \bibfield  {author} {\bibinfo {author} {\bibfnamefont {G.}~\bibnamefont
  {Aad}} \emph {et~al.} (\bibinfo {collaboration} {ATLAS}),\ }\href
  {https://doi.org/10.1007/JHEP07(2023)166} {\bibfield  {journal} {\bibinfo
  {journal} {JHEP}\ }\textbf {\bibinfo {volume} {07}},\ \bibinfo {pages}
  {166}},\ \Eprint {https://arxiv.org/abs/2302.05225} {arXiv:2302.05225
  [hep-ex]} \BibitemShut {NoStop}%
\bibitem [{\citenamefont {Efron}(1979)}]{10.1214/aos/1176344552}%
  \BibitemOpen
  \bibfield  {author} {\bibinfo {author} {\bibfnamefont {B.}~\bibnamefont
  {Efron}},\ }\href {https://doi.org/10.1214/aos/1176344552} {\bibfield
  {journal} {\bibinfo  {journal} {The Annals of Statistics}\ }\textbf {\bibinfo
  {volume} {7}},\ \bibinfo {pages} {1 } (\bibinfo {year} {1979})}\BibitemShut
  {NoStop}%
\bibitem [{\citenamefont {Kingma}\ and\ \citenamefont
  {Ba}(2017)}]{kingma2017adam}%
  \BibitemOpen
  \bibfield  {author} {\bibinfo {author} {\bibfnamefont {D.~P.}\ \bibnamefont
  {Kingma}}\ and\ \bibinfo {author} {\bibfnamefont {J.}~\bibnamefont {Ba}},\
  }\href@noop {} {\bibinfo {title} {Adam: A method for stochastic
  optimization}} (\bibinfo {year} {2017}),\ \Eprint
  {https://arxiv.org/abs/1412.6980} {arXiv:1412.6980 [cs.LG]} \BibitemShut
  {NoStop}%
\bibitem [{\citenamefont {Grosso}(2021)}]{Grosso_New_Physics_Learning_2021}%
  \BibitemOpen
  \bibfield  {author} {\bibinfo {author} {\bibfnamefont {G.}~\bibnamefont
  {Grosso}},\ }\href {https://github.com/GaiaGrosso/NPLM_package} {\bibinfo
  {title} {{New Physics Learning Machine (NPLM): package}}} (\bibinfo {year}
  {2021})\BibitemShut {NoStop}%
\bibitem [{\citenamefont {Sirunyan}\ \emph {et~al.}(2021)\citenamefont
  {Sirunyan} \emph {et~al.}}]{CMS:2021rsq}%
  \BibitemOpen
  \bibfield  {author} {\bibinfo {author} {\bibfnamefont {A.~M.}\ \bibnamefont
  {Sirunyan}} \emph {et~al.} (\bibinfo {collaboration} {CMS}),\ }\href
  {https://doi.org/10.1103/PhysRevD.104.032013} {\bibfield  {journal} {\bibinfo
   {journal} {Phys. Rev. D}\ }\textbf {\bibinfo {volume} {104}},\ \bibinfo
  {pages} {032013} (\bibinfo {year} {2021})},\ \Eprint
  {https://arxiv.org/abs/2105.03007} {arXiv:2105.03007 [hep-ex]} \BibitemShut
  {NoStop}%
\end{thebibliography}%

\end{document}